\documentclass[superscriptaddress,prx, longbibliography]{revtex4-2}
\usepackage[utf8]{inputenc}
\usepackage{physics}
\usepackage{amsmath,amsfonts}
\usepackage{indentfirst}
\usepackage{array}
\usepackage{amsthm}
\usepackage{amssymb}
\usepackage{comment}
\usepackage{geometry}
\usepackage{float}
\usepackage{makecell,xcolor}

\newcommand{\mh}[1]{\textcolor{red}{#1}}
\newcommand{\yu}[1]{\textcolor{blue}{#1}}

\usepackage{chemformula} %% $\ch{H2O}$ etc
\usepackage{qcircuit}
\usepackage{subcaption}

\begin{document}
\title{Quantum Simulation of Preferred Tautomeric State Prediction}

\author{Yu Shee}
\thanks{These authors contributed equally to this work.}
\affiliation{Bleximo Corp., Berkeley, California 94720, USA}
\affiliation{Department of Chemistry, Yale University, New Haven, Connecticut 06511, USA}
\author{Tzu-Lan Yeh}
\thanks{These authors contributed equally to this work.}
\affiliation{Insilico Medicine Taiwan Ltd., Taipei City 110208, Taiwan}
\author{Jen-Yueh Hsiao}
\affiliation{Hon Hai (Foxconn) Research Institute, Taipei City 114, Taiwan}
\author{Ann Yang}
\affiliation{Insilico Medicine Taiwan Ltd., Taipei City 110208, Taiwan}
\author{Yen-Chu Lin}
\thanks{Corresponding e-mail: jimmy.lin@insilicomedicine.com}
\affiliation{Insilico Medicine Taiwan Ltd., Taipei City 110208, Taiwan}
\affiliation{Department of Pharmacy, National Yang Ming Chiao Tung University, Taipei City 112304, Taiwan}
\author{Min-Hsiu Hsieh}
\thanks{Corresponding e-mail: min-hsiu.hsieh@foxocnn.com}
\affiliation{Hon Hai (Foxconn) Research Institute, Taipei City 114, Taiwan}

\date{\today}

\begin{abstract}
\begin{comment}
\yu{1. topic sentence and question underlying this research study.}
Simulating molecules is known to be computationally difficult where quantum devices are expected to bring advantages.
\yu{2. previous research and rationale for new research.}
Quantum simulations on quantum devices often lack real world applications under current quantum hardware limitations. We propose a hybrid quantum-classical workflow to predict the stability of tautomeric states and investigate its applicability on two tautomeric systems: acetone and Edaravone.
\yu{3. methods for this study.}
To realize the algorithm on noisy intermediate-scale quantum (NISQ) devices,  we resort to quantum chemistry and efficient encoding methods to reduce the qubit resources and circuit depth. We select active-space molecular orbitals based on quantum chemistry methods and map the Hamiltonian onto quantum devices qubit-efficiently. The variational quantum eigensolver (VQE) algorithm is then employed for ground state estimation where hardware-efficient ansatz circuits are used. 
\yu{4. findings and significance}
We run the numerical experiments on simulators where the tautomeric state prediction agree with our benchmarks. This study aims to provide a practical application for both current and future quantum devices.
\end{comment}

%\tly{Insilico's Abstract, perhaps we can combine}
Prediction of tautomers plays an essential role in computer-aided drug discovery. However, it remains a challenging task nowadays to accurately predict the canonical tautomeric form of a given drug-like molecule. Lack of extensive tautomer databases, most likely due to the difficulty in experimental studies, hampers the development of effective empirical methods for tautomer predictions. A more accurate estimation of the stable tautomeric form can be achieved by quantum chemistry calculations. Yet, the computational cost required prevents quantum chemistry calculation as a standard tool for tautomer prediction in computer-aided drug discovery. In this paper we propose a hybrid quantum chemistry-quantum computation workflow to efficiently predict the dominant tautomeric form. Specifically, we select active-space molecular orbitals based on quantum chemistry methods. Then we utilize efficient encoding methods to map the Hamiltonian onto quantum devices to reduce the qubit resources and circuit depth. Finally, variational quantum eigensolver (VQE) algorithms are employed for ground state estimation where hardware-efficient ansatz circuits are used. To demonstrate the applicability of our methodology, we perform  experiments on two tautomeric systems: acetone and Edaravone, each having 52 and 150 spin-orbitals in the STO-3G basis set, respectively. Our numerical results show that their tautomeric state prediction agrees with the CCSD benchmarks. Moreover, the required quantum resources are efficient: in the example of Edaravone, we could achieve chemical accuracy with only eight qubits and 80 two-qubit gates.  

%We are able to accurately predict the tautomeric states of The required quantum resources 
%We are able to accurate predict 
%\yu{4. findings and significance}

%We run the numerical experiments on simulators where the tautomeric state prediction agree with our benchmarks. %This study aims to provide a practical application for both current and future quantum devices.

\end{abstract}

\maketitle

\section{Introduction}

%\mh{@TLY: Define Tautomers/Tautomerism. Provide examples of significance.}
% Define Tautomerization
Tautomers are constitutional isomers that spontaneously convert to one another in dynamic equilibrium. The process of this interconversion is called tautomerization. Typical tautomerization involves the movement of a proton from one position to another and rearrangement of a double bond within the molecule. Other types of tautomerisms include annular, ring-chain, and valence tautomerisms \cite{Antonov2016-vb, Muller1994-xf, Alkorta2007-tn}. One well-known example of tautomerization, and quite often involved in the field of drug development, is the keto-enol tautomerism, in which the carbonyl double bond (keto form) is interconverted to an alkene double bond (enol form). This is accompanied by the shift of the alpha proton in the keto form to the hydroxyl group in the enol form. 

Tautomerization plays an important role in biological systems. Non-Watson-Crick base pairing can occur due to tautomerization of nucleic acid base pairs. Such nucleic acid mismatches induced by tautomerization result in spontaneous mutagenesis and hence genetic instability \cite{Wang2011-lf, Bebenek2011-as}. In the case of drug molecules, the movement of proton within the molecules will lead to the conversion from a hydrogen bond donor to a hydrogen bond acceptor or vice versa, which is essential for the analysis of structure-activity relationship (SAR). Moreover, it has been estimated that more than a quarter of marketed drugs can exhibit tautomerism \cite{Martin2009-yy}. Prediction of tautomeric states of compounds of interest is therefore an important subject in the field of computer aided drug design.

% Will elaborate more on the previous approaches for the tautomer ratio problem (TLY) %
State of the art algorithms for tautomeric prediction usually involve enumeration of possible tautomers, followed by prediction of the dominant form or estimation of the tautomer populations \cite{Warr2010-pz}. The latter is usually achieved using empirical rules in combination with scoring functions or relies on the estimation of pKa, etc. Such algorithms usually aim to provide a list of possible tautomeric forms with ratios of the corresponding species, since it is important to estimate the population of the tautomers due to the small differences in their free energies that could be easily compensated by the interaction with proteins. Lack of extensive data on free energies between tautomers and that in different solvent environments render a great challenge for accurate prediction by empirical methods. It is thus our hope that quantum chemical approaches could provide better accuracy and fill the gap.  

{Computational quantum chemistry algorithms are heavily used in drug design and discovery to simulate chemical reactions and properties. These quantum chemistry computational strategies can accelerate the drug discovery processes and reduce the costs for wet lab experiments. While quantum chemistry approaches have brought accuracy to the calculations, their computational cost  makes it not suitable as a daily used tool in computer-aided drug discovery (CADD). Instead, empirical or semi-empirical methods are used as standard tools. Emerging use of density functional calculations brings balance between accuracy and computation cost in the field. As such, density functional theory (DFT) methods are most widely used for the calculation of electron density of molecular systems.} However, their accuracy depends on the electronic structures of the molecular systems; hence, DFT is not universally applicable \cite{cohen_insights_2008}. Instead, wave function based techniques, such as the configuration interaction (CI) and coupled cluster (CC) methods, are used. The runtime of the full CI (FCI) method, which is equivalent to diagonalizing the Hamiltonian matrix, scales exponentially with respect to system sizes. The coupled cluster singles and doubles (CCSD) method, on the other hand, provides reasonable trade-offs between accuracy and runtime. Even though the aforementioned classical algorithms are well-designed, they still have unreasonable time and space requirements for medium and large molecules \cite{dykstra_theory_2011}. This poses severe limitation on the use of  quantum chemistry methods in drug discovery, e.g., prediction of dominant product in tautomerism, understanding the dynamics of protein folding, and ligand binding free energy calculations.

%\mh{@Yu Shee: Quantum simulation/hamiltonian simulation is the method to do it.}

Quantum simulation is considered one of the most promising applications of quantum technology since it has the potential of overcoming the exponential barrier of solving electronic structure problems \cite{feynman_simulating_1982}. Jordon-Wigner (JW) transformation \cite{jordan_1928} was found to give mappings from fermionic creation and annihilation operators to qubit operators \cite{somma_simulating_2002}. This result paved the road to generically simulate physical systems on well-controlled quantum computers. Moreover, the quantum phase estimation (QPE) algorithm \cite{kitaev_quantum_1995, du_nmr_2010, lanyon_towards_2010, li_solving_2011, omalley_scalable_2016, paesani_experimental_2017, santagati_witnessing_2018, wang_quantum_2015, abrams_quantum_1999, aspuru-guzik_simulated_2005} provides accurate spectral calculations for molecular Hamiltonian with the potential of exponential speedup. However, the QPE method is not applicable on current quantum devices because it requires long coherence time and high gate fidelity. Instead, several variational methods \cite{peruzzo_variational_2014, omalley_scalable_2016, mcclean_theory_2016, kandala_hardware-efficient_2017, kandala_error_2019, jiang_quantum_2018, kivlichan_quantum_2018, wecker_solving_2015, babbush_chemical_2015, sugisaki_quantum_2016, sugisaki_quantum_2019, PRXQuantum.3.030901} suitable for noisy intermediate-scale quantum (NISQ) devices were proposed. One of the most practical approaches is the variational quantum eigensolver (VQE) algorithm \cite{peruzzo_variational_2014, mcclean_theory_2016}. VQE is a hybrid quantum-classical method that approximates ground states of molecular Hamiltonian by variationally tuning the ansatz parameters and is expected to give better accuracy than CCSD results with polynomial costs.

Quantum simulation of small molecules has since then been widely implemented. Researchers have worked towards molecules (\ch{BeH2}, \ch{H2O}, and \ch{H12}) with improved algorithms and hardware design \cite{kandala_hardware-efficient_2017, google_ai_quantum_and_collaborators_hartree-fock_2020, nam_ground-state_2020}. Others also provide strategies to simulate slightly larger systems, such as \ch{CO2}, \ch{C2H4}, \ch{C_18}, and the nitrogenase iron-sulfur molecular clusters \cite{cao_towards_2021, li_toward_2021, tazhigulov_simulating_2022}, using symmetries, fragmentation of molecules, spin-model simplifications, or novel qubit encoding methods \cite{bravyi_tapering_2017, moll_optimizing_2016, babbush_exponentially_2018, steudtner_fermion--qubit_2018, kirby_second-quantized_2021, shee_qubit-efficient_2021}. 
%\mh{@Yu Shee: Highlight limitations of these outcomes to justify the research significance of our works.} 
However, these quantum simulation results are still quite limited in the problem size and often lack real world applications. 

In this work, we aim to design a general methodology for a pharmaceutical application; namely, the prediction of preferred tautomeric states. We expect  that our scheme could work for both current and future quantum devices. 
{Specifically, we presented a hybrid quantum chemistry-quantum computation approach to predict the dominant form of tautomers, where quantum chemistry methods are used to construct the system of interest and reduce the size of the Hamiltonian to meet current quantum hardware requirement, and the reduced Hamiltonian was mapped to qubits using qubit-efficient encoding (QEE) \cite{shee_qubit-efficient_2021} and subsequently simulated with variational quantum eigensolver (VQE). 
%While enumeration of possible tautomeric states is not within the scope of this work, 
We focus on the problem of tautomer ratio estimation and prediction of the preferred tautomeric state by considering the energetics (relative free energy) of the states using quantum chemistry approaches to make the system applicable on current quantum cumputing schemes. This approach was practiced on two illustrative examples, acetone and Edaravone. Our simulation results showed reasonable agreement with the calculations using CCSD.} 

\section{Results}

\begin{figure}[h]
\centering
\includegraphics[width=0.7\textwidth]{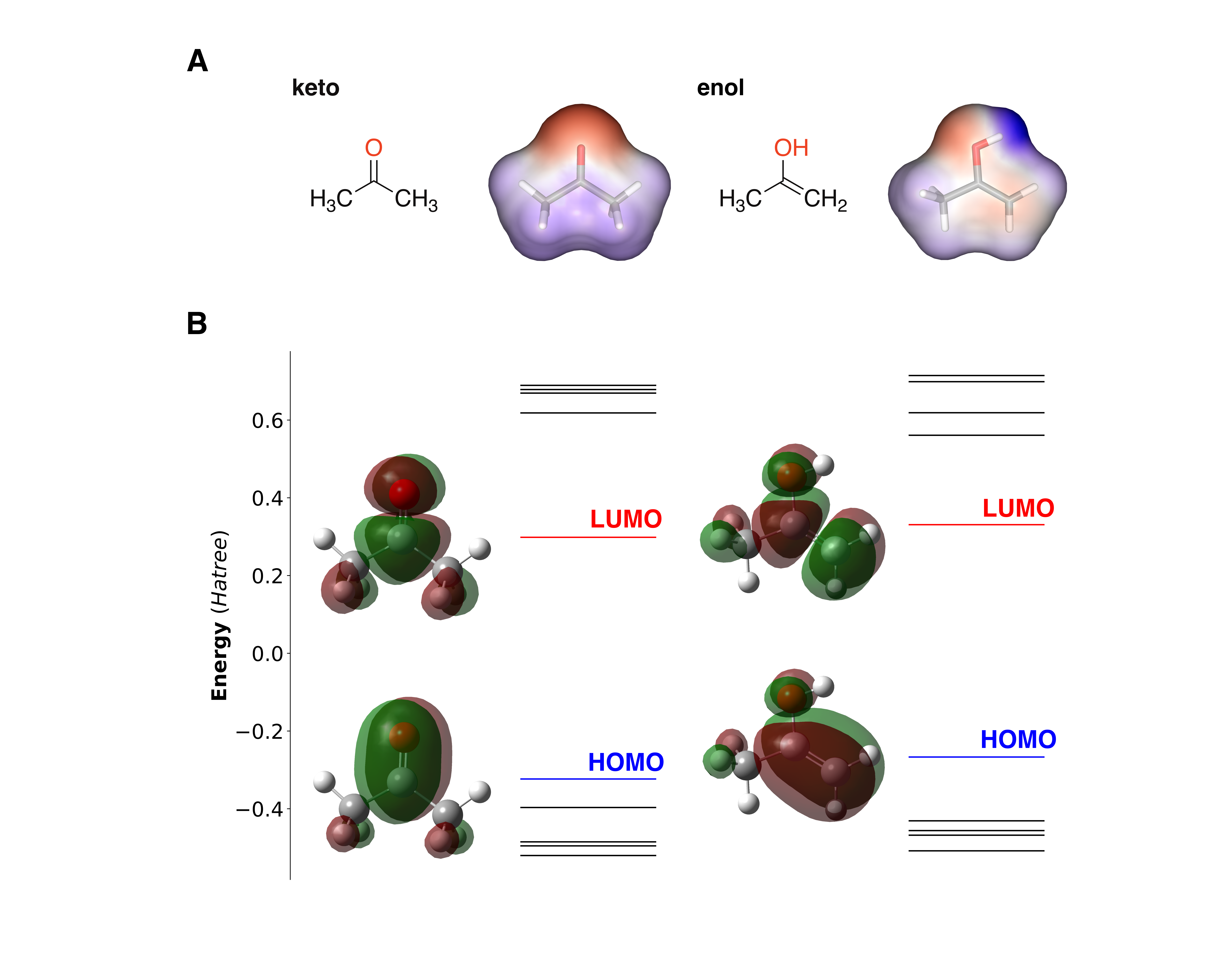}

\caption{(A) Tautomeric isomers of acetone. The geometries were optimized at the B3LYP/6-311++G(d,p) level and the electrostatic potential surface generated by ESP\_DNN\cite{Rathi2020-ki}. (B) Energy levels of molecular orbitals around HOMO/LUMO as calculated using MP2/STO-3g. The full list molecular orbital eignevalues can be found in Appendix~\ref{appendix:Molecular Orbital Occupancy}.}
    \label{fig:acetone_system}
\end{figure}

The most simple and intuitive way to estimate the relative stability of tautomers is perhaps by considering the difference in bond energy, since tautomerization usually involves rearrangement of double bonds. Take the well-known keto-enol tautomerism for example. The bond dissociation energy difference between C=O, C-C, C-H (keto form) and O-H, C-O, C=C (enol form) is 66 kJ/mol (15.8 kcal/mol), favoring the keto form. However, as the complexity of the molecules increases, the relative stability could not be attributed so easily. 
%For example, the tautomerism of Warfarin involves ring-chain tautomerism. Moreover, 
Factors like electrostatic effect, steric hindrance from the other parts of the molecule and intra-molecular hydrogen bonds need careful consideration as well. Commercial or publicly available CADD tools deal with the tautomerism problem using empirical or rule-based chemoinformatics methodologies, e.g., scoring tautomers based on the prediction of microstate and microstate pKa values. There is, however, still room for improvement with such an approach due to the lack of extensive databases and over-parameterization of the prediction models.

In principle, two important factors decide the tautomeric equilibrium. The relative potential energy difference, which determines the equilibrium direction, accounts for the stability between the isomers. On the other hand, the rate of isomer interconversion depends on the activation energy. As tautomerism involves bond breaking and bond formation, the best way to handle this problem should be quantum mechanics-based approaches with the system described by molecular orbitals. However, such first-principle calculations are impractical even with the computation resources nowadays. 
%Commercial or publicly available CADD tools deal with the tautomerism problem using empirical or rule-based chemoinformatics methodologies. For example, one commonly used approach to score tautomers involves the prediction of microstate and microstate pKa values. There are, however, still room for improvement with such approach due to the lack of extensive database and over-parameterization of the prediction models. 
With the advent of quantum computing technologies, theoretical quantum chemical approaches could now play an important and ‘applicable’ role for the tautomer prediction for computer aided drug discovery. 
%In this work, we aim to devise an ‘workable’ quantum mechanics based computational scheme with current quantum computing simulators to predict the relative stability of the tautomeric forms.

%\mh{YU: please provide a summarizing paragraph here.}
In the following, we first present the quantum chemistry overview of the two tautomeric systems: (1) acetone and its enol form, and (2) Edaravone's keto, enol, and amine forms in the STO-3G basis set. Next, we introduce current challenges of simulating medium-to-large molecules and our workflow that could contribute to resolving the problems. Besides, we provide the background of the qubit-efficient encoding methods for quantum simulation. Lastly, we show the numerical results for the implementation of our workflow with quantum simulations of the two tautomeric systems. To the best of our knowledge, these are the largest molecules in terms of the number of spin-orbitals (without the fragmentation of molecules using quantum embedding theory or the application of spin models from spectroscopic studies) for numerical simulation using a quantum algorithm so far.

\subsection*{Overview of Systems}

\textbf{Acetone.}
We start with a relatively simple example of tautomerization. Acetone exhibits keto-enol tautomerization, in which the acetone and its enol form, propen-2-ol, interconvert to each other (Fig.~\ref{fig:acetone_system}(A)). The relative stability of these two isomers can be easily assessed by comparing their bond energy, and the keto from is more stable than the enol form by about 16 kcal/mol. Fig.~\ref{fig:acetone_system}(A) shows the optimized geometries of the keto and enol forms of the acetone at the level of 6-311++G(d,p) using B3LYP in the Gaussian 16 program. To enable quantum computing of the system at a reasonable computational cost, we reconstituted the molecules with the minimal STO-3G at the second-order Møller–Plesset perturbation theory (MP2) level with 52 spin-orbitals. The corresponding energy diagrams of the molecular orbitals near HOMO are shown in Fig~\ref{fig:acetone_system}(B) for the keto and enol form, respectively. The keto form is more stable than the enol form in most cases because the C=O double bond is stronger than the C=C bond. Polarization of the C=O double bond, as depicted in electrostatic potential surface of the acetone, gives it a relatively higher bond energy. The rearrangement of the double bond between the keto and enol form is apparent from the shift of the $\pi$ bonding patterns of the HOMOs for the corresponding tautomers.

\begin{figure}[h]
\centering
\includegraphics[width=0.65\textwidth]{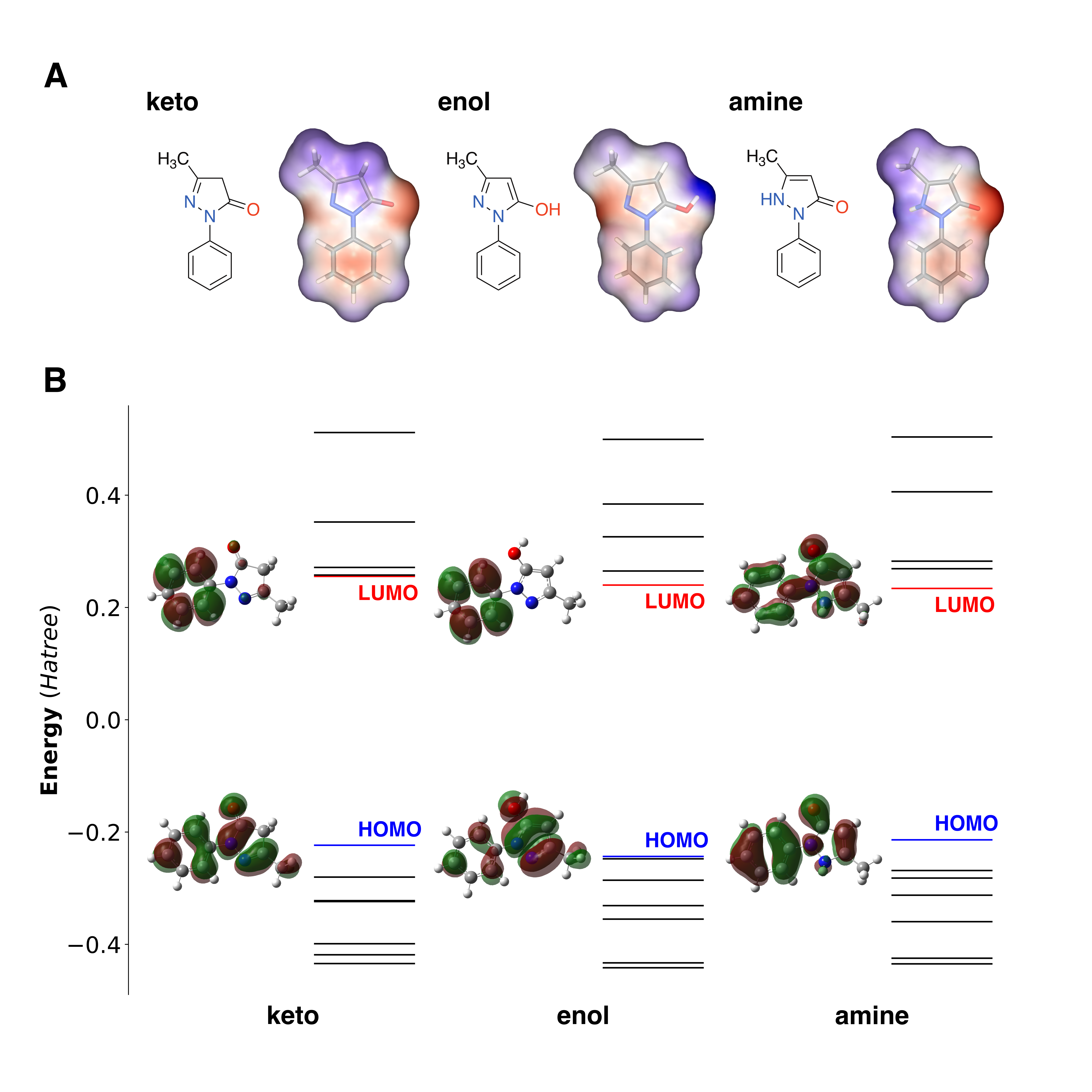}
\caption{(A)Tautomeric isomers of Edaravone. The geometries were optimized at the B3LYP/6-311++G(d,p) level and the electrostatic potential surface generated by ESP\_DNN \cite{Rathi2020-ki}. (B) Energy levels of molecular orbitals around HOMO/LUMO as calculated using MP2/STO-3g. The full list molecular orbital eignevalues can be found in Appendix~\ref{appendix:Molecular Orbital Occupancy}.}
    \label{fig:Edavarone_system}
\end{figure}

\textbf{Edaravone.}
Edaravone is an FDA-approved drug for the treatment of amyotrophic lateral sclerosis (ALS). The mechanism of the action of Edaravone involves its radical scavenging activity.  The idea originated from a research program in Mitsubishi Yuka Pharmaceutical Corporation, in which the scientists try to utilize the radical-scavenging activity of phenol and avoid related toxicity. The researchers  believed that an aromatic heterocyclic system with the potential of keto-enol tautomerization could exist in the form with a hydroxyl group \cite{Watanabe2018-lp}. Edaravone exists in three forms of tautomeric isomers – keto, enol, and amine (Fig.~\ref{fig:Edavarone_system}A), with varying physicochemical properties. Moreover, the three forms of Edaravone exhibit different antioxidant activity. Therefore, it is important to know which form among the three dominates to better understand the pharmacological effect of Edaravone. Estimation of relative stability among the three tautomers for Edaravone is not as trivial as for acetone, since the molecular structure of Edaravone is much more complicated. For example, the aromaticity of the \textit{N}-substituted pyrazolone core in the three isomers varies. Moreover the surface charge distribution of the three tautomers differs quite a lot and the twisted angles between the pyrazolone and the \textit{N}-substituted benzene are distinct among the three according to the optimized geometries of the forms (at the level of B3LYP/6-311++G(d,p) theory) as depicted in Fig~\ref{fig:Edavarone_system}(A). Reconstitution of the electronic configuration using the minimal STO-3g basis at MP2 level yields 150 molecular orbitals. Energy diagrams of the molecular orbitals near HOMO or the HOMOs and LUMOs of the three tautomeric species do not provide clear pictures for the relative stabilities of the three species as in the case of acetone (Fig.~\ref{fig:Edavarone_system}(B)).

Notice that it is already considered computationally expensive to construct the electronic structures of Edavarone even with classical quantum chemistry approaches. However, in the field of medicinal chemistry, there are more compounds with larger sizes. It is therefore not an efficient way to predict the thermodynamic stability of different tautomer states by comparing the energies of the isomers using quantum chemistry directly.

In order to leverage the power of quantum computation, we proceed as follows. 
Firstly, the second-quantized electronic Hamiltonian can be written as 
\begin{equation}
\label{eq:second_hamil_1}
  H
  =\sum_{pq}
  {h_{pq}a_p^\dag a_q}
  +\frac{1}{2}\sum_{pqrs}
  {h_{pqrs}a_p^\dag a_q^\dag a_ra_s},
\end{equation}
where $h_{pq}$ and $h_{pqrs}$ are the overlap and exchange integrals. The indices of the spin-orbitals are represented by $p$, $q$, $r$, and $s$ in the summation of Eq.~(\ref{eq:second_hamil_1}). For example, the indices run from 0 to 51 for acetone and from 0 to 149 for Edaravone because the molecules in the STO-3G basis set have 52 and 150 spin-orbitals respectively. To simulate electronic structure problems on quantum computers, the creation and annihilation operators and the electronic states have to be mapped to qubits. 

Common qubit encoding methods are Jordan-Wigner, parity, and Bravyi-Kitaev encoding where the qubit requirements are $\mathcal O(N)$ where $N$ is the number of spin-orbitals. For example, Jordan-Wigner (JW) encoding uses $N$ qubits to store the $N$ spin-orbital occupation number. In this encoding scheme, the $\ket{0}$ qubit state represents that the spin-orbital is not occupied while the $\ket{1}$ qubit state represents that the spin-orbital is occupied. The mappings of the creation and annihilation operators for JW are
\begin{align}
    a_p^\dag 
    & = \frac12 (X_p-iY_p) \otimes Z_{p-1}\otimes\cdots\otimes Z_0, \\
    a_p 
    & = \frac12 (X_p+iY_p) \otimes Z_{p-1}\otimes\cdots\otimes Z_0,
\end{align}
where the $Z$ Pauli strings address the fact that electronic wavefunctions are anti-symmetric. For acetone and Edaravone in the STO-3G basis set, JW encoding will map the electronic structure problems to 52 qubits and 150 qubits. No meaningful data would be collected for VQE using that many qubits on NISQ devices. It is, therefore, important to reduce the problem size.

%\textbf{{Current Challenge.}} 
%As discussed in the previous section, it is difficult to construct the electronic structures of Acetone (52 spin-orbitals) and Edaravone (150 spin-orbitals) even with a minimal basis set under the current quantum computing scheme. For the case of Edaravone, it is also considered computationally expensive even with classical quantum chemistry approaches. However, in the field of medicinal chemistry, there are more compounds with larger sizes. It is therefore not an efficient way to predict the thermodynamic stability of different tautomer states by comparing the energies of the isomers using quantum chemistry.

%One rational approach to reduce the size of the problem would be reducing the MO space. The selection of active space was achieved mostly by excluding the inner core and the outer virtual orbitals. \testcolor{red}{need to add more review/} In practice, the natural orbitals with occupancy closest to 0 (completely virtual) or 2 (fully occupied) were frozen or removed.

\subsection*{Workflow}

%Here we proposed a hybrid computational scheme for the prediction of the preferred tautomeric state, comprising classical quantum chemistry (using classical computers) and quantum computation.  The workflow involves (1) geometry optimization of the isomers using density functional theory, (2) reconstruct the molecular orbital space in a minimal basis set, (3) reduction of the MO space, (4) single point energy calculations of active MO sets, (5) select the most representative active MO set by comparison to the unreduced set, and finally (6) energy calculation in Quantum computing scheme using QEE, as illustrated in Fig (ref workflow).

\begin{figure}[h]
\centering
\includegraphics[width=0.6\textwidth]{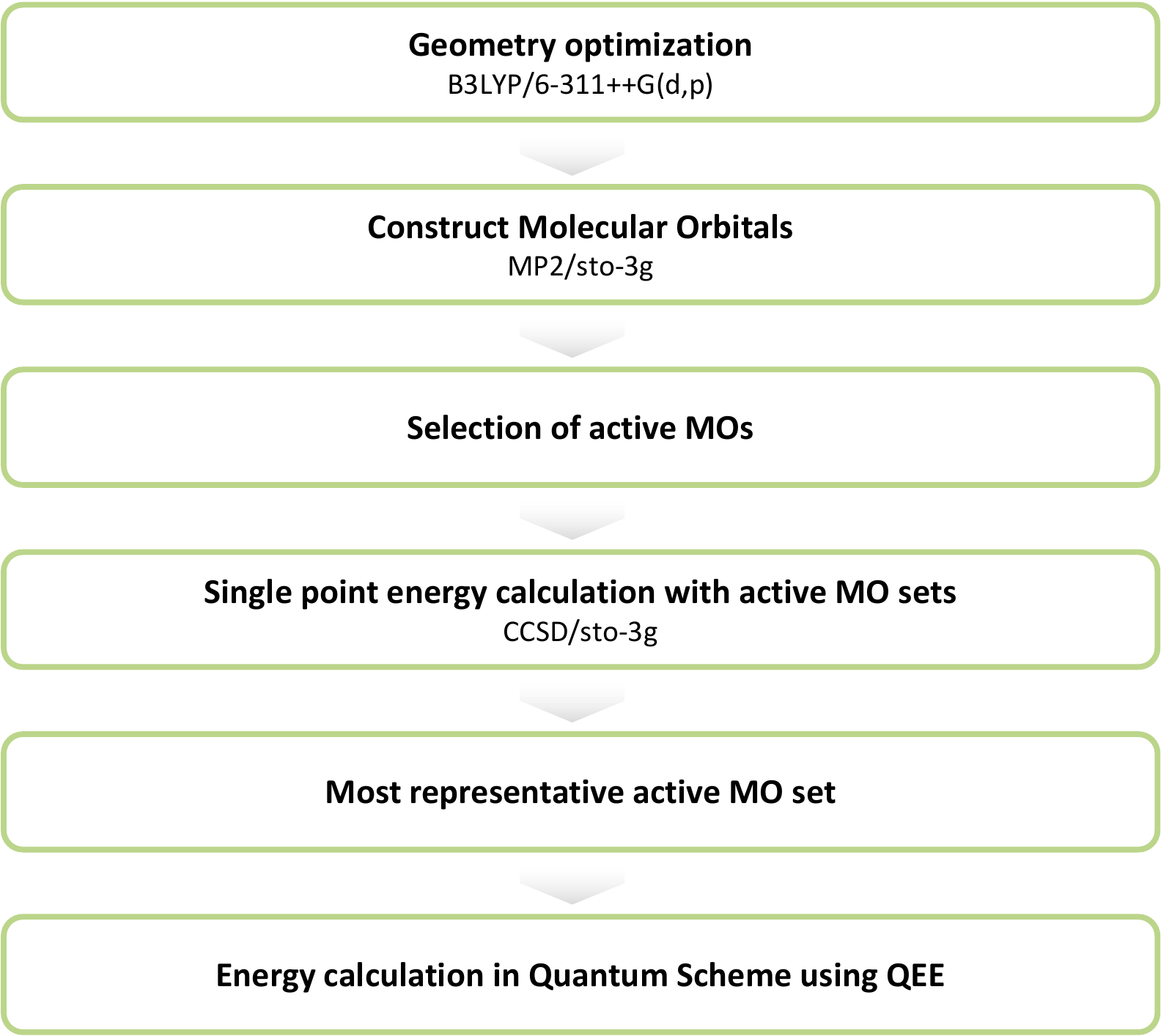}

\caption{Proposed workflow of preferred tautomeric state prediction}
    \label{fig:workflow}
\end{figure}

Here we propose a hybrid computational scheme for the prediction of the preferred tautomeric state, comprising classical quantum chemistry (using classical computers) and quantum computation.  The workflow involves (1) geometry optimization of the isomers using density functional theory, (2) reconstruction of the molecular orbital space in a minimal basis set, (3) reduction of the MO space, (4) single point energy calculations of active MO sets, (5) selection of the most representative active MO set, and finally (6) energy calculation in Quantum computing scheme using QEE, as illustrated in Fig. \ref{fig:workflow}. We use the CCSD energies of the full systems as the benchmark of our workflow performance.
% \yu{There are two ways to describe the workflow: 1. If we want to say that we select the active MO set based on comparison between CCSD full set and the active set energies calculated using quantum computer, we should change our aim of this study into providing a filtering method for MO selection criteria. This means that we tell the readers that we do not have a selection criteria, but we rather provide a method/workflow for future scientist that wants to develop a selection criteria. 2. If the aim of this study is to let readers know what current/near term quantum devices could do, we should just write "select MO based on quantum chemistry methods (MP2 and natural orbital occupancy) and quantum hardware limitations." This means that we have a selection criteria or at least experience in how to select active MO, but we would not disclose it. And if the referees want us to prove that our selection criteria/experience is useful and correct, we could select one active MO set for larger molecules and compare it with CCSD full set energy.}

Specifically, to reduce the problem size, we use quantum chemistry methods to select active molecular orbitals (MOs) and map the Hamiltonian onto fewer qubits using a qubit-efficient encoding method. The geometries of the tautomeric isomers are prepared and initially optimized using Avogadro \cite{Hanwell2012-nf} with the MMFF94 force field \cite{Halgren1996-hg, Halgren1996-jm, Halgren1996-nn, Halgren1996-ql, Halgren1996-zp} and further optimized at the level of 6-311++G(d,p) using B3LYP \cite{Becke1993-mi, Stephens1994-nk} in the Gaussian 16 program. Due to limited resources for quantum computing, the molecules are reconstituted with the minimal STO-3g at the MP2 level \cite{Head-Gordon1988-ez}, followed by the calculation of occupancies of natural orbitals (NOs) using Gaussian 16. The STO-3g basis set has been widely used in several pioneering works of computation chemistry with quantum computing scheme \cite{kandala_hardware-efficient_2017, google_ai_quantum_and_collaborators_hartree-fock_2020, nam_ground-state_2020}. Although the computational cost has been greatly reduced with the minimal basis set, the number of required qubits still exceeds the affordable amount of current QC simulators or real quantum devices. 
%\textcolor{red}{/explicit numbers: from google’s results in discussion: 8 qubit and 82 two-qubit gates are about the limitation for quantitative meaningful results /} 
The reduction of the MO space is then achieved by considering the occupation numbers and quantum hardware limitations. The selection of active space was achieved mostly by excluding the inner core and the outer virtual orbitals. In practice, the natural orbitals with occupancy closest to 0 (completely virtual) or 2 (fully occupied) were frozen or removed. Besides, quantum hardware constraints and the efficiencies of the qubit-efficient encoding method are also considered in the selection process. The reduced second-quantized Hamiltonian is then encoded onto qubits. % (with qubit counts fewer than those from using the JW encoding and other common encoding methods).

The natural orbitals for Acetone and Edaravone tatutomers calculated at the MP2/STO-3G level can be found in Appendix~\ref{appendix:Molecular Orbital Occupancy}. Selected combinations of active MOs are listed in Appendix~\ref{appendix:Active_MO} for acetone and Edaravone, respectively. After the active space selection from quantum chemistry methods, we have 12 spin-orbitals (6 molecular orbitals) for acetone, propen-2-ol, and the three tautomers of Edaravone. With Jordan-Wigner (JW) encoding, the qubit count to simulate the quantum systems is 12 qubits which is still a difficult task for noisy intermediate-scale quantum devices without error mitigation. Therefore, we use qubit-efficient encoding (QEE) to encode the systems onto 8 qubits shown in Table~\ref{tab:JW_QEE_Comparison} because it provides fewer qubit counts and renders the systems to be suitable for the use of hardware-efficient ansatzes. We then use a hardware-efficient ansatz to simulate the systems with variational quantum eigensolver (VQE).

\begin{table}[t]
\centering
\begin{ruledtabular}
\begin{tabular}{ccccc}
Molecule & Active Space Orbitals & (Electrons, MO) & JW Qubit Count & QEE Qubit Count \\
\hline
acetone     & 14-19  & (4e, 6o) & 12 & 8 \\
propen-2-ol & 14-19  & (4e, 6o) & 12 & 8 \\
Edaravone (keto) & 44-49  & (4e, 6o) & 12 & 8 \\
Edaravone (enol) & 44-49  & (4e, 6o) & 12 & 8 \\
Edaravone (amine)  & 44-49  & (4e, 6o) & 12 & 8 \\
  
\end{tabular}
\end{ruledtabular}
\caption{Comparison of qubit counts for Jordan-Wigner encoding and qubit-efficient encoding for acetone, propen-2-ol, and the three tautomers of Edaravone (in the STO-3G basis sets and equilibrium bond distances) with some molecular orbitals being frozen/removed. Note that the molecular orbital indices are ordered from the lowest to the highest energies and the indices start from 0.}

\label{tab:JW_QEE_Comparison}
\end{table}

Even though the qubit requirement is reduced by classical active space selection methods,
%(12 spin orbitals left, 12 qubits for JW), 
it is still hard to be implemented on a quantum device as the circuit depth would not permit. For JW encoding, some unphysical (e.g. not particle conserving or violating other symmetries) electronic configurations are also encoded. This necessitates the usage of chemical-inspired ansatzes (e.g. the unitary couple cluster (UCC) ansatz) that often have larger circuit depth so that the trial wavefunction only represents the states in the chemical subspace. It is still possible to use a hardware-efficient ansatz for JW encoded problems, but the number of entangling layers has to be large. With more entangling layers, the ansatz circuit would have better expressibility \cite{PhysRevResearch.2.033125} and entanglement to include the solution space \cite{sim_expressibility_2019}, but it may suffer from the vanishing gradient and the barren plateau problem \cite{mcclean_barren_2018, https://doi.org/10.48550/arxiv.2011.06258, https://doi.org/10.48550/arxiv.2112.15002}. Thus, we resort to a qubit-efficient encoding (QEE) method proposed in \cite{shee_qubit-efficient_2021} which provides an logarithmic saving of qubit counts (8 qubits for QEE). Additionally, QEE only encodes the physical (significant) electronic configurations such that low depth hardware-efficient ansatzes are suitable for QEE encoded problems.

\textbf{Qubit Efficient Encoding.}
%\subsection{Qubit Efficient Encoding}
As we want to work on a larger electronic system, an encoding scheme using $\mathcal O(N)$ qubits would not be practical on current and near-future devices. Therefore, it would be suitable to resort to encoding methods using fewer qubits. Here, we choose Qubit Efficient Encoding (QEE) from \cite{shee_qubit-efficient_2021} where the qubit scales logarithmically with respect to $N$. This is done by only mapping particle conserving or other eletronic configurations with certain symmetries to the qubit basis states which exploits the exponential growth of qubit Hilbert space.

Since the encoding scheme works in a subspace of the space spanned by the occupation basis, we will not be able to map a single creation or annihilation operator as the operators change the number of electrons in the system. Nevertheless, the second-quantized Hamiltonian can be written as a linear combination of excitation operators, which conserve the number of particles, $E_{pq}\equiv a_p^\dag a_q$ yielding
\begin{equation}
    H = \sum_{pq}{h_{pq}E_{pq}}
    +\frac12 \sum_{pqrs} h_{pqrs}(
    \delta_{qr}E_{ps}
    -E_{pr}E_{qs})
    \label{eq:double_excitation_identity}
\end{equation}
by using the anti-commutation relations of fermionic operators. Any excitation operator $E_{pq}$ can be written as $E_{pq} = \sum_{k,k'=0}^{|\mathcal F _m|-1} 
c_{k'k}^{pq} \ket{\mathbf f_{k'}}_{\text{f}} \bra{\mathbf f_k}_{\text{f}}$, where $\mathcal F _m$ is the set of all particle conserving (or with other symmetries) electronic configurations $\ket{\mathbf f_{k}}_{\text{f}}$ and $c^{pq}_{k'k} = \bra{\mathbf f_{k'}}_{\text{f}} E_{pq} \ket{\mathbf f_k}_{\text{f}}$. $c^{pq}_{k'k}$ is zero if the transition from $\ket{\mathbf f_k}_{\text{f}}$ to $\ket{\mathbf f_{k'}}_{\text{f}}$ via excitation operator $E_{pq}$ is impossible. The coefficient $c^{pq}_{k'k}$ can be $\pm 1$ according to the antisymmetric property of fermions where $c^{pq}_{k'k} = \prod_{i=\min(p,q)+1}^{\max(p,q)-1} (-1)^{f_i}$. With a transformation $\mathcal E$ that maps the selected electronic configurations one-to-one to qubit basis states in an arbitrary order, the corresponding qubit operator of $E_{pq}$ can be written as  
\begin{equation}
    \Tilde{E}_{pq} = \mathcal E \circ E_{pq} \circ \mathcal E^{-1} = \sum_{k, k'=0}^{|\mathcal F _m|-1} c^{pq}_{k'k} \ket{\mathbf q_{k'}}_{\text{q}}\bra{\mathbf q_k}_{\text{q}}, 
\end{equation}
where $\ket{\mathbf q_k}_{\text{q}} = \mathcal E \ket{\mathbf f_k}_{\text{f}}$ is the encoded qubit state of $\ket{\mathbf f_k}_{\text{f}}$ under the map $\mathcal E$. The transition between the two basis states $\ket{\mathbf q}_{\text{q}} = \ket{q_{Q-1},...,q_0}_{\text{q}}$ 
and $\ket{\mathbf q'}_{\text{q}} = \ket{q'_{Q-1},...,q_0'}_{\text{q}}$ can be factorized as $\ket{\mathbf q'}_{\text{q}}\bra{\mathbf q}_{\text{q}} = {\bigotimes_{k=0}^{Q-1}} \ket{q'_k}_{\text{q}} \bra{q_k}_{\text{q}}$. Therefore, the encoded excitation operator can be expressed as 
\begin{equation}\label{eq: Epq as T}
    \Tilde{E}_{pq} 
    = \sum_{k,k'=0}^{|\mathcal F _m|-1} 
    c^{pq}_{k'k} 
    \ket{\mathbf q_{k'}}_{\text{q}}
    \bra{\mathbf q_k}_{\text{q}}
    = \sum_{k,k'=0}^{|\mathcal F _m|-1}
    \bigotimes_{w=0}^{Q-1} c^{pq}_{k'k}  T_{k'k,w},
\end{equation}
where $T_{k'k,w}$ could be one of the following operators
\begin{align}
    Q^+ & = \ket{1}_{\text{q}} \bra{0}_{\text{q}} = \frac12 (X-iY), \\
    Q^- & = \ket{0}_{\text{q}} \bra{1}_{\text{q}} = \frac12 (X+iY), \\
    N^{(0)} & = \ket{0}_{\text{q}} \bra{0}_{\text{q}} = \frac12 (I+Z), \\
    N^{(1)} & = \ket{1}_{\text{q}} \bra{1}_{\text{q}} = \frac12 (I-Z)
\end{align}
according to the qubit basis state transitions.

As each operator $T_{k'k,w}$ is a sum of two Pauli operators, this allows us to express the qubit Hamiltonian
\begin{equation}
{H}_{\text{q}} = \sum_{pq}{h_{pq}\Tilde E_{pq}}
    +\frac12 \sum_{pqrs} h_{pqrs}(
    \delta_{qr}\Tilde E_{ps}
    -\Tilde E_{pr} \Tilde E_{qs})
\end{equation}
as a sum of Pauli operator strings. With the Hamiltonian in Pauli strings, the energy expectation value can be obtained using a quantum computer. In this work, we select electronic configurations that are particle-conserving and singlet so we reduce the qubit requirement from 12 qubits to 8 qubits.

\begin{figure}[hbt!]
    \centering
    \begin{subfigure}[t]{0.4\textwidth}
        \Qcircuit @C=0.7em @R=0.5em {
        & \gate{R_y(\theta_0)} & \ctrl{1} & \qw & \gate{R_y(\theta_{15})} & \qw & \targ & \qw \\
        & \gate{R_y(\theta_1)} & \targ    & \qw & \gate{R_y(\theta_8)} & \ctrl{1} & \qw & \qw \\
        & \gate{R_y(\theta_2)} & \ctrl{1} & \qw & \gate{R_y(\theta_9)} & \targ & \qw & \qw \\
        & \gate{R_y(\theta_3)} & \targ    & \qw & \gate{R_y(\theta_{10})} & \ctrl{1} & \qw & \qw \\
        & \gate{R_y(\theta_4)} & \ctrl{1} & \qw & \gate{R_y(\theta_{11})} & \targ & \qw & \qw \\
        & \gate{R_y(\theta_5)} & \targ    & \qw & \gate{R_y(\theta_{12})} & \ctrl{1} & \qw & \qw \\
        & \gate{R_y(\theta_6)} & \ctrl{1} & \qw & \gate{R_y(\theta_{13})} & \targ & \qw & \qw \\
        & \gate{R_y(\theta_7)} & \targ    & \qw & \gate{R_y(\theta_{14})} & \qw & \ctrl{-7} & \qw \\
        }
        \caption{Two hardware-efficient layers of the 8-qubit ansatz circuit. Note that there are single-qubit parametrized $R_y$ rotations on all the qubits at the end of this ansatz circuit.}
        \label{fig:AnsatzCircuitLayer}
    \end{subfigure}
    \begin{subfigure}[t]{0.4\textwidth}
        \Qcircuit @C=0.7em @R=0.5em {
        & \gate{R_y(\theta_0)} & \ctrl{1} & \qw      & \qw      & \qw      & \qw      & \qw      & \qw      & \qw \\
        & \gate{R_y(\theta_1)} & \targ    & \ctrl{1} & \qw      & \qw      & \qw      & \qw      & \qw      & \qw \\
        & \gate{R_y(\theta_2)} & \qw      & \targ    & \ctrl{1} & \qw      & \qw      & \qw      & \qw      & \qw \\
        & \gate{R_y(\theta_3)} & \qw      & \qw      & \targ    & \ctrl{1} & \qw      & \qw      & \qw      & \qw\\
        & \gate{R_y(\theta_4)} & \qw      & \qw      & \qw      & \targ    & \ctrl{1} & \qw      & \qw      & \qw\\
        & \gate{R_y(\theta_5)} & \qw      & \qw      & \qw      & \qw      & \targ    & \ctrl{1} & \qw      & \qw\\
        & \gate{R_y(\theta_6)} & \qw      & \qw      & \qw      & \qw      & \qw      & \targ    & \ctrl{1} & \qw\\
        & \gate{R_y(\theta_7)} & \qw      & \qw      & \qw      & \qw      & \qw      & \qw      & \targ    & \qw\\
        }
        \caption{One hardware-efficient layer of the alternative 8-qubit ansatz circuit. Note that there is no single-qubit rotation at the end of this ansatz circuit.}
        \label{fig:AlternativeAnsatzCircuitLayer}
    \end{subfigure}
        \caption{The hardware-efficient ansatz circuit layers used in this work.}
        \label{fig:ansatz_circuit_layers}
\end{figure}
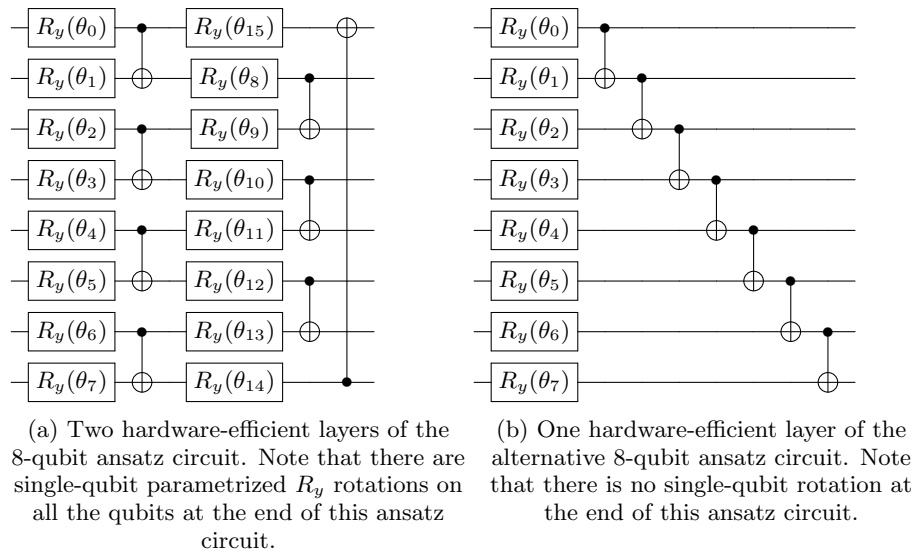

\textbf{Variational quantum circuits and parameter optimization.}
After obtaining the qubit Hamiltonian from QEE, we employ two different hardware-efficient ansatz, as illustrated in Fig.~\ref{fig:ansatz_circuit_layers}, for the VQE algorithm. The first hardware-efficient ansatz consists of four two-qubit entangling building blocks 
%shown in Fig.~\ref{fig:AnsatzBlock}
, and the layers are arranged in a staggered form where Fig.~\ref{fig:AnsatzCircuitLayer} gives an example of two hardware-efficient layers. The first ansatz used in this work consists of at most 20 layers 
%(stack 10 circuits in Fig.~\ref{fig:AnsatzCircuitLayer})
and eight parameterized $R_y$ rotations at each layer in the end. Therefore, the largest circuit in this form has 80 CNOT gates and 168 parameters. While for the second/alternative ansatz, each layer consists of seven CNOT gates and eight parameterized $R_y$ rotations shown in Fig.~\ref{fig:AlternativeAnsatzCircuitLayer}. The largest circuit in this form consists of 10 layers and has 70 CNOT gates and 80 parameters. We have also run this alternative ansatz circuit on a noisy simulator where the circuit consists of four layers and has 28 CNOT gates and 32 parameters.

For the initial state, we have compared two initialization strategies. First, we have used the Hartree-Fock (HF) state as the initial state where all the parameters are set to zero initially \cite{shee_qubit-efficient_2021}.
%(we have encoded the zeroth qubit basis state $\ket{00...0}$ as the HF state indicated in \cite{shee_qubit-efficient_2021}). 
We have also compared HF initialization with the Gaussian initialization strategy proposed in \cite{kay_2022}. Zhang et al. \cite{kay_2022} proposed this Gaussian initialization strategy to escape from barren plateau where the initial parameters are sampled from a Gaussian distribution. The Gaussian distribution used in this work has a mean of 0 and a variance of 0.3. %\yu{I have added Gaussian initialization parameters here.} 
The classical optimization of the variational parameters was done using the Sequential Least Squares Programming (SLSQP) method.

\begin{comment}
\begin{figure}[hbt!]
\centerline{
\Qcircuit @C=0.7em @R=0.5em {
& \gate{R_y(\theta_0)} & \ctrl{1} & \qw \\
& \gate{R_y(\theta_1)} & \targ    & \qw \\
}
}
\caption{A single ansatz building block.}
\label{fig:AnsatzBlock}
\end{figure}
\end{comment}

%\mh{TLY updated from here}
%\subsection{Geometry Optimization}

\subsection*{Numerical Results}

We investigate the performance of our workflow with several numerical experiments. First, the CCSD energies for the full systems of the tautomers in the STO-3G basis set are calculated and are considered our benchmark for energy estimation performance. Second, energy calculations of VQE with hardware-efficient ansatz on a noiseless quantum simulator are performed for both the acetone and the Edaravone tautomers. 
%using different numbers of hardware-efficient circuit layers %(Fig.~\ref{fig:AnsatzCircuitLayer} gives an example of two hardware-efficient layers of the ansatz we used) with parameterized $R_y$ rotations at the end of the circuits (results shown in Fig.~\ref{fig:acetone_convergence} and Fig.~\ref{fig:edaravone_convergence}). 
Third, VQE energies on a noiseless quantum simulator for the acetone system using an alternative ansatz circuit %(Appendix~\ref{appendix:Alternative Ansatz Form}) 
are obtained.
%with the results shown in Fig.~\ref{fig:acetone_convergence_alternative}. 
Lastly, a noisy simulation (with the noise model detailed in Appendix~\ref{appendix:Noise Model}) of the acetone system has been done using four layers of the alternative ansatz circuit.

%\yu{I talk about full system CCSD here}
The CCSD energies of the full systems and the active-space systems in the STO-3G basis set are computed. For the acetone system (Table~\ref{tab:Gaussian_RelEnergy_Acetone}), the full system CCSD energy of the enol form relative to the keto form is 24.070 kcal/mol, and the active-space system CCSD energy of the enol form relative to the keto form is 23.766 kcal/mol. For the Edaravone system (Table~\ref{tab:Gaussian_RelEnergy_Edaravone}), the full system CCSD energy of the enol form relative to the keto form is 13.726 kcal/mol and the CCSD energy of the amine form relative to the keto form is 25.947 kcal/mol. While for active-space system of Edaravone, the CCSD energy of the enol form relative to the keto form is 9.197 kcal/mol and the CCSD energy of the amine form relative to the keto form is 25.398 kcal/mol.

\begin{comment}% by MH Jul 24, 2022.
% need to rewrite this part. Just use it to benchmark the performance of our VQC results.
The ground state energies calculated with complete MOs using CCSD/STO-3g in Gaussian 16 were used as benchmarks for the selection of active MO sets. Among the combinations of the active MOs calculated, the set with the two MOs below HOMO) and 4 MOs above LUMO (i.e. HOMO-1, HOMO, LUMO, LUMO+1, LUMO+2, and LUMO+3) were shown to have relative ground state energy values closest to the set without any MO reduction (see supplemental information for more detail \mh{use ref tag!}). Therefore, these 6 MOs (12 spin orbitals with 4 electrons) were used for VQE calculations. The relative ground state energies calculated with this MO set using CCSD/STO-3g in Gaussian for acetone and Edaravone are shown in Table \ref{tab:Gaussian_RelEnergy_Acetone} and Table \ref{tab:Gaussian_RelEnergy_Edaravone}, respectively.
\mh{TLY: need to make elaborate or make argument why this set will give best results}
\yu{I agree, might need to argue the selection process instead of saying that the MOs are selected because of their great prediction of energy gaps (check out the last sentences I wrote in the 1st paragraph of "Molecular Orbital Reduction" in "Workflow")}
\end{comment}

\begin{table}[t]
    \centering\ 
\begin{ruledtabular}
\begin{tabular}{ccc}
 & acetone & Propen-2-ol \\
\hline
Full system CCSD energies relative to acetone &  0 & 24.070\\
Active set CCSD energies relative to acetone & 0 & 23.766 \\
\end{tabular}
\end{ruledtabular}
\caption{Relative ground state energy of acetone and propen-2-ol in kcal/mol. Relative energies for both the full MO and the active set (HOMO-1, HOMO, LUMO, LUMO+1, LUMO+2, and LUMO+3) were shown. Various combinations of orbital reduction are provided in Supplementary Information Table~\ref{tab:Acetone_activeMO}.}
\label{tab:Gaussian_RelEnergy_Acetone}
\end{table}

\begin{table}[t]
    \centering\ 
\begin{ruledtabular}
\begin{tabular}{cccc}
 & Keto & Enol & Amine\\
\hline
Full system CCSD energies relative to keto form & 0.000 & 13.726 & 25.947\\
Active set CCSD energies relative to keto form & 0.000 & 9.197 & 25.398 \\
\end{tabular}
\end{ruledtabular}
\caption{Relative ground state energy of the keto, enol, and amine form of Edavarone in kcal/mol. Relative energies for both the full MO and the active set (HOMO-1, HOMO, LUMO, LUMO+1, LUMO+2, and LUMO+3) were shown. Various combinations of orbital reduction are provided in Supplementary Information Table~\ref{tab:Edavarone_activeMO}.}
\label{tab:Gaussian_RelEnergy_Edaravone}
\end{table}

%\mh{Yu: Then discuss the numerical outcomes here. Go over the subplots one by one. Avoid jumping between figures. Things to discuss here are: accuracy, convergence speed, etc. }
%\mh{Next, we consider the quantum simulation of acetone. The numerical outcomes are plotted in Fig. 5.} 
%noiseless simulation of VQE has been done. 
% The VQE results with Gaussian initialization from a noiseless simulator in Fig.~\ref{fig:acetone_vqe_errors_gaussian} and Fig.~\ref{fig:edaravone_vqe_errors_gaussian} show that most of the VQE energies reached chemical accuracy (1 kcal/mol). For acetone in Fig.~\ref{fig:acetone_vqe_errors_gaussian} with 20 circuit layers, the keto form has an error of 0.135 kcal/mol and the enol form has an error of 0.258 kcal/mol. For Edaravone in Fig.~\ref{fig:edaravone_vqe_errors_gaussian} with 20 circuit layers, the keto form has an error of 0.730 kcal/mol, the enol form has an error of 1.006 kcal/mol, the amine form has an error of 1.168 kcal/mol. For both acetone and Edaravone tautomerism, most of the energy gaps between the tautomers from VQE calculation agree with the CCSD results of the full systems (where orbitals are not frozen or removed) within chemical accuracy. Besides, VQE calculations of active-space systems have the same results with CCSD calculations of the full systems for the predictions of stability between the tautomers (see the 
% dotted lines in Fig.~\ref{fig:acetone_vqe_energy_levels_gaussian} and Fig.~\ref{fig:edaravone_vqe_energy_levels_gaussian}). %\yu{I compared CCSD and VQE here}

\begin{figure}[h]
     \centering
     \begin{subfigure}[b]{0.4\textwidth}
         \centering
         \includegraphics[width=\textwidth]{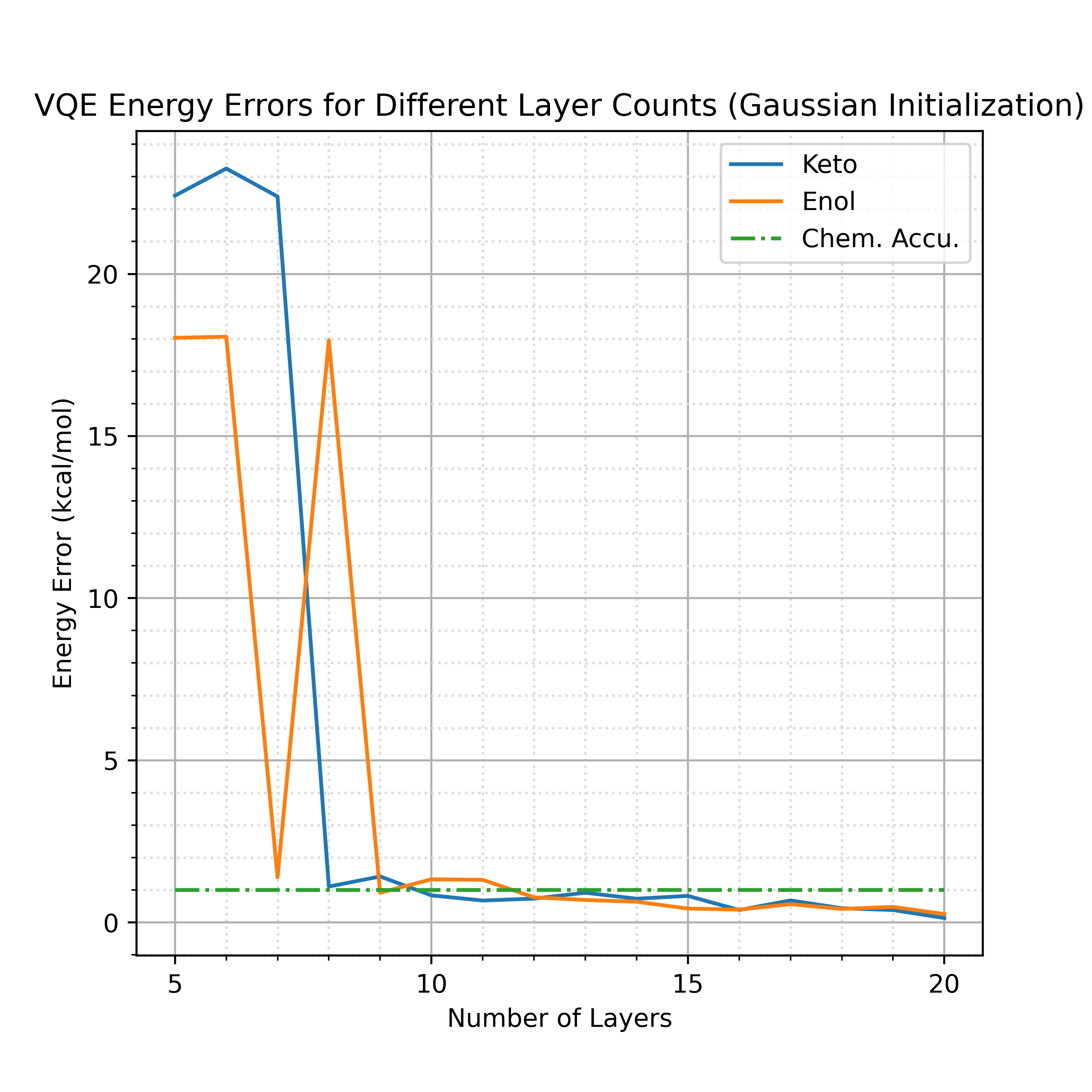}
         \caption{VQE energy errors (Gaussian Init.).}
         \label{fig:acetone_vqe_errors_gaussian}
     \end{subfigure}
     \begin{subfigure}[b]{0.4\textwidth}
         \centering
         \includegraphics[width=\textwidth]{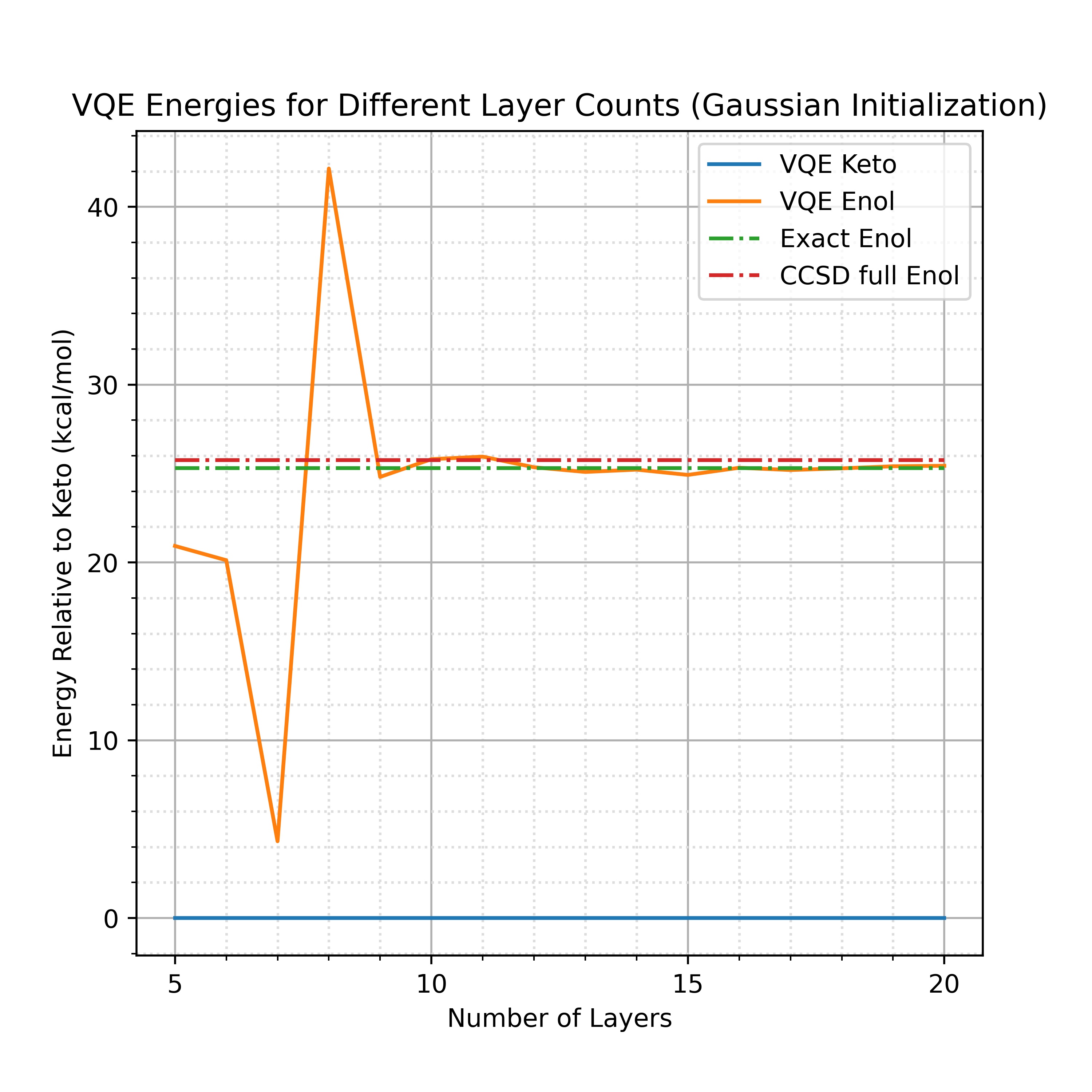}
         \caption{VQE energies (Gaussian Init.).}
         \label{fig:acetone_vqe_energy_levels_gaussian}
     \end{subfigure}
     \begin{subfigure}[b]{0.4\textwidth}
         \centering
         \includegraphics[width=\textwidth]{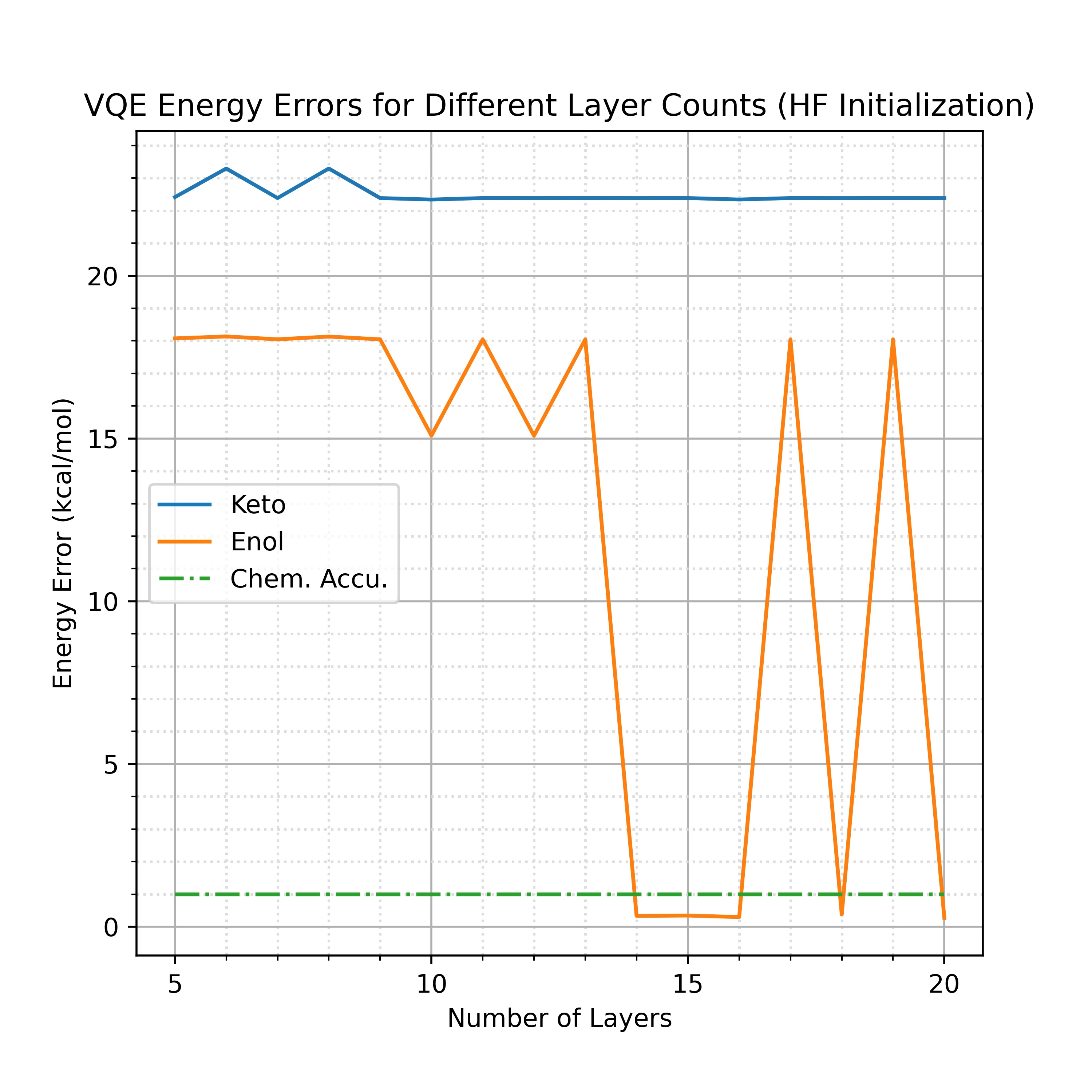}
         \caption{VQE energy errors (HF Init.).}
         \label{fig:acetone_vqe_errors_hf}
     \end{subfigure}
     \begin{subfigure}[b]{0.4\textwidth}
         \centering
         \includegraphics[width=\textwidth]{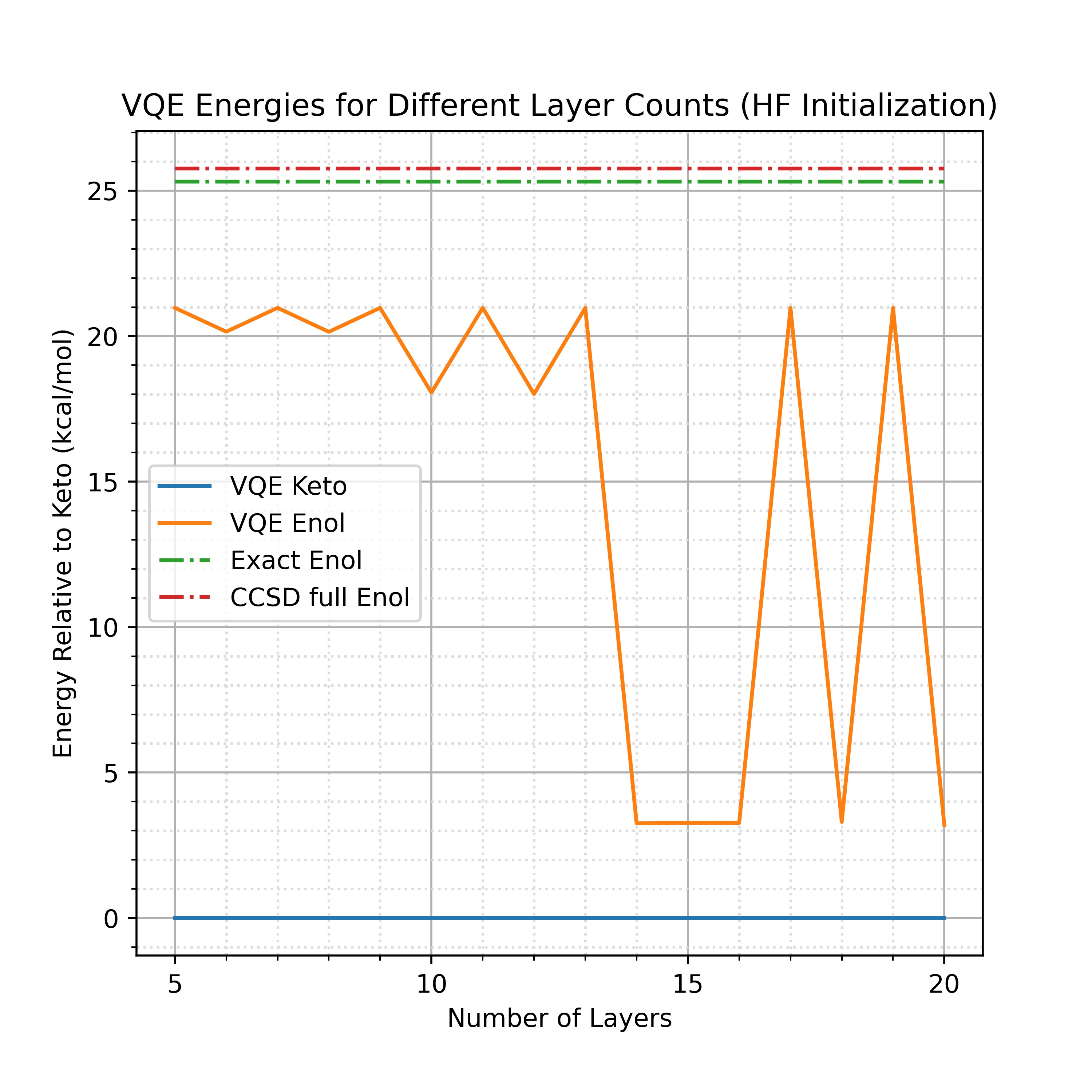}
         \caption{VQE energies (HF Init.).}
         \label{fig:acetone_vqe_energy_levels_hf}
     \end{subfigure}
        \caption{The relationship between acetone VQE results and the number of hardware-efficient ansatz layers. (a) and (b) are the results when Gaussian-initialized parameters are used while (c) and (d) are the results when initial states are the HF states. All the x-axes represent the number of hardware-efficient circuit layers. For (a) and (c), the y-axes represent VQE energy errors of the keto or enol form of acetone, and the dashed lines represent chemical accuracy (Chem. Accu.). For (b) and (d), the y-axes represent the energies relative to the keto form (so VQE energies, CCSD energies, and exact energies of the keto form are set to zero). Note that the exact energies (Exact in the legend) are the results from exact diagonalization of the active-space Hamiltonian for each tautomer, and CCSD full in the legend represents the CCSD energy of the full system relative to the keto form.}
        \label{fig:acetone_convergence}
\end{figure}

For the VQE results of acetone shown in Fig.~\ref{fig:acetone_convergence}, we use two different parameter initialization strategies (Gaussian and HF) and investigate how different number of ansatz circuit layers affect VQE calculations. For acetone in Fig.~\ref{fig:acetone_vqe_errors_gaussian} with 20 circuit layers using Gaussian initialization, the keto form has an error of 0.135 kcal/mol and the enol form has an error of 0.258 kcal/mol which are well below chemical accuracy (1 kcal/mol). It can be seen that more entangling ansatz layers provides lower VQE error (see Fig.~\ref{fig:acetone_vqe_errors_gaussian}) and better estimations of the relative energy gaps between the keto and enol forms of acetone (see Fig.~\ref{fig:acetone_vqe_energy_levels_gaussian}). Besides, VQE calculations of active-space systems have the same results with CCSD calculations of the full systems for the predictions of stability between the tautomers (see the dotted lines in Fig.~\ref{fig:acetone_vqe_energy_levels_gaussian}). However, for the case where initial states are HF states, the VQE results are often found to be trapped in local minimum or barren plateau. This can be clearly observed from the keto form of acetone in Fig.~\ref{fig:acetone_vqe_errors_hf} where the final states are almost the same as the initial HF state. While for the case where Gaussian initialized parameters are used, the VQE shows faster convergence to chemical accuracy (require lower circuit depth) where only 12 circuit layers are needed to reach chemical accuracy. Also, the relative energy gaps between the keto and enol forms of acetone are not accurate when using HF intialization (see Fig.~\ref{fig:acetone_vqe_energy_levels_hf}).

\begin{figure}[h]
     \centering
     \begin{subfigure}[b]{0.4\textwidth}
         \centering
         \includegraphics[width=\textwidth]{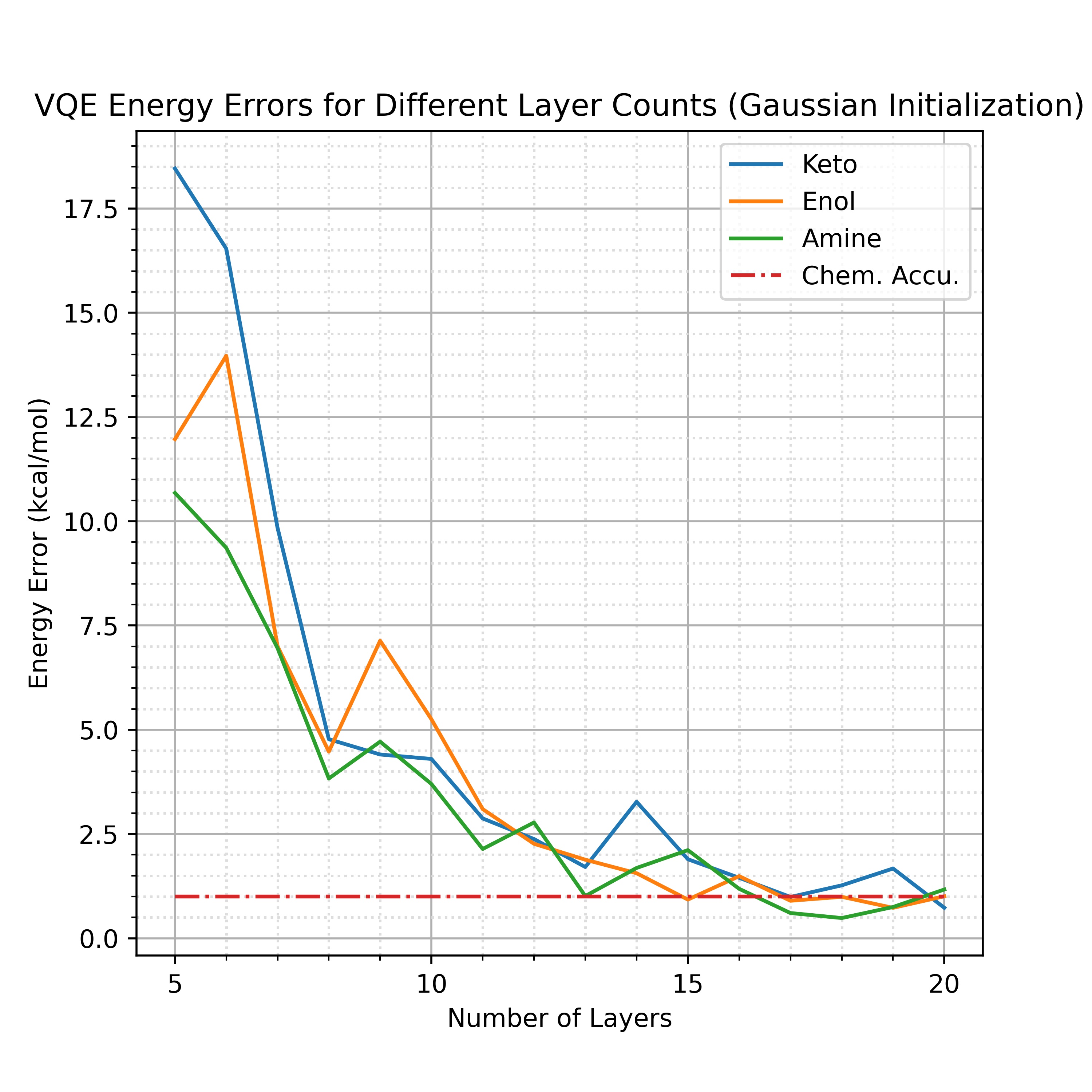}
         \caption{VQE energy errors (Gaussian Init.).}
         \label{fig:edaravone_vqe_errors_gaussian}
     \end{subfigure}
     \begin{subfigure}[b]{0.4\textwidth}
         \centering
         \includegraphics[width=\textwidth]{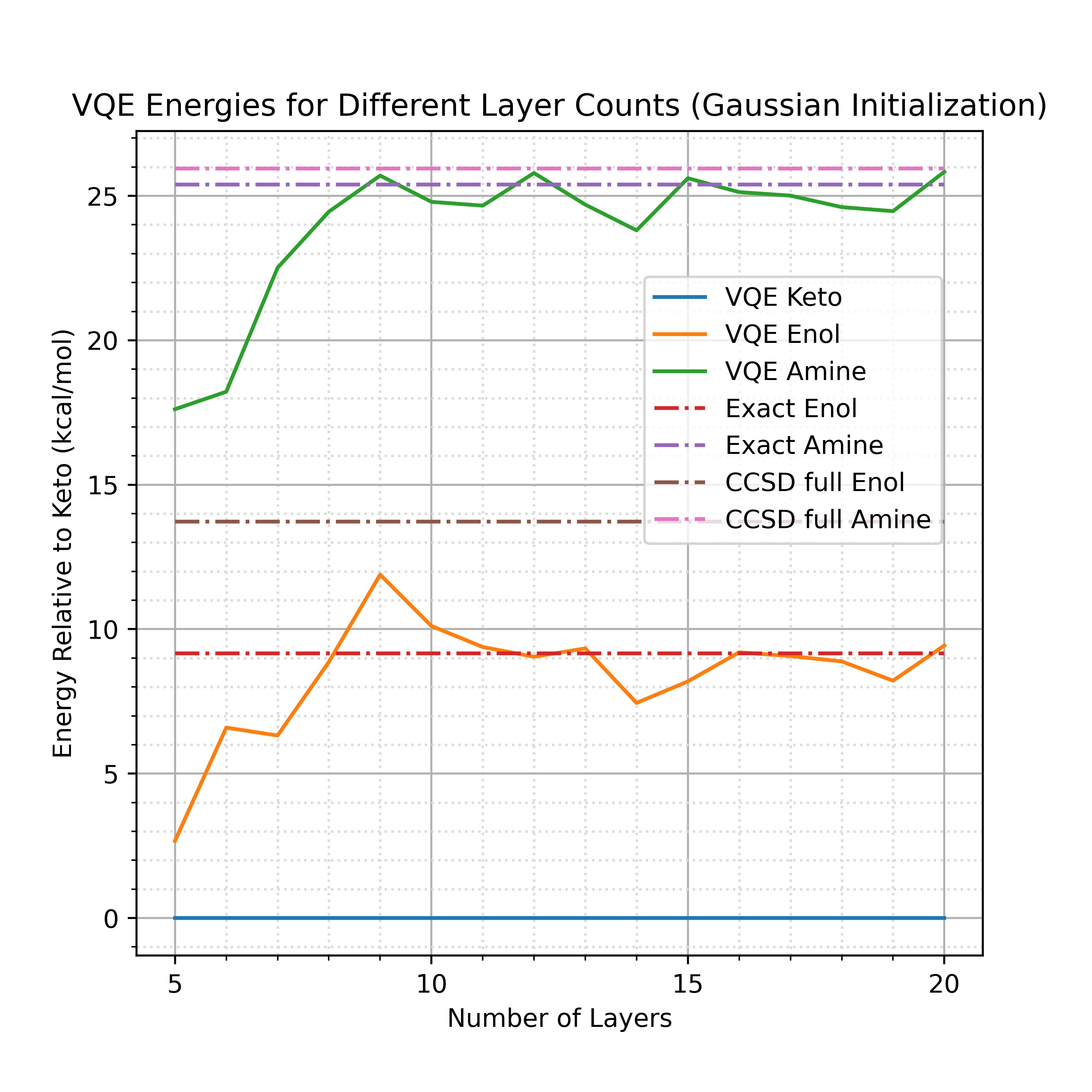}
         \caption{VQE energies (Gaussian Init.).}
         \label{fig:edaravone_vqe_energy_levels_gaussian}
     \end{subfigure}
     \begin{subfigure}[b]{0.4\textwidth}
         \centering
         \includegraphics[width=\textwidth]{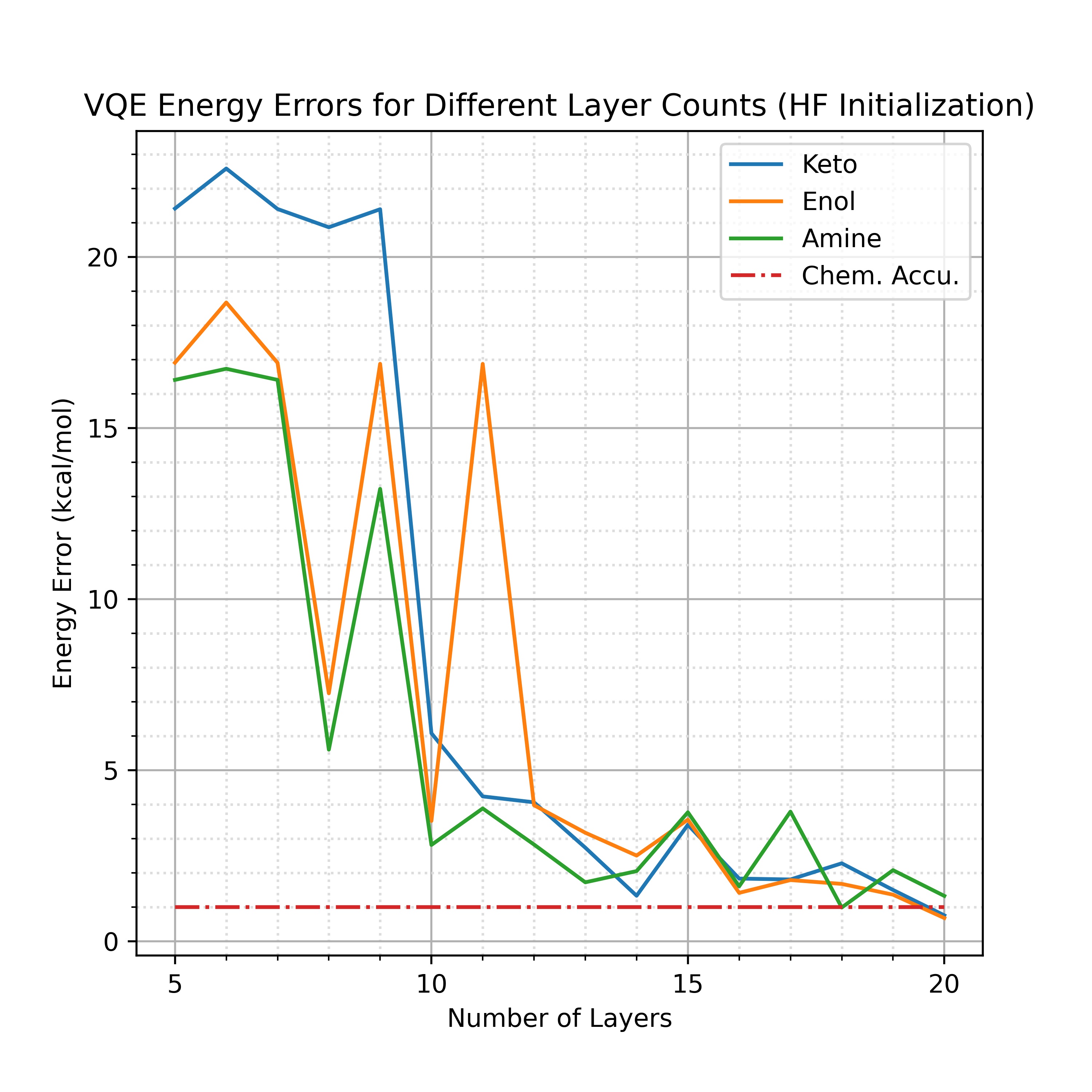}
         \caption{VQE energy errors (HF Init.).}
         \label{fig:edaravone_vqe_errors_hf}
     \end{subfigure}
     \begin{subfigure}[b]{0.4\textwidth}
         \centering
         \includegraphics[width=\textwidth]{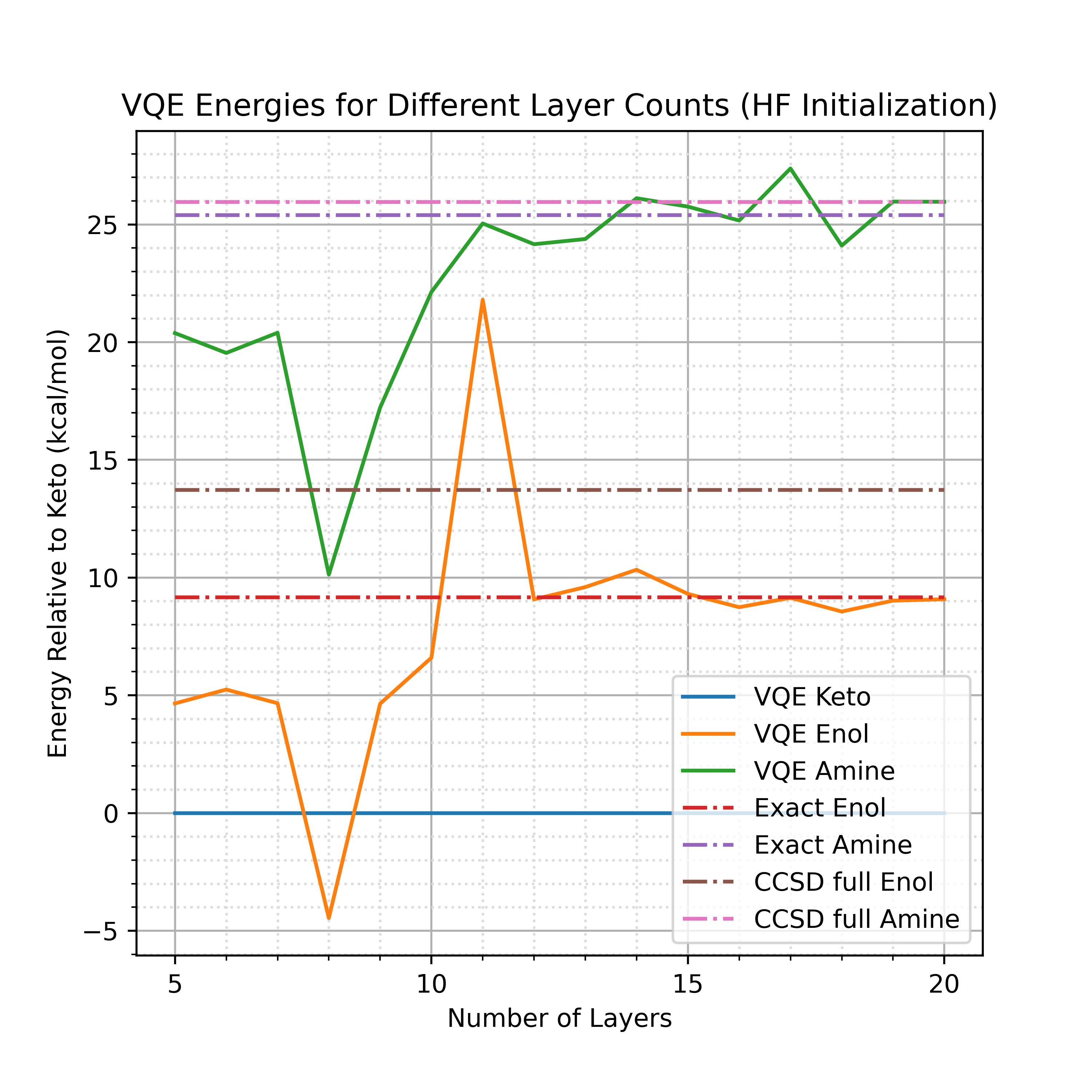}
         \caption{VQE energies (HF Init.).}
         \label{fig:edaravone_vqe_energy_levels_hf}
     \end{subfigure}
        \caption{The relationship between Edaravone VQE results and the number of hardware-efficient ansatz layers. (a) and (b) are the results when Gaussian-initialized parameters are used while (c) and (d) are the results when initial states are the HF states. All the x-axes represent the number of hardware-efficient circuit layers. For (a) and (c), the y-axes represent VQE energy errors of the keto, enol, or amine form of Edaravone, and the dashed lines represent chemical accuracy (Chem. Accu.). For (b) and (d), the y-axes represent the energies relative to the keto form (so VQE energies, CCSD energies, and exact energies of the keto form are set to zero). Note that the exact energies (Exact in the legend) are the results from exact diagonalization of the active-space Hamiltonian for each tautomer, and CCSD full in the legend represents the CCSD energies of the full systems relative to the keto form.}
        \label{fig:edaravone_convergence}
\end{figure}

Similar VQE results can be found in Fig.~\ref{fig:edaravone_convergence} for the Edavarone system. For Edaravone in Fig.~\ref{fig:edaravone_vqe_errors_gaussian} with 20 circuit layers, the keto form has an error of 0.730 kcal/mol, the enol form has an error of 1.006 kcal/mol, the amine form has an error of 1.168 kcal/mol which are fairly close to chemical accuracy. Fig.~\ref{fig:edaravone_vqe_errors_gaussian} also shows that more entangling ansatz layers often provides lower VQE error and better estimations of the relative energy (Fig.~\ref{fig:edaravone_vqe_energy_levels_gaussian}). For the case of HF initialization for Edavarone, the VQE errors and calculations of relative energy gaps also reach similar accuracy in    Fig.~\ref{fig:edaravone_vqe_errors_hf} and Fig.~\ref{fig:edaravone_vqe_energy_levels_hf}. However, it can still be observed that Gaussian initialization has a better convergence with respect to the number of circuit layers.

\begin{figure}[h]
     \centering
     \begin{subfigure}[b]{0.4\textwidth}
         \centering
         \includegraphics[width=\textwidth]{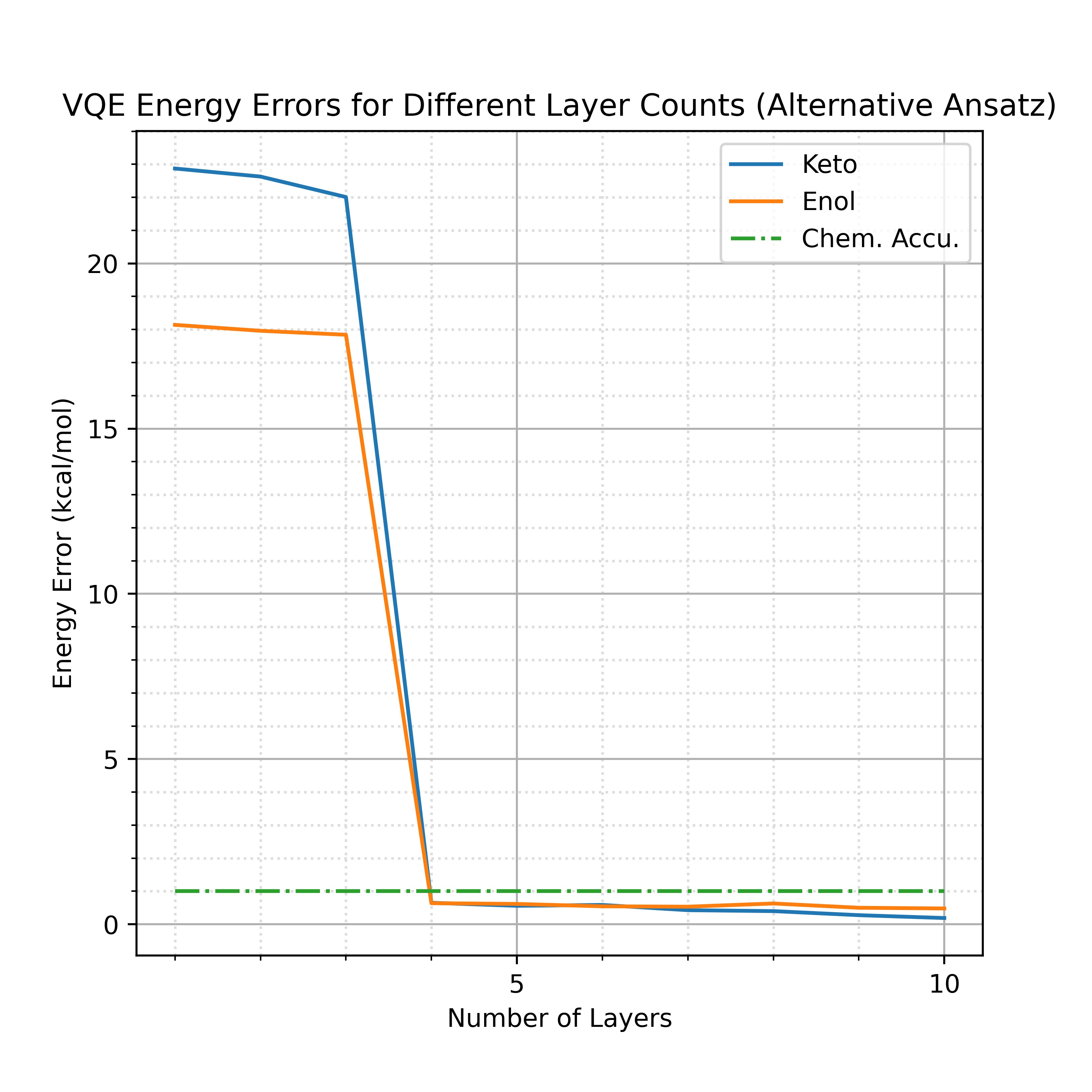}
         \caption{VQE energy errors (Gaussian Init.).}
         \label{fig:acetone_vqe_errors_alternative}
     \end{subfigure}
     \begin{subfigure}[b]{0.4\textwidth}
         \centering
         \includegraphics[width=\textwidth]{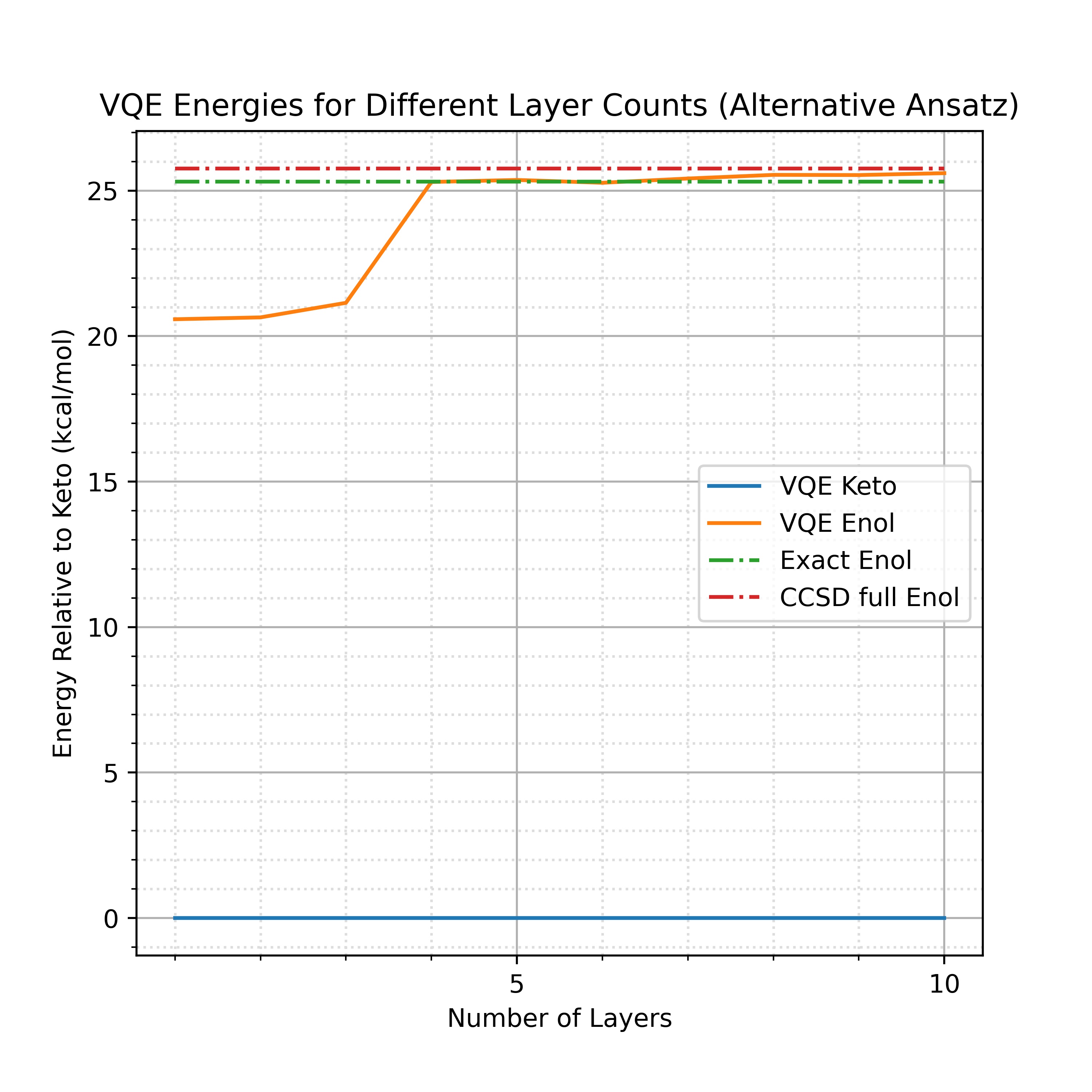}
         \caption{VQE energies (Gaussian Init.).}
         \label{fig:acetone_vqe_energy_levels_alternative}
     \end{subfigure}
        \caption{The relationship between acetone VQE results and the number of the alternative ansatz layers from Fig.~\ref{fig:AlternativeAnsatzCircuitLayer}. Gaussian-initialized parameters are used. All the x-axes represent the number of hardware-efficient circuit layers. For (a), the y-axis represents VQE energy errors of the keto or enol form of acetone, and the dashed line represents chemical accuracy (Chem. Accu.). For (b), the y-axis represents the energies relative to the keto form (so VQE energies, CCSD energies, and exact energies of the keto form are set to zero). Note that the exact energies (Exact in the legend) are the results from exact diagonalization of the active-space Hamiltonian for each tautomer, and CCSD full in the legend represents the CCSD energy of the full system relative to the keto form.}
        \label{fig:acetone_convergence_alternative}
\end{figure}

Additionally, we tested the performance of an alternative ansatz (see Fig.~\ref{fig:AlternativeAnsatzCircuitLayer} for a single layer) on the acetone system. The results in Fig.~\ref{fig:acetone_convergence_alternative} show that the energy gap estimation has reached chemical accuracy with just 4 entangling layers (28 two-qubit gates). The circuit depth and the two-qubit gate count are reasonable for near-term quantum devices. Thus, we have run the circuit with the parameters being set at the optimal point on a noisy simulator (the noise model is based on the thermal relaxation channel from Appendix~\ref{appendix:Noise Model}). The results for noisy simulation can be found in Table~\ref{tab:EnegyDifferenceAcetoneNoisy} where the VQE errors and the energy gaps are found to be close to chemical accuracy.

\begin{table}[t]
%\centering
\begin{ruledtabular}
\begin{tabular}{ccc}
 & acetone & propen-2-ol \\
\hline
Full system CCSD energies relative to acetone & 0 & 24.070 \\
Exact active-space system energies relative to acetone & 0 & 23.773 \\
active-space system VQE energies relative to acetone & 0 & 22.801 \\
active-space system VQE Error & 2.787 & 1.805 \\
\end{tabular}
\end{ruledtabular}
\caption{Energy calculations of acetone and propen-2-ol in kcal/mol with the VQE results calculated on a noisy simulator. The energies are presented in the form of energy gaps (relative to acetone). Exact energies of the active-space systems were calculated by the diagonalization of the Hamiltonian. VQE energies were calculated using a 4-layer hardware-efficient ansatz (Fig.~\ref{fig:AlternativeAnsatzCircuitLayer}) with Gaussian initialization for the parameters on a noisy simulator (Appendix~\ref{appendix:Noise Model}). VQE errors were calculated from the energy difference between VQE energies and the exact energies of the active-space systems.}
\label{tab:EnegyDifferenceAcetoneNoisy}
\end{table}

\section{Discussion}

%\mh{What we did and motivation}
In this work, we have proposed a hybrid quantum-classical workflow for the prediction of the relative stability of tautomers. It is of the pharmaceutical industry's interest to accurately predict the preferred tautomeric state of a given molecules and its computation methodology. However, the computational requirements using classical approaches for quantum chemistry are too expensive, so we have made use of quantum simulation for electronic structure calculations.

%\mh{Previous work: Google's approach and results}
We are aware of some studies that are making the effort to realize quantum simulation as a practical application for quantum chemistry. Recently, Tazhigulov et al. \cite{tazhigulov_simulating_2022} simulated strongly-correlated molecules that are more relevant to real world problems on Google’s Sycamore quantum processor. They mapped the electronic structures of iron-sulfur molecular clusters and $\alpha$-\ch{RuCl3} into low-energy spin models using the results from theoretical and spectroscopic studies. This simplification reduced the qubit requirements and rendered the possibility of simulating larger molecules. They then used  the finite-temperature version of quantum imaginary time evolution to capture physical observables. However, the circuit depth for the evolution were too large for the quantum device. Thus, they recompiled the circuit on a classical noiseless emulator, and numerous error mitigation techniques were employed. One of their experiments that used 9 qubits and 82 two-qubit gates achieved acceptable accuracy, while the data for the experiment that used 11 qubits and 310 two-qubit gates were not quantitatively meaningful. This research benchmarked the performance and limitation of quantum simulation on quantum hardware. It is, therefore, important to reduce hardware requirements of physical simulation to this limitation (e.g. using similar qubit and two-qubit gate counts). 

%\mh{Our approach: Natural orbital selection + maximizing the power of qubit reduction for QEE} 
Instead of utilizing empirical information, we have used \textit{ab initio} quantum chemistry methods to reduce quantum hardware requirements. Besides, the problem of interest is to find the relative ground state energy differences rather than absolute energies or other observables, so there are more rooms for orbital reduction. First, we optimize the geometry using B3LYP. Next, we use the natural orbitals from MP2 to select active orbitals. The selection criteria includes the occupation numbers and hardware limitation. To exploit as much quantum resources, we map the reduced Hamiltonian onto fewer qubits using QEE. Finally, we simulate the molecules/tautomers with 12 active spin-orbitals using VQE with 8 qubits and 80 two-qubit gates, which is fairly close to the quantum hardware limitation from \cite{tazhigulov_simulating_2022} (9 qubits and 82 two-qubit gates). In this work, not much efforts were done on exploring the best ansatz circuit for the VQE simulation, but we have found that an alternative ansatz circuit with 28 two-qubit gates can sufficiently provide accurate VQE calculations for the acetone system. This two-qubit gate count is much fewer than the quantum hardware limitation from \cite{tazhigulov_simulating_2022} which shows the applicability of our methodology. % \mh{28 two-qubit gates for what? doing the same task as the 82 two-qbuit gates?}
% \yu{I modified so that it's clear now}

%\mh{Result interpretation}
The VQE results for predicted stability agrees with those from full system CCSD results. For the prediction of acetone, its keto form is more stable than its enol form. For the case of Edaravone, its keto form is the most stable and its amine form is the most unstable. Note that these results did not incorporate solvent effects, which is an important factor in consideration of tautomerism in biological systems. Continuum solvation models, such as Polarizable Continuum Model (PCM)\cite{Amovilli1998-ii}, the Solvation Model based on Density (SMD)\cite{Marenich2009-vg} or the Conductor-like Screening Model (COSMO)\cite{Klamt1993-ra}, can be used to take into account the solvent effects. For example the PCM-VQE algorithm provides a self-consistent way to include the solvent effect which does not require extra quantum resources.
%but \cite{castaldo_quantum_2021} provides a polarizable continuum model (PCM) VQE method to take the effects into account in a self-consistent way that do not require extra quantum resources.

The CCSD results are considered to be the solution for the comparison with VQE results since FCI is too expensive for such large systems. Nevertheless, there could be even larger systems where CCSD starts becoming intractable. This is where our hybrid quantum-classical workflow could come into play because MP2 (which provides natural orbital occupancy) is computationally cheaper than CCSD and quantum simulation has the potential of becoming a more advantageous approach. 

Note that even though the VQE data are from noiseless simulator, there are still some VQE errors. The reasons could be that either the ansatz circuit could not capture the numerically true ground states or the parameter optimization stopped at local minimum or got stuck on barren plateau. Nevertheless, the errors are within chemical accuracy and does not affect the stability prediction. 

%\mh{Our method's value in NISQ era}
The inclusion of active space selection process in this work alleviates the burden on quantum resources. Of course a quantum device with more qubits that are less error-prone can bring value into quantum chemistry, but we are not close to achieving large-scale fault-tolerant in any architecture. Therefore, the workflow and methodology for quantum chemistry presented here is valuable in the NISQ era. As quantum computers will certainly be improved, the active orbital selection criteria can be adjusted such that more orbitals can be included. Such versatility stems from the fact that the limitation of quantum hardware is included in the selection criteria.

%\mh{Towards useful quantum computing applications in prediction of chemical events and drug discovery/Closing notes}
Given the importance of tautomerism in pharmaceutical research, our work aims to extend the applications of quantum computing in drug discovery and predictions of chemical events. To realize useful applications for quantum computers in near term, hybrid methods and reasonable classical preprocessing are necessary. On the other hand, the classical methods should be adaptable to the advancement of future quantum computers. Certain modifications on the quantum subroutines can also be done to improve the whole picture. We have employed QEE to increase the maximal number of spin-orbitals that can be simulated on limited quantum resources. Besides, to avoid the local minimum/barren plateau problem seen in this work, one could employ some parameter optimization strategies such as the initialization strategy proposed in \cite{kay_2022} and \cite{grant_initialization_2019} or one could change the VQE subroutine to methods such as imaginary time evolution \cite{mcardle_variational_2019} for better convergence when future hardware admits.

%\section{Method}

\bibliographystyle{apsrev4-2}
\bibliography{bib.bib}

\begin{thebibliography}{66}%
\makeatletter
\providecommand \@ifxundefined [1]{%
 \@ifx{#1\undefined}
}%
\providecommand \@ifnum [1]{%
 \ifnum #1\expandafter \@firstoftwo
 \else \expandafter \@secondoftwo
 \fi
}%
\providecommand \@ifx [1]{%
 \ifx #1\expandafter \@firstoftwo
 \else \expandafter \@secondoftwo
 \fi
}%
\providecommand \natexlab [1]{#1}%
\providecommand \enquote  [1]{``#1''}%
\providecommand \bibnamefont  [1]{#1}%
\providecommand \bibfnamefont [1]{#1}%
\providecommand \citenamefont [1]{#1}%
\providecommand \href@noop [0]{\@secondoftwo}%
\providecommand \href [0]{\begingroup \@sanitize@url \@href}%
\providecommand \@href[1]{\@@startlink{#1}\@@href}%
\providecommand \@@href[1]{\endgroup#1\@@endlink}%
\providecommand \@sanitize@url [0]{\catcode `\\12\catcode `\$12\catcode
  `\&12\catcode `\#12\catcode `\^12\catcode `\_12\catcode `\%12\relax}%
\providecommand \@@startlink[1]{}%
\providecommand \@@endlink[0]{}%
\providecommand \url  [0]{\begingroup\@sanitize@url \@url }%
\providecommand \@url [1]{\endgroup\@href {#1}{\urlprefix }}%
\providecommand \urlprefix  [0]{URL }%
\providecommand \Eprint [0]{\href }%
\providecommand \doibase [0]{https://doi.org/}%
\providecommand \selectlanguage [0]{\@gobble}%
\providecommand \bibinfo  [0]{\@secondoftwo}%
\providecommand \bibfield  [0]{\@secondoftwo}%
\providecommand \translation [1]{[#1]}%
\providecommand \BibitemOpen [0]{}%
\providecommand \bibitemStop [0]{}%
\providecommand \bibitemNoStop [0]{.\EOS\space}%
\providecommand \EOS [0]{\spacefactor3000\relax}%
\providecommand \BibitemShut  [1]{\csname bibitem#1\endcsname}%
\let\auto@bib@innerbib\@empty
%</preamble>
\bibitem [{\citenamefont {Antonov}(2016)}]{Antonov2016-vb}%
  \BibitemOpen
  \bibfield  {author} {\bibinfo {author} {\bibfnamefont {L.}~\bibnamefont
  {Antonov}},\ }\href@noop {} {\emph {\bibinfo {title} {Tautomerism : Concepts
  and Applications in Science and Technology}}}\ (\bibinfo  {publisher}
  {Wiley},\ \bibinfo {year} {2016})\BibitemShut {NoStop}%
\bibitem [{\citenamefont {Muller}(1994)}]{Muller1994-xf}%
  \BibitemOpen
  \bibfield  {author} {\bibinfo {author} {\bibfnamefont {P.}~\bibnamefont
  {Muller}},\ }\href@noop {} {\bibfield  {journal} {\bibinfo  {journal} {J.
  Macromol. Sci. Part A Pure Appl. Chem.}\ }\textbf {\bibinfo {volume} {66}},\
  \bibinfo {pages} {1077} (\bibinfo {year} {1994})}\BibitemShut {NoStop}%
\bibitem [{\citenamefont {Alkorta}\ \emph {et~al.}(2007)\citenamefont
  {Alkorta}, \citenamefont {Goya}, \citenamefont {Elguero},\ and\ \citenamefont
  {Singh}}]{Alkorta2007-tn}%
  \BibitemOpen
  \bibfield  {author} {\bibinfo {author} {\bibfnamefont {I.}~\bibnamefont
  {Alkorta}}, \bibinfo {author} {\bibfnamefont {P.}~\bibnamefont {Goya}},
  \bibinfo {author} {\bibfnamefont {J.}~\bibnamefont {Elguero}},\ and\ \bibinfo
  {author} {\bibfnamefont {S.~P.}\ \bibnamefont {Singh}},\ }\href@noop {}
  {\bibfield  {journal} {\bibinfo  {journal} {National Academy Science
  Letters}\ }\textbf {\bibinfo {volume} {30}},\ \bibinfo {pages} {139}
  (\bibinfo {year} {2007})}\BibitemShut {NoStop}%
\bibitem [{\citenamefont {Wang}\ \emph {et~al.}(2011)\citenamefont {Wang},
  \citenamefont {Hellinga},\ and\ \citenamefont {Beese}}]{Wang2011-lf}%
  \BibitemOpen
  \bibfield  {author} {\bibinfo {author} {\bibfnamefont {W.}~\bibnamefont
  {Wang}}, \bibinfo {author} {\bibfnamefont {H.~W.}\ \bibnamefont {Hellinga}},\
  and\ \bibinfo {author} {\bibfnamefont {L.~S.}\ \bibnamefont {Beese}},\
  }\href@noop {} {\bibfield  {journal} {\bibinfo  {journal} {Proceedings of the
  National Academy of Sciences}\ }\textbf {\bibinfo {volume} {108}},\ \bibinfo
  {pages} {17644} (\bibinfo {year} {2011})}\BibitemShut {NoStop}%
\bibitem [{\citenamefont {Bebenek}\ \emph {et~al.}(2011)\citenamefont
  {Bebenek}, \citenamefont {Pedersen},\ and\ \citenamefont
  {Kunkel}}]{Bebenek2011-as}%
  \BibitemOpen
  \bibfield  {author} {\bibinfo {author} {\bibfnamefont {K.}~\bibnamefont
  {Bebenek}}, \bibinfo {author} {\bibfnamefont {L.~C.}\ \bibnamefont
  {Pedersen}},\ and\ \bibinfo {author} {\bibfnamefont {T.~A.}\ \bibnamefont
  {Kunkel}},\ }\href@noop {} {\bibfield  {journal} {\bibinfo  {journal}
  {Proceedings of the National Academy of Sciences}\ }\textbf {\bibinfo
  {volume} {108}},\ \bibinfo {pages} {1862} (\bibinfo {year}
  {2011})}\BibitemShut {NoStop}%
\bibitem [{\citenamefont {Martin}(2009)}]{Martin2009-yy}%
  \BibitemOpen
  \bibfield  {author} {\bibinfo {author} {\bibfnamefont {Y.~C.}\ \bibnamefont
  {Martin}},\ }\href@noop {} {\bibfield  {journal} {\bibinfo  {journal} {J.
  Comput. Aided Mol. Des.}\ }\textbf {\bibinfo {volume} {23}},\ \bibinfo
  {pages} {693} (\bibinfo {year} {2009})}\BibitemShut {NoStop}%
\bibitem [{\citenamefont {Warr}(2010)}]{Warr2010-pz}%
  \BibitemOpen
  \bibfield  {author} {\bibinfo {author} {\bibfnamefont {W.~A.}\ \bibnamefont
  {Warr}},\ }\href@noop {} {\bibfield  {journal} {\bibinfo  {journal} {J.
  Comput. Aided Mol. Des.}\ }\textbf {\bibinfo {volume} {24}},\ \bibinfo
  {pages} {497} (\bibinfo {year} {2010})}\BibitemShut {NoStop}%
\bibitem [{\citenamefont {Cohen}\ \emph {et~al.}(2008)\citenamefont {Cohen},
  \citenamefont {Mori-Sánchez},\ and\ \citenamefont
  {Yang}}]{cohen_insights_2008}%
  \BibitemOpen
  \bibfield  {author} {\bibinfo {author} {\bibfnamefont {A.~J.}\ \bibnamefont
  {Cohen}}, \bibinfo {author} {\bibfnamefont {P.}~\bibnamefont
  {Mori-Sánchez}},\ and\ \bibinfo {author} {\bibfnamefont {W.}~\bibnamefont
  {Yang}},\ }\href {https://doi.org/10.1126/science.1158722} {\bibfield
  {journal} {\bibinfo  {journal} {Science}\ }\textbf {\bibinfo {volume}
  {321}},\ \bibinfo {pages} {792} (\bibinfo {year} {2008})}\BibitemShut
  {NoStop}%
\bibitem [{\citenamefont {Dykstra}\ \emph {et~al.}(2011)\citenamefont
  {Dykstra}, \citenamefont {Frenking},\ and\ \citenamefont
  {Kim}}]{dykstra_theory_2011}%
  \BibitemOpen
  \bibfield  {author} {\bibinfo {author} {\bibfnamefont {C.}~\bibnamefont
  {Dykstra}}, \bibinfo {author} {\bibfnamefont {G.}~\bibnamefont {Frenking}},\
  and\ \bibinfo {author} {\bibfnamefont {K.}~\bibnamefont {Kim}},\ }\href
  {http://qut.eblib.com.au/patron/FullRecord.aspx?p=269993} {\emph {\bibinfo
  {title} {Theory and {Applications} of {Computational} {Chemistry}: the
  {First} {Forty} {Years}.}}}\ (\bibinfo  {publisher} {Elsevier Science},\
  \bibinfo {address} {Amsterdam},\ \bibinfo {year} {2011})\BibitemShut
  {NoStop}%
\bibitem [{\citenamefont {Feynman}(1982)}]{feynman_simulating_1982}%
  \BibitemOpen
  \bibfield  {author} {\bibinfo {author} {\bibfnamefont {R.~P.}\ \bibnamefont
  {Feynman}},\ }\href {https://doi.org/10.1007/BF02650179} {\bibfield
  {journal} {\bibinfo  {journal} {Int J Theor Phys}\ }\textbf {\bibinfo
  {volume} {21}},\ \bibinfo {pages} {467} (\bibinfo {year} {1982})}\BibitemShut
  {NoStop}%
\bibitem [{\citenamefont {Jordan}\ and\ \citenamefont
  {Wigner}(1928)}]{jordan_1928}%
  \BibitemOpen
  \bibfield  {author} {\bibinfo {author} {\bibfnamefont {P.}~\bibnamefont
  {Jordan}}\ and\ \bibinfo {author} {\bibfnamefont {E.}~\bibnamefont
  {Wigner}},\ }\href {https://doi.org/10.1007/BF01331938} {\bibfield  {journal}
  {\bibinfo  {journal} {Z. Physik}\ }\textbf {\bibinfo {volume} {47}},\
  \bibinfo {pages} {631} (\bibinfo {year} {1928})}\BibitemShut {NoStop}%
\bibitem [{\citenamefont {Somma}\ \emph {et~al.}(2002)\citenamefont {Somma},
  \citenamefont {Ortiz}, \citenamefont {Gubernatis}, \citenamefont {Knill},\
  and\ \citenamefont {Laflamme}}]{somma_simulating_2002}%
  \BibitemOpen
  \bibfield  {author} {\bibinfo {author} {\bibfnamefont {R.}~\bibnamefont
  {Somma}}, \bibinfo {author} {\bibfnamefont {G.}~\bibnamefont {Ortiz}},
  \bibinfo {author} {\bibfnamefont {J.~E.}\ \bibnamefont {Gubernatis}},
  \bibinfo {author} {\bibfnamefont {E.}~\bibnamefont {Knill}},\ and\ \bibinfo
  {author} {\bibfnamefont {R.}~\bibnamefont {Laflamme}},\ }\href
  {https://doi.org/10.1103/PhysRevA.65.042323} {\bibfield  {journal} {\bibinfo
  {journal} {Phys. Rev. A}\ }\textbf {\bibinfo {volume} {65}},\ \bibinfo
  {pages} {042323} (\bibinfo {year} {2002})}\BibitemShut {NoStop}%
\bibitem [{\citenamefont {Kitaev}(1995)}]{kitaev_quantum_1995}%
  \BibitemOpen
  \bibfield  {author} {\bibinfo {author} {\bibfnamefont {A.~Y.}\ \bibnamefont
  {Kitaev}},\ }\href {http://arxiv.org/abs/quant-ph/9511026} {\bibfield
  {journal} {\bibinfo  {journal} {arXiv:quant-ph/9511026}\ } (\bibinfo {year}
  {1995})},\ \bibinfo {note} {arXiv: quant-ph/9511026}\BibitemShut {NoStop}%
\bibitem [{\citenamefont {Du}\ \emph {et~al.}(2010)\citenamefont {Du},
  \citenamefont {Xu}, \citenamefont {Peng}, \citenamefont {Wang}, \citenamefont
  {Wu},\ and\ \citenamefont {Lu}}]{du_nmr_2010}%
  \BibitemOpen
  \bibfield  {author} {\bibinfo {author} {\bibfnamefont {J.}~\bibnamefont
  {Du}}, \bibinfo {author} {\bibfnamefont {N.}~\bibnamefont {Xu}}, \bibinfo
  {author} {\bibfnamefont {X.}~\bibnamefont {Peng}}, \bibinfo {author}
  {\bibfnamefont {P.}~\bibnamefont {Wang}}, \bibinfo {author} {\bibfnamefont
  {S.}~\bibnamefont {Wu}},\ and\ \bibinfo {author} {\bibfnamefont
  {D.}~\bibnamefont {Lu}},\ }\href
  {https://doi.org/10.1103/PhysRevLett.104.030502} {\bibfield  {journal}
  {\bibinfo  {journal} {Phys. Rev. Lett.}\ }\textbf {\bibinfo {volume} {104}},\
  \bibinfo {pages} {030502} (\bibinfo {year} {2010})}\BibitemShut {NoStop}%
\bibitem [{\citenamefont {Lanyon}\ \emph {et~al.}(2010)\citenamefont {Lanyon},
  \citenamefont {Whitfield}, \citenamefont {Gillett}, \citenamefont {Goggin},
  \citenamefont {Almeida}, \citenamefont {Kassal}, \citenamefont {Biamonte},
  \citenamefont {Mohseni}, \citenamefont {Powell}, \citenamefont {Barbieri},
  \citenamefont {Aspuru-Guzik},\ and\ \citenamefont
  {White}}]{lanyon_towards_2010}%
  \BibitemOpen
  \bibfield  {author} {\bibinfo {author} {\bibfnamefont {B.~P.}\ \bibnamefont
  {Lanyon}}, \bibinfo {author} {\bibfnamefont {J.~D.}\ \bibnamefont
  {Whitfield}}, \bibinfo {author} {\bibfnamefont {G.~G.}\ \bibnamefont
  {Gillett}}, \bibinfo {author} {\bibfnamefont {M.~E.}\ \bibnamefont {Goggin}},
  \bibinfo {author} {\bibfnamefont {M.~P.}\ \bibnamefont {Almeida}}, \bibinfo
  {author} {\bibfnamefont {I.}~\bibnamefont {Kassal}}, \bibinfo {author}
  {\bibfnamefont {J.~D.}\ \bibnamefont {Biamonte}}, \bibinfo {author}
  {\bibfnamefont {M.}~\bibnamefont {Mohseni}}, \bibinfo {author} {\bibfnamefont
  {B.~J.}\ \bibnamefont {Powell}}, \bibinfo {author} {\bibfnamefont
  {M.}~\bibnamefont {Barbieri}}, \bibinfo {author} {\bibfnamefont
  {A.}~\bibnamefont {Aspuru-Guzik}},\ and\ \bibinfo {author} {\bibfnamefont
  {A.~G.}\ \bibnamefont {White}},\ }\href {https://doi.org/10.1038/nchem.483}
  {\bibfield  {journal} {\bibinfo  {journal} {Nature Chem}\ }\textbf {\bibinfo
  {volume} {2}},\ \bibinfo {pages} {106} (\bibinfo {year} {2010})}\BibitemShut
  {NoStop}%
\bibitem [{\citenamefont {Li}\ \emph {et~al.}(2011)\citenamefont {Li},
  \citenamefont {Yung}, \citenamefont {Chen}, \citenamefont {Lu}, \citenamefont
  {Whitfield}, \citenamefont {Peng}, \citenamefont {Aspuru-Guzik},\ and\
  \citenamefont {Du}}]{li_solving_2011}%
  \BibitemOpen
  \bibfield  {author} {\bibinfo {author} {\bibfnamefont {Z.}~\bibnamefont
  {Li}}, \bibinfo {author} {\bibfnamefont {M.-H.}\ \bibnamefont {Yung}},
  \bibinfo {author} {\bibfnamefont {H.}~\bibnamefont {Chen}}, \bibinfo {author}
  {\bibfnamefont {D.}~\bibnamefont {Lu}}, \bibinfo {author} {\bibfnamefont
  {J.~D.}\ \bibnamefont {Whitfield}}, \bibinfo {author} {\bibfnamefont
  {X.}~\bibnamefont {Peng}}, \bibinfo {author} {\bibfnamefont {A.}~\bibnamefont
  {Aspuru-Guzik}},\ and\ \bibinfo {author} {\bibfnamefont {J.}~\bibnamefont
  {Du}},\ }\href {https://doi.org/10.1038/srep00088} {\bibfield  {journal}
  {\bibinfo  {journal} {Sci Rep}\ }\textbf {\bibinfo {volume} {1}},\ \bibinfo
  {pages} {88} (\bibinfo {year} {2011})}\BibitemShut {NoStop}%
\bibitem [{\citenamefont {O{\textquoteright}Malley}\ \emph
  {et~al.}(2016)\citenamefont {O{\textquoteright}Malley}, \citenamefont
  {Babbush}, \citenamefont {Kivlichan}, \citenamefont {Romero}, \citenamefont
  {McClean}, \citenamefont {Barends}, \citenamefont {Kelly}, \citenamefont
  {Roushan}, \citenamefont {Tranter}, \citenamefont {Ding}, \citenamefont
  {Campbell}, \citenamefont {Chen}, \citenamefont {Chen}, \citenamefont
  {Chiaro}, \citenamefont {Dunsworth}, \citenamefont {Fowler}, \citenamefont
  {Jeffrey}, \citenamefont {Lucero}, \citenamefont {Megrant}, \citenamefont
  {Mutus}, \citenamefont {Neeley}, \citenamefont {Neill}, \citenamefont
  {Quintana}, \citenamefont {Sank}, \citenamefont {Vainsencher}, \citenamefont
  {Wenner}, \citenamefont {White}, \citenamefont {Coveney}, \citenamefont
  {Love}, \citenamefont {Neven}, \citenamefont {Aspuru-Guzik},\ and\
  \citenamefont {Martinis}}]{omalley_scalable_2016}%
  \BibitemOpen
  \bibfield  {author} {\bibinfo {author} {\bibfnamefont {P.~J.~J.}\
  \bibnamefont {O{\textquoteright}Malley}}, \bibinfo {author} {\bibfnamefont
  {R.}~\bibnamefont {Babbush}}, \bibinfo {author} {\bibfnamefont {I.~D.}\
  \bibnamefont {Kivlichan}}, \bibinfo {author} {\bibfnamefont {J.}~\bibnamefont
  {Romero}}, \bibinfo {author} {\bibfnamefont {J.~R.}\ \bibnamefont {McClean}},
  \bibinfo {author} {\bibfnamefont {R.}~\bibnamefont {Barends}}, \bibinfo
  {author} {\bibfnamefont {J.}~\bibnamefont {Kelly}}, \bibinfo {author}
  {\bibfnamefont {P.}~\bibnamefont {Roushan}}, \bibinfo {author} {\bibfnamefont
  {A.}~\bibnamefont {Tranter}}, \bibinfo {author} {\bibfnamefont
  {N.}~\bibnamefont {Ding}}, \bibinfo {author} {\bibfnamefont {B.}~\bibnamefont
  {Campbell}}, \bibinfo {author} {\bibfnamefont {Y.}~\bibnamefont {Chen}},
  \bibinfo {author} {\bibfnamefont {Z.}~\bibnamefont {Chen}}, \bibinfo {author}
  {\bibfnamefont {B.}~\bibnamefont {Chiaro}}, \bibinfo {author} {\bibfnamefont
  {A.}~\bibnamefont {Dunsworth}}, \bibinfo {author} {\bibfnamefont {A.~G.}\
  \bibnamefont {Fowler}}, \bibinfo {author} {\bibfnamefont {E.}~\bibnamefont
  {Jeffrey}}, \bibinfo {author} {\bibfnamefont {E.}~\bibnamefont {Lucero}},
  \bibinfo {author} {\bibfnamefont {A.}~\bibnamefont {Megrant}}, \bibinfo
  {author} {\bibfnamefont {J.~Y.}\ \bibnamefont {Mutus}}, \bibinfo {author}
  {\bibfnamefont {M.}~\bibnamefont {Neeley}}, \bibinfo {author} {\bibfnamefont
  {C.}~\bibnamefont {Neill}}, \bibinfo {author} {\bibfnamefont
  {C.}~\bibnamefont {Quintana}}, \bibinfo {author} {\bibfnamefont
  {D.}~\bibnamefont {Sank}}, \bibinfo {author} {\bibfnamefont {A.}~\bibnamefont
  {Vainsencher}}, \bibinfo {author} {\bibfnamefont {J.}~\bibnamefont {Wenner}},
  \bibinfo {author} {\bibfnamefont {T.~C.}\ \bibnamefont {White}}, \bibinfo
  {author} {\bibfnamefont {P.~V.}\ \bibnamefont {Coveney}}, \bibinfo {author}
  {\bibfnamefont {P.~J.}\ \bibnamefont {Love}}, \bibinfo {author}
  {\bibfnamefont {H.}~\bibnamefont {Neven}}, \bibinfo {author} {\bibfnamefont
  {A.}~\bibnamefont {Aspuru-Guzik}},\ and\ \bibinfo {author} {\bibfnamefont
  {J.~M.}\ \bibnamefont {Martinis}},\ }\href
  {https://doi.org/10.1103/PhysRevX.6.031007} {\bibfield  {journal} {\bibinfo
  {journal} {Phys. Rev. X}\ }\textbf {\bibinfo {volume} {6}},\ \bibinfo {pages}
  {031007} (\bibinfo {year} {2016})}\BibitemShut {NoStop}%
\bibitem [{\citenamefont {Paesani}\ \emph {et~al.}(2017)\citenamefont
  {Paesani}, \citenamefont {Gentile}, \citenamefont {Santagati}, \citenamefont
  {Wang}, \citenamefont {Wiebe}, \citenamefont {Tew}, \citenamefont
  {O{\textquoteright}Brien},\ and\ \citenamefont
  {Thompson}}]{paesani_experimental_2017}%
  \BibitemOpen
  \bibfield  {author} {\bibinfo {author} {\bibfnamefont {S.}~\bibnamefont
  {Paesani}}, \bibinfo {author} {\bibfnamefont {A.~A.}\ \bibnamefont
  {Gentile}}, \bibinfo {author} {\bibfnamefont {R.}~\bibnamefont {Santagati}},
  \bibinfo {author} {\bibfnamefont {J.}~\bibnamefont {Wang}}, \bibinfo {author}
  {\bibfnamefont {N.}~\bibnamefont {Wiebe}}, \bibinfo {author} {\bibfnamefont
  {D.~P.}\ \bibnamefont {Tew}}, \bibinfo {author} {\bibfnamefont {J.~L.}\
  \bibnamefont {O{\textquoteright}Brien}},\ and\ \bibinfo {author}
  {\bibfnamefont {M.~G.}\ \bibnamefont {Thompson}},\ }\href
  {https://doi.org/10.1103/PhysRevLett.118.100503} {\bibfield  {journal}
  {\bibinfo  {journal} {Phys. Rev. Lett.}\ }\textbf {\bibinfo {volume} {118}},\
  \bibinfo {pages} {100503} (\bibinfo {year} {2017})}\BibitemShut {NoStop}%
\bibitem [{\citenamefont {Santagati}\ \emph {et~al.}(2018)\citenamefont
  {Santagati}, \citenamefont {Wang}, \citenamefont {Gentile}, \citenamefont
  {Paesani}, \citenamefont {Wiebe}, \citenamefont {McClean}, \citenamefont
  {Morley-Short}, \citenamefont {Shadbolt}, \citenamefont {Bonneau},
  \citenamefont {Silverstone}, \citenamefont {Tew}, \citenamefont {Zhou},
  \citenamefont {O’Brien},\ and\ \citenamefont
  {Thompson}}]{santagati_witnessing_2018}%
  \BibitemOpen
  \bibfield  {author} {\bibinfo {author} {\bibfnamefont {R.}~\bibnamefont
  {Santagati}}, \bibinfo {author} {\bibfnamefont {J.}~\bibnamefont {Wang}},
  \bibinfo {author} {\bibfnamefont {A.~A.}\ \bibnamefont {Gentile}}, \bibinfo
  {author} {\bibfnamefont {S.}~\bibnamefont {Paesani}}, \bibinfo {author}
  {\bibfnamefont {N.}~\bibnamefont {Wiebe}}, \bibinfo {author} {\bibfnamefont
  {J.~R.}\ \bibnamefont {McClean}}, \bibinfo {author} {\bibfnamefont
  {S.}~\bibnamefont {Morley-Short}}, \bibinfo {author} {\bibfnamefont {P.~J.}\
  \bibnamefont {Shadbolt}}, \bibinfo {author} {\bibfnamefont {D.}~\bibnamefont
  {Bonneau}}, \bibinfo {author} {\bibfnamefont {J.~W.}\ \bibnamefont
  {Silverstone}}, \bibinfo {author} {\bibfnamefont {D.~P.}\ \bibnamefont
  {Tew}}, \bibinfo {author} {\bibfnamefont {X.}~\bibnamefont {Zhou}}, \bibinfo
  {author} {\bibfnamefont {J.~L.}\ \bibnamefont {O’Brien}},\ and\ \bibinfo
  {author} {\bibfnamefont {M.~G.}\ \bibnamefont {Thompson}},\ }\href
  {https://doi.org/10.1126/sciadv.aap9646} {\bibfield  {journal} {\bibinfo
  {journal} {Sci. Adv.}\ }\textbf {\bibinfo {volume} {4}},\ \bibinfo {pages}
  {eaap9646} (\bibinfo {year} {2018})}\BibitemShut {NoStop}%
\bibitem [{\citenamefont {Wang}\ \emph {et~al.}(2015)\citenamefont {Wang},
  \citenamefont {Dolde}, \citenamefont {Biamonte}, \citenamefont {Babbush},
  \citenamefont {Bergholm}, \citenamefont {Yang}, \citenamefont {Jakobi},
  \citenamefont {Neumann}, \citenamefont {Aspuru-Guzik}, \citenamefont
  {Whitfield},\ and\ \citenamefont {Wrachtrup}}]{wang_quantum_2015}%
  \BibitemOpen
  \bibfield  {author} {\bibinfo {author} {\bibfnamefont {Y.}~\bibnamefont
  {Wang}}, \bibinfo {author} {\bibfnamefont {F.}~\bibnamefont {Dolde}},
  \bibinfo {author} {\bibfnamefont {J.}~\bibnamefont {Biamonte}}, \bibinfo
  {author} {\bibfnamefont {R.}~\bibnamefont {Babbush}}, \bibinfo {author}
  {\bibfnamefont {V.}~\bibnamefont {Bergholm}}, \bibinfo {author}
  {\bibfnamefont {S.}~\bibnamefont {Yang}}, \bibinfo {author} {\bibfnamefont
  {I.}~\bibnamefont {Jakobi}}, \bibinfo {author} {\bibfnamefont
  {P.}~\bibnamefont {Neumann}}, \bibinfo {author} {\bibfnamefont
  {A.}~\bibnamefont {Aspuru-Guzik}}, \bibinfo {author} {\bibfnamefont {J.~D.}\
  \bibnamefont {Whitfield}},\ and\ \bibinfo {author} {\bibfnamefont
  {J.}~\bibnamefont {Wrachtrup}},\ }\href
  {https://doi.org/10.1021/acsnano.5b01651} {\bibfield  {journal} {\bibinfo
  {journal} {ACS Nano}\ }\textbf {\bibinfo {volume} {9}},\ \bibinfo {pages}
  {7769} (\bibinfo {year} {2015})}\BibitemShut {NoStop}%
\bibitem [{\citenamefont {Abrams}\ and\ \citenamefont
  {Lloyd}(1999)}]{abrams_quantum_1999}%
  \BibitemOpen
  \bibfield  {author} {\bibinfo {author} {\bibfnamefont {D.~S.}\ \bibnamefont
  {Abrams}}\ and\ \bibinfo {author} {\bibfnamefont {S.}~\bibnamefont {Lloyd}},\
  }\href {https://doi.org/10.1103/PhysRevLett.83.5162} {\bibfield  {journal}
  {\bibinfo  {journal} {Phys. Rev. Lett.}\ }\textbf {\bibinfo {volume} {83}},\
  \bibinfo {pages} {5162} (\bibinfo {year} {1999})}\BibitemShut {NoStop}%
\bibitem [{\citenamefont {Aspuru-Guzik}(2005)}]{aspuru-guzik_simulated_2005}%
  \BibitemOpen
  \bibfield  {author} {\bibinfo {author} {\bibfnamefont {A.}~\bibnamefont
  {Aspuru-Guzik}},\ }\href {https://doi.org/10.1126/science.1113479} {\bibfield
   {journal} {\bibinfo  {journal} {Science}\ }\textbf {\bibinfo {volume}
  {309}},\ \bibinfo {pages} {1704} (\bibinfo {year} {2005})}\BibitemShut
  {NoStop}%
\bibitem [{\citenamefont {Peruzzo}\ \emph {et~al.}(2014)\citenamefont
  {Peruzzo}, \citenamefont {McClean}, \citenamefont {Shadbolt}, \citenamefont
  {Yung}, \citenamefont {Zhou}, \citenamefont {Love}, \citenamefont
  {Aspuru-Guzik},\ and\ \citenamefont {O’Brien}}]{peruzzo_variational_2014}%
  \BibitemOpen
  \bibfield  {author} {\bibinfo {author} {\bibfnamefont {A.}~\bibnamefont
  {Peruzzo}}, \bibinfo {author} {\bibfnamefont {J.}~\bibnamefont {McClean}},
  \bibinfo {author} {\bibfnamefont {P.}~\bibnamefont {Shadbolt}}, \bibinfo
  {author} {\bibfnamefont {M.-H.}\ \bibnamefont {Yung}}, \bibinfo {author}
  {\bibfnamefont {X.-Q.}\ \bibnamefont {Zhou}}, \bibinfo {author}
  {\bibfnamefont {P.~J.}\ \bibnamefont {Love}}, \bibinfo {author}
  {\bibfnamefont {A.}~\bibnamefont {Aspuru-Guzik}},\ and\ \bibinfo {author}
  {\bibfnamefont {J.~L.}\ \bibnamefont {O’Brien}},\ }\href
  {https://doi.org/10.1038/ncomms5213} {\bibfield  {journal} {\bibinfo
  {journal} {Nat Commun}\ }\textbf {\bibinfo {volume} {5}},\ \bibinfo {pages}
  {4213} (\bibinfo {year} {2014})}\BibitemShut {NoStop}%
\bibitem [{\citenamefont {McClean}\ \emph {et~al.}(2016)\citenamefont
  {McClean}, \citenamefont {Romero}, \citenamefont {Babbush},\ and\
  \citenamefont {Aspuru-Guzik}}]{mcclean_theory_2016}%
  \BibitemOpen
  \bibfield  {author} {\bibinfo {author} {\bibfnamefont {J.~R.}\ \bibnamefont
  {McClean}}, \bibinfo {author} {\bibfnamefont {J.}~\bibnamefont {Romero}},
  \bibinfo {author} {\bibfnamefont {R.}~\bibnamefont {Babbush}},\ and\ \bibinfo
  {author} {\bibfnamefont {A.}~\bibnamefont {Aspuru-Guzik}},\ }\href
  {https://doi.org/10.1088/1367-2630/18/2/023023} {\bibfield  {journal}
  {\bibinfo  {journal} {New J. Phys.}\ }\textbf {\bibinfo {volume} {18}},\
  \bibinfo {pages} {023023} (\bibinfo {year} {2016})}\BibitemShut {NoStop}%
\bibitem [{\citenamefont {Kandala}\ \emph {et~al.}(2017)\citenamefont
  {Kandala}, \citenamefont {Mezzacapo}, \citenamefont {Temme}, \citenamefont
  {Takita}, \citenamefont {Brink}, \citenamefont {Chow},\ and\ \citenamefont
  {Gambetta}}]{kandala_hardware-efficient_2017}%
  \BibitemOpen
  \bibfield  {author} {\bibinfo {author} {\bibfnamefont {A.}~\bibnamefont
  {Kandala}}, \bibinfo {author} {\bibfnamefont {A.}~\bibnamefont {Mezzacapo}},
  \bibinfo {author} {\bibfnamefont {K.}~\bibnamefont {Temme}}, \bibinfo
  {author} {\bibfnamefont {M.}~\bibnamefont {Takita}}, \bibinfo {author}
  {\bibfnamefont {M.}~\bibnamefont {Brink}}, \bibinfo {author} {\bibfnamefont
  {J.~M.}\ \bibnamefont {Chow}},\ and\ \bibinfo {author} {\bibfnamefont
  {J.~M.}\ \bibnamefont {Gambetta}},\ }\href
  {https://doi.org/10.1038/nature23879} {\bibfield  {journal} {\bibinfo
  {journal} {Nature}\ }\textbf {\bibinfo {volume} {549}},\ \bibinfo {pages}
  {242} (\bibinfo {year} {2017})}\BibitemShut {NoStop}%
\bibitem [{\citenamefont {Kandala}\ \emph {et~al.}(2019)\citenamefont
  {Kandala}, \citenamefont {Temme}, \citenamefont {Córcoles}, \citenamefont
  {Mezzacapo}, \citenamefont {Chow},\ and\ \citenamefont
  {Gambetta}}]{kandala_error_2019}%
  \BibitemOpen
  \bibfield  {author} {\bibinfo {author} {\bibfnamefont {A.}~\bibnamefont
  {Kandala}}, \bibinfo {author} {\bibfnamefont {K.}~\bibnamefont {Temme}},
  \bibinfo {author} {\bibfnamefont {A.~D.}\ \bibnamefont {Córcoles}}, \bibinfo
  {author} {\bibfnamefont {A.}~\bibnamefont {Mezzacapo}}, \bibinfo {author}
  {\bibfnamefont {J.~M.}\ \bibnamefont {Chow}},\ and\ \bibinfo {author}
  {\bibfnamefont {J.~M.}\ \bibnamefont {Gambetta}},\ }\href
  {https://doi.org/10.1038/s41586-019-1040-7} {\bibfield  {journal} {\bibinfo
  {journal} {Nature}\ }\textbf {\bibinfo {volume} {567}},\ \bibinfo {pages}
  {491} (\bibinfo {year} {2019})}\BibitemShut {NoStop}%
\bibitem [{\citenamefont {Jiang}\ \emph {et~al.}(2018)\citenamefont {Jiang},
  \citenamefont {Sung}, \citenamefont {Kechedzhi}, \citenamefont
  {Smelyanskiy},\ and\ \citenamefont {Boixo}}]{jiang_quantum_2018}%
  \BibitemOpen
  \bibfield  {author} {\bibinfo {author} {\bibfnamefont {Z.}~\bibnamefont
  {Jiang}}, \bibinfo {author} {\bibfnamefont {K.~J.}\ \bibnamefont {Sung}},
  \bibinfo {author} {\bibfnamefont {K.}~\bibnamefont {Kechedzhi}}, \bibinfo
  {author} {\bibfnamefont {V.~N.}\ \bibnamefont {Smelyanskiy}},\ and\ \bibinfo
  {author} {\bibfnamefont {S.}~\bibnamefont {Boixo}},\ }\href
  {https://doi.org/10.1103/PhysRevApplied.9.044036} {\bibfield  {journal}
  {\bibinfo  {journal} {Phys. Rev. Applied}\ }\textbf {\bibinfo {volume} {9}},\
  \bibinfo {pages} {044036} (\bibinfo {year} {2018})}\BibitemShut {NoStop}%
\bibitem [{\citenamefont {Kivlichan}\ \emph {et~al.}(2018)\citenamefont
  {Kivlichan}, \citenamefont {McClean}, \citenamefont {Wiebe}, \citenamefont
  {Gidney}, \citenamefont {Aspuru-Guzik}, \citenamefont {Chan},\ and\
  \citenamefont {Babbush}}]{kivlichan_quantum_2018}%
  \BibitemOpen
  \bibfield  {author} {\bibinfo {author} {\bibfnamefont {I.~D.}\ \bibnamefont
  {Kivlichan}}, \bibinfo {author} {\bibfnamefont {J.}~\bibnamefont {McClean}},
  \bibinfo {author} {\bibfnamefont {N.}~\bibnamefont {Wiebe}}, \bibinfo
  {author} {\bibfnamefont {C.}~\bibnamefont {Gidney}}, \bibinfo {author}
  {\bibfnamefont {A.}~\bibnamefont {Aspuru-Guzik}}, \bibinfo {author}
  {\bibfnamefont {G.~K.-L.}\ \bibnamefont {Chan}},\ and\ \bibinfo {author}
  {\bibfnamefont {R.}~\bibnamefont {Babbush}},\ }\href
  {https://doi.org/10.1103/PhysRevLett.120.110501} {\bibfield  {journal}
  {\bibinfo  {journal} {Phys. Rev. Lett.}\ }\textbf {\bibinfo {volume} {120}},\
  \bibinfo {pages} {110501} (\bibinfo {year} {2018})}\BibitemShut {NoStop}%
\bibitem [{\citenamefont {Wecker}\ \emph {et~al.}(2015)\citenamefont {Wecker},
  \citenamefont {Hastings}, \citenamefont {Wiebe}, \citenamefont {Clark},
  \citenamefont {Nayak},\ and\ \citenamefont {Troyer}}]{wecker_solving_2015}%
  \BibitemOpen
  \bibfield  {author} {\bibinfo {author} {\bibfnamefont {D.}~\bibnamefont
  {Wecker}}, \bibinfo {author} {\bibfnamefont {M.~B.}\ \bibnamefont
  {Hastings}}, \bibinfo {author} {\bibfnamefont {N.}~\bibnamefont {Wiebe}},
  \bibinfo {author} {\bibfnamefont {B.~K.}\ \bibnamefont {Clark}}, \bibinfo
  {author} {\bibfnamefont {C.}~\bibnamefont {Nayak}},\ and\ \bibinfo {author}
  {\bibfnamefont {M.}~\bibnamefont {Troyer}},\ }\href
  {https://doi.org/10.1103/PhysRevA.92.062318} {\bibfield  {journal} {\bibinfo
  {journal} {Phys. Rev. A}\ }\textbf {\bibinfo {volume} {92}},\ \bibinfo
  {pages} {062318} (\bibinfo {year} {2015})}\BibitemShut {NoStop}%
\bibitem [{\citenamefont {Babbush}\ \emph {et~al.}(2015)\citenamefont
  {Babbush}, \citenamefont {McClean}, \citenamefont {Wecker}, \citenamefont
  {Aspuru-Guzik},\ and\ \citenamefont {Wiebe}}]{babbush_chemical_2015}%
  \BibitemOpen
  \bibfield  {author} {\bibinfo {author} {\bibfnamefont {R.}~\bibnamefont
  {Babbush}}, \bibinfo {author} {\bibfnamefont {J.}~\bibnamefont {McClean}},
  \bibinfo {author} {\bibfnamefont {D.}~\bibnamefont {Wecker}}, \bibinfo
  {author} {\bibfnamefont {A.}~\bibnamefont {Aspuru-Guzik}},\ and\ \bibinfo
  {author} {\bibfnamefont {N.}~\bibnamefont {Wiebe}},\ }\href
  {https://doi.org/10.1103/PhysRevA.91.022311} {\bibfield  {journal} {\bibinfo
  {journal} {Phys. Rev. A}\ }\textbf {\bibinfo {volume} {91}},\ \bibinfo
  {pages} {022311} (\bibinfo {year} {2015})}\BibitemShut {NoStop}%
\bibitem [{\citenamefont {Sugisaki}\ \emph {et~al.}(2016)\citenamefont
  {Sugisaki}, \citenamefont {Yamamoto}, \citenamefont {Nakazawa}, \citenamefont
  {Toyota}, \citenamefont {Sato}, \citenamefont {Shiomi},\ and\ \citenamefont
  {Takui}}]{sugisaki_quantum_2016}%
  \BibitemOpen
  \bibfield  {author} {\bibinfo {author} {\bibfnamefont {K.}~\bibnamefont
  {Sugisaki}}, \bibinfo {author} {\bibfnamefont {S.}~\bibnamefont {Yamamoto}},
  \bibinfo {author} {\bibfnamefont {S.}~\bibnamefont {Nakazawa}}, \bibinfo
  {author} {\bibfnamefont {K.}~\bibnamefont {Toyota}}, \bibinfo {author}
  {\bibfnamefont {K.}~\bibnamefont {Sato}}, \bibinfo {author} {\bibfnamefont
  {D.}~\bibnamefont {Shiomi}},\ and\ \bibinfo {author} {\bibfnamefont
  {T.}~\bibnamefont {Takui}},\ }\href
  {https://doi.org/10.1021/acs.jpca.6b04932} {\bibfield  {journal} {\bibinfo
  {journal} {J. Phys. Chem. A}\ }\textbf {\bibinfo {volume} {120}},\ \bibinfo
  {pages} {6459} (\bibinfo {year} {2016})}\BibitemShut {NoStop}%
\bibitem [{\citenamefont {Sugisaki}\ \emph {et~al.}(2019)\citenamefont
  {Sugisaki}, \citenamefont {Nakazawa}, \citenamefont {Toyota}, \citenamefont
  {Sato}, \citenamefont {Shiomi},\ and\ \citenamefont
  {Takui}}]{sugisaki_quantum_2019}%
  \BibitemOpen
  \bibfield  {author} {\bibinfo {author} {\bibfnamefont {K.}~\bibnamefont
  {Sugisaki}}, \bibinfo {author} {\bibfnamefont {S.}~\bibnamefont {Nakazawa}},
  \bibinfo {author} {\bibfnamefont {K.}~\bibnamefont {Toyota}}, \bibinfo
  {author} {\bibfnamefont {K.}~\bibnamefont {Sato}}, \bibinfo {author}
  {\bibfnamefont {D.}~\bibnamefont {Shiomi}},\ and\ \bibinfo {author}
  {\bibfnamefont {T.}~\bibnamefont {Takui}},\ }\href
  {https://doi.org/10.1021/acscentsci.8b00788} {\bibfield  {journal} {\bibinfo
  {journal} {ACS Cent. Sci.}\ }\textbf {\bibinfo {volume} {5}},\ \bibinfo
  {pages} {167} (\bibinfo {year} {2019})}\BibitemShut {NoStop}%
\bibitem [{\citenamefont {Du}\ \emph {et~al.}(2022)\citenamefont {Du},
  \citenamefont {Hsieh}, \citenamefont {Liu}, \citenamefont {You},\ and\
  \citenamefont {Tao}}]{PRXQuantum.3.030901}%
  \BibitemOpen
  \bibfield  {author} {\bibinfo {author} {\bibfnamefont {Y.}~\bibnamefont
  {Du}}, \bibinfo {author} {\bibfnamefont {M.-H.}\ \bibnamefont {Hsieh}},
  \bibinfo {author} {\bibfnamefont {T.}~\bibnamefont {Liu}}, \bibinfo {author}
  {\bibfnamefont {S.}~\bibnamefont {You}},\ and\ \bibinfo {author}
  {\bibfnamefont {D.}~\bibnamefont {Tao}},\ }\href
  {https://doi.org/10.1103/PRXQuantum.3.030901} {\bibfield  {journal} {\bibinfo
   {journal} {PRX Quantum}\ }\textbf {\bibinfo {volume} {3}},\ \bibinfo {pages}
  {030901} (\bibinfo {year} {2022})}\BibitemShut {NoStop}%
\bibitem [{\citenamefont {{Google AI Quantum and Collaborators*†}}\ \emph
  {et~al.}(2020)\citenamefont {{Google AI Quantum and Collaborators*†}},
  \citenamefont {Arute}, \citenamefont {Arya}, \citenamefont {Babbush},
  \citenamefont {Bacon}, \citenamefont {Bardin}, \citenamefont {Barends},
  \citenamefont {Boixo}, \citenamefont {Broughton}, \citenamefont {Buckley},
  \citenamefont {Buell}, \citenamefont {Burkett}, \citenamefont {Bushnell},
  \citenamefont {Chen}, \citenamefont {Chen}, \citenamefont {Chiaro},
  \citenamefont {Collins}, \citenamefont {Courtney}, \citenamefont {Demura},
  \citenamefont {Dunsworth}, \citenamefont {Farhi}, \citenamefont {Fowler},
  \citenamefont {Foxen}, \citenamefont {Gidney}, \citenamefont {Giustina},
  \citenamefont {Graff}, \citenamefont {Habegger}, \citenamefont {Harrigan},
  \citenamefont {Ho}, \citenamefont {Hong}, \citenamefont {Huang},
  \citenamefont {Huggins}, \citenamefont {Ioffe}, \citenamefont {Isakov},
  \citenamefont {Jeffrey}, \citenamefont {Jiang}, \citenamefont {Jones},
  \citenamefont {Kafri}, \citenamefont {Kechedzhi}, \citenamefont {Kelly},
  \citenamefont {Kim}, \citenamefont {Klimov}, \citenamefont {Korotkov},
  \citenamefont {Kostritsa}, \citenamefont {Landhuis}, \citenamefont {Laptev},
  \citenamefont {Lindmark}, \citenamefont {Lucero}, \citenamefont {Martin},
  \citenamefont {Martinis}, \citenamefont {McClean}, \citenamefont {McEwen},
  \citenamefont {Megrant}, \citenamefont {Mi}, \citenamefont {Mohseni},
  \citenamefont {Mruczkiewicz}, \citenamefont {Mutus}, \citenamefont {Naaman},
  \citenamefont {Neeley}, \citenamefont {Neill}, \citenamefont {Neven},
  \citenamefont {Niu}, \citenamefont {O’Brien}, \citenamefont {Ostby},
  \citenamefont {Petukhov}, \citenamefont {Putterman}, \citenamefont
  {Quintana}, \citenamefont {Roushan}, \citenamefont {Rubin}, \citenamefont
  {Sank}, \citenamefont {Satzinger}, \citenamefont {Smelyanskiy}, \citenamefont
  {Strain}, \citenamefont {Sung}, \citenamefont {Szalay}, \citenamefont
  {Takeshita}, \citenamefont {Vainsencher}, \citenamefont {White},
  \citenamefont {Wiebe}, \citenamefont {Yao}, \citenamefont {Yeh},\ and\
  \citenamefont
  {Zalcman}}]{google_ai_quantum_and_collaborators_hartree-fock_2020}%
  \BibitemOpen
  \bibfield  {author} {\bibinfo {author} {\bibnamefont {{Google AI Quantum and
  Collaborators*†}}}, \bibinfo {author} {\bibfnamefont {F.}~\bibnamefont
  {Arute}}, \bibinfo {author} {\bibfnamefont {K.}~\bibnamefont {Arya}},
  \bibinfo {author} {\bibfnamefont {R.}~\bibnamefont {Babbush}}, \bibinfo
  {author} {\bibfnamefont {D.}~\bibnamefont {Bacon}}, \bibinfo {author}
  {\bibfnamefont {J.~C.}\ \bibnamefont {Bardin}}, \bibinfo {author}
  {\bibfnamefont {R.}~\bibnamefont {Barends}}, \bibinfo {author} {\bibfnamefont
  {S.}~\bibnamefont {Boixo}}, \bibinfo {author} {\bibfnamefont
  {M.}~\bibnamefont {Broughton}}, \bibinfo {author} {\bibfnamefont {B.~B.}\
  \bibnamefont {Buckley}}, \bibinfo {author} {\bibfnamefont {D.~A.}\
  \bibnamefont {Buell}}, \bibinfo {author} {\bibfnamefont {B.}~\bibnamefont
  {Burkett}}, \bibinfo {author} {\bibfnamefont {N.}~\bibnamefont {Bushnell}},
  \bibinfo {author} {\bibfnamefont {Y.}~\bibnamefont {Chen}}, \bibinfo {author}
  {\bibfnamefont {Z.}~\bibnamefont {Chen}}, \bibinfo {author} {\bibfnamefont
  {B.}~\bibnamefont {Chiaro}}, \bibinfo {author} {\bibfnamefont
  {R.}~\bibnamefont {Collins}}, \bibinfo {author} {\bibfnamefont
  {W.}~\bibnamefont {Courtney}}, \bibinfo {author} {\bibfnamefont
  {S.}~\bibnamefont {Demura}}, \bibinfo {author} {\bibfnamefont
  {A.}~\bibnamefont {Dunsworth}}, \bibinfo {author} {\bibfnamefont
  {E.}~\bibnamefont {Farhi}}, \bibinfo {author} {\bibfnamefont
  {A.}~\bibnamefont {Fowler}}, \bibinfo {author} {\bibfnamefont
  {B.}~\bibnamefont {Foxen}}, \bibinfo {author} {\bibfnamefont
  {C.}~\bibnamefont {Gidney}}, \bibinfo {author} {\bibfnamefont
  {M.}~\bibnamefont {Giustina}}, \bibinfo {author} {\bibfnamefont
  {R.}~\bibnamefont {Graff}}, \bibinfo {author} {\bibfnamefont
  {S.}~\bibnamefont {Habegger}}, \bibinfo {author} {\bibfnamefont {M.~P.}\
  \bibnamefont {Harrigan}}, \bibinfo {author} {\bibfnamefont {A.}~\bibnamefont
  {Ho}}, \bibinfo {author} {\bibfnamefont {S.}~\bibnamefont {Hong}}, \bibinfo
  {author} {\bibfnamefont {T.}~\bibnamefont {Huang}}, \bibinfo {author}
  {\bibfnamefont {W.~J.}\ \bibnamefont {Huggins}}, \bibinfo {author}
  {\bibfnamefont {L.}~\bibnamefont {Ioffe}}, \bibinfo {author} {\bibfnamefont
  {S.~V.}\ \bibnamefont {Isakov}}, \bibinfo {author} {\bibfnamefont
  {E.}~\bibnamefont {Jeffrey}}, \bibinfo {author} {\bibfnamefont
  {Z.}~\bibnamefont {Jiang}}, \bibinfo {author} {\bibfnamefont
  {C.}~\bibnamefont {Jones}}, \bibinfo {author} {\bibfnamefont
  {D.}~\bibnamefont {Kafri}}, \bibinfo {author} {\bibfnamefont
  {K.}~\bibnamefont {Kechedzhi}}, \bibinfo {author} {\bibfnamefont
  {J.}~\bibnamefont {Kelly}}, \bibinfo {author} {\bibfnamefont
  {S.}~\bibnamefont {Kim}}, \bibinfo {author} {\bibfnamefont {P.~V.}\
  \bibnamefont {Klimov}}, \bibinfo {author} {\bibfnamefont {A.}~\bibnamefont
  {Korotkov}}, \bibinfo {author} {\bibfnamefont {F.}~\bibnamefont {Kostritsa}},
  \bibinfo {author} {\bibfnamefont {D.}~\bibnamefont {Landhuis}}, \bibinfo
  {author} {\bibfnamefont {P.}~\bibnamefont {Laptev}}, \bibinfo {author}
  {\bibfnamefont {M.}~\bibnamefont {Lindmark}}, \bibinfo {author}
  {\bibfnamefont {E.}~\bibnamefont {Lucero}}, \bibinfo {author} {\bibfnamefont
  {O.}~\bibnamefont {Martin}}, \bibinfo {author} {\bibfnamefont {J.~M.}\
  \bibnamefont {Martinis}}, \bibinfo {author} {\bibfnamefont {J.~R.}\
  \bibnamefont {McClean}}, \bibinfo {author} {\bibfnamefont {M.}~\bibnamefont
  {McEwen}}, \bibinfo {author} {\bibfnamefont {A.}~\bibnamefont {Megrant}},
  \bibinfo {author} {\bibfnamefont {X.}~\bibnamefont {Mi}}, \bibinfo {author}
  {\bibfnamefont {M.}~\bibnamefont {Mohseni}}, \bibinfo {author} {\bibfnamefont
  {W.}~\bibnamefont {Mruczkiewicz}}, \bibinfo {author} {\bibfnamefont
  {J.}~\bibnamefont {Mutus}}, \bibinfo {author} {\bibfnamefont
  {O.}~\bibnamefont {Naaman}}, \bibinfo {author} {\bibfnamefont
  {M.}~\bibnamefont {Neeley}}, \bibinfo {author} {\bibfnamefont
  {C.}~\bibnamefont {Neill}}, \bibinfo {author} {\bibfnamefont
  {H.}~\bibnamefont {Neven}}, \bibinfo {author} {\bibfnamefont {M.~Y.}\
  \bibnamefont {Niu}}, \bibinfo {author} {\bibfnamefont {T.~E.}\ \bibnamefont
  {O’Brien}}, \bibinfo {author} {\bibfnamefont {E.}~\bibnamefont {Ostby}},
  \bibinfo {author} {\bibfnamefont {A.}~\bibnamefont {Petukhov}}, \bibinfo
  {author} {\bibfnamefont {H.}~\bibnamefont {Putterman}}, \bibinfo {author}
  {\bibfnamefont {C.}~\bibnamefont {Quintana}}, \bibinfo {author}
  {\bibfnamefont {P.}~\bibnamefont {Roushan}}, \bibinfo {author} {\bibfnamefont
  {N.~C.}\ \bibnamefont {Rubin}}, \bibinfo {author} {\bibfnamefont
  {D.}~\bibnamefont {Sank}}, \bibinfo {author} {\bibfnamefont {K.~J.}\
  \bibnamefont {Satzinger}}, \bibinfo {author} {\bibfnamefont {V.}~\bibnamefont
  {Smelyanskiy}}, \bibinfo {author} {\bibfnamefont {D.}~\bibnamefont {Strain}},
  \bibinfo {author} {\bibfnamefont {K.~J.}\ \bibnamefont {Sung}}, \bibinfo
  {author} {\bibfnamefont {M.}~\bibnamefont {Szalay}}, \bibinfo {author}
  {\bibfnamefont {T.~Y.}\ \bibnamefont {Takeshita}}, \bibinfo {author}
  {\bibfnamefont {A.}~\bibnamefont {Vainsencher}}, \bibinfo {author}
  {\bibfnamefont {T.}~\bibnamefont {White}}, \bibinfo {author} {\bibfnamefont
  {N.}~\bibnamefont {Wiebe}}, \bibinfo {author} {\bibfnamefont {Z.~J.}\
  \bibnamefont {Yao}}, \bibinfo {author} {\bibfnamefont {P.}~\bibnamefont
  {Yeh}},\ and\ \bibinfo {author} {\bibfnamefont {A.}~\bibnamefont {Zalcman}},\
  }\href {https://doi.org/10.1126/science.abb9811} {\bibfield  {journal}
  {\bibinfo  {journal} {Science}\ }\textbf {\bibinfo {volume} {369}},\ \bibinfo
  {pages} {1084} (\bibinfo {year} {2020})}\BibitemShut {NoStop}%
\bibitem [{\citenamefont {Nam}\ \emph {et~al.}(2020)\citenamefont {Nam},
  \citenamefont {Chen}, \citenamefont {Pisenti}, \citenamefont {Wright},
  \citenamefont {Delaney}, \citenamefont {Maslov}, \citenamefont {Brown},
  \citenamefont {Allen}, \citenamefont {Amini}, \citenamefont {Apisdorf},
  \citenamefont {Beck}, \citenamefont {Blinov}, \citenamefont {Chaplin},
  \citenamefont {Chmielewski}, \citenamefont {Collins}, \citenamefont
  {Debnath}, \citenamefont {Hudek}, \citenamefont {Ducore}, \citenamefont
  {Keesan}, \citenamefont {Kreikemeier}, \citenamefont {Mizrahi}, \citenamefont
  {Solomon}, \citenamefont {Williams}, \citenamefont {Wong-Campos},
  \citenamefont {Moehring}, \citenamefont {Monroe},\ and\ \citenamefont
  {Kim}}]{nam_ground-state_2020}%
  \BibitemOpen
  \bibfield  {author} {\bibinfo {author} {\bibfnamefont {Y.}~\bibnamefont
  {Nam}}, \bibinfo {author} {\bibfnamefont {J.-S.}\ \bibnamefont {Chen}},
  \bibinfo {author} {\bibfnamefont {N.~C.}\ \bibnamefont {Pisenti}}, \bibinfo
  {author} {\bibfnamefont {K.}~\bibnamefont {Wright}}, \bibinfo {author}
  {\bibfnamefont {C.}~\bibnamefont {Delaney}}, \bibinfo {author} {\bibfnamefont
  {D.}~\bibnamefont {Maslov}}, \bibinfo {author} {\bibfnamefont {K.~R.}\
  \bibnamefont {Brown}}, \bibinfo {author} {\bibfnamefont {S.}~\bibnamefont
  {Allen}}, \bibinfo {author} {\bibfnamefont {J.~M.}\ \bibnamefont {Amini}},
  \bibinfo {author} {\bibfnamefont {J.}~\bibnamefont {Apisdorf}}, \bibinfo
  {author} {\bibfnamefont {K.~M.}\ \bibnamefont {Beck}}, \bibinfo {author}
  {\bibfnamefont {A.}~\bibnamefont {Blinov}}, \bibinfo {author} {\bibfnamefont
  {V.}~\bibnamefont {Chaplin}}, \bibinfo {author} {\bibfnamefont
  {M.}~\bibnamefont {Chmielewski}}, \bibinfo {author} {\bibfnamefont
  {C.}~\bibnamefont {Collins}}, \bibinfo {author} {\bibfnamefont
  {S.}~\bibnamefont {Debnath}}, \bibinfo {author} {\bibfnamefont {K.~M.}\
  \bibnamefont {Hudek}}, \bibinfo {author} {\bibfnamefont {A.~M.}\ \bibnamefont
  {Ducore}}, \bibinfo {author} {\bibfnamefont {M.}~\bibnamefont {Keesan}},
  \bibinfo {author} {\bibfnamefont {S.~M.}\ \bibnamefont {Kreikemeier}},
  \bibinfo {author} {\bibfnamefont {J.}~\bibnamefont {Mizrahi}}, \bibinfo
  {author} {\bibfnamefont {P.}~\bibnamefont {Solomon}}, \bibinfo {author}
  {\bibfnamefont {M.}~\bibnamefont {Williams}}, \bibinfo {author}
  {\bibfnamefont {J.~D.}\ \bibnamefont {Wong-Campos}}, \bibinfo {author}
  {\bibfnamefont {D.}~\bibnamefont {Moehring}}, \bibinfo {author}
  {\bibfnamefont {C.}~\bibnamefont {Monroe}},\ and\ \bibinfo {author}
  {\bibfnamefont {J.}~\bibnamefont {Kim}},\ }\href
  {https://doi.org/10.1038/s41534-020-0259-3} {\bibfield  {journal} {\bibinfo
  {journal} {npj Quantum Inf}\ }\textbf {\bibinfo {volume} {6}},\ \bibinfo
  {pages} {33} (\bibinfo {year} {2020})}\BibitemShut {NoStop}%
\bibitem [{\citenamefont {Cao}\ \emph {et~al.}(2021)\citenamefont {Cao},
  \citenamefont {Hu}, \citenamefont {Zhang}, \citenamefont {Xu}, \citenamefont
  {Chen}, \citenamefont {Yu}, \citenamefont {Li}, \citenamefont {Hu},
  \citenamefont {Lv},\ and\ \citenamefont {Yung}}]{cao_towards_2021}%
  \BibitemOpen
  \bibfield  {author} {\bibinfo {author} {\bibfnamefont {C.}~\bibnamefont
  {Cao}}, \bibinfo {author} {\bibfnamefont {J.}~\bibnamefont {Hu}}, \bibinfo
  {author} {\bibfnamefont {W.}~\bibnamefont {Zhang}}, \bibinfo {author}
  {\bibfnamefont {X.}~\bibnamefont {Xu}}, \bibinfo {author} {\bibfnamefont
  {D.}~\bibnamefont {Chen}}, \bibinfo {author} {\bibfnamefont {F.}~\bibnamefont
  {Yu}}, \bibinfo {author} {\bibfnamefont {J.}~\bibnamefont {Li}}, \bibinfo
  {author} {\bibfnamefont {H.}~\bibnamefont {Hu}}, \bibinfo {author}
  {\bibfnamefont {D.}~\bibnamefont {Lv}},\ and\ \bibinfo {author}
  {\bibfnamefont {M.-H.}\ \bibnamefont {Yung}},\ }\href
  {http://arxiv.org/abs/2109.02110} {\bibfield  {journal} {\bibinfo  {journal}
  {arXiv:2109.02110 [quant-ph]}\ } (\bibinfo {year} {2021})},\ \bibinfo {note}
  {arXiv: 2109.02110}\BibitemShut {NoStop}%
\bibitem [{\citenamefont {Li}\ \emph {et~al.}(2021)\citenamefont {Li},
  \citenamefont {Huang}, \citenamefont {Cao}, \citenamefont {Huang},
  \citenamefont {Shuai}, \citenamefont {Sun}, \citenamefont {Sun},
  \citenamefont {Yuan},\ and\ \citenamefont {Lv}}]{li_toward_2021}%
  \BibitemOpen
  \bibfield  {author} {\bibinfo {author} {\bibfnamefont {W.}~\bibnamefont
  {Li}}, \bibinfo {author} {\bibfnamefont {Z.}~\bibnamefont {Huang}}, \bibinfo
  {author} {\bibfnamefont {C.}~\bibnamefont {Cao}}, \bibinfo {author}
  {\bibfnamefont {Y.}~\bibnamefont {Huang}}, \bibinfo {author} {\bibfnamefont
  {Z.}~\bibnamefont {Shuai}}, \bibinfo {author} {\bibfnamefont
  {X.}~\bibnamefont {Sun}}, \bibinfo {author} {\bibfnamefont {J.}~\bibnamefont
  {Sun}}, \bibinfo {author} {\bibfnamefont {X.}~\bibnamefont {Yuan}},\ and\
  \bibinfo {author} {\bibfnamefont {D.}~\bibnamefont {Lv}},\ }\href
  {http://arxiv.org/abs/2109.08062} {\bibfield  {journal} {\bibinfo  {journal}
  {arXiv:2109.08062 [quant-ph]}\ } (\bibinfo {year} {2021})},\ \bibinfo {note}
  {arXiv: 2109.08062}\BibitemShut {NoStop}%
\bibitem [{\citenamefont {Tazhigulov}\ \emph {et~al.}(2022)\citenamefont
  {Tazhigulov}, \citenamefont {Sun}, \citenamefont {Haghshenas}, \citenamefont
  {Zhai}, \citenamefont {Tan}, \citenamefont {Rubin}, \citenamefont {Babbush},
  \citenamefont {Minnich},\ and\ \citenamefont
  {Chan}}]{tazhigulov_simulating_2022}%
  \BibitemOpen
  \bibfield  {author} {\bibinfo {author} {\bibfnamefont {R.~N.}\ \bibnamefont
  {Tazhigulov}}, \bibinfo {author} {\bibfnamefont {S.-N.}\ \bibnamefont {Sun}},
  \bibinfo {author} {\bibfnamefont {R.}~\bibnamefont {Haghshenas}}, \bibinfo
  {author} {\bibfnamefont {H.}~\bibnamefont {Zhai}}, \bibinfo {author}
  {\bibfnamefont {A.~T.~K.}\ \bibnamefont {Tan}}, \bibinfo {author}
  {\bibfnamefont {N.~C.}\ \bibnamefont {Rubin}}, \bibinfo {author}
  {\bibfnamefont {R.}~\bibnamefont {Babbush}}, \bibinfo {author} {\bibfnamefont
  {A.~J.}\ \bibnamefont {Minnich}},\ and\ \bibinfo {author} {\bibfnamefont
  {G.~K.-L.}\ \bibnamefont {Chan}},\ }\href {http://arxiv.org/abs/2203.15291}
  {\bibfield  {journal} {\bibinfo  {journal} {arXiv:2203.15291 [quant-ph]}\ }
  (\bibinfo {year} {2022})},\ \bibinfo {note} {arXiv: 2203.15291}\BibitemShut
  {NoStop}%
\bibitem [{\citenamefont {Bravyi}\ \emph {et~al.}(2017)\citenamefont {Bravyi},
  \citenamefont {Gambetta}, \citenamefont {Mezzacapo},\ and\ \citenamefont
  {Temme}}]{bravyi_tapering_2017}%
  \BibitemOpen
  \bibfield  {author} {\bibinfo {author} {\bibfnamefont {S.}~\bibnamefont
  {Bravyi}}, \bibinfo {author} {\bibfnamefont {J.~M.}\ \bibnamefont
  {Gambetta}}, \bibinfo {author} {\bibfnamefont {A.}~\bibnamefont
  {Mezzacapo}},\ and\ \bibinfo {author} {\bibfnamefont {K.}~\bibnamefont
  {Temme}},\ }\href {http://arxiv.org/abs/1701.08213} {\bibfield  {journal}
  {\bibinfo  {journal} {arXiv:1701.08213 [quant-ph]}\ } (\bibinfo {year}
  {2017})},\ \bibinfo {note} {arXiv: 1701.08213}\BibitemShut {NoStop}%
\bibitem [{\citenamefont {Moll}\ \emph {et~al.}(2016)\citenamefont {Moll},
  \citenamefont {Fuhrer}, \citenamefont {Staar},\ and\ \citenamefont
  {Tavernelli}}]{moll_optimizing_2016}%
  \BibitemOpen
  \bibfield  {author} {\bibinfo {author} {\bibfnamefont {N.}~\bibnamefont
  {Moll}}, \bibinfo {author} {\bibfnamefont {A.}~\bibnamefont {Fuhrer}},
  \bibinfo {author} {\bibfnamefont {P.}~\bibnamefont {Staar}},\ and\ \bibinfo
  {author} {\bibfnamefont {I.}~\bibnamefont {Tavernelli}},\ }\href
  {https://doi.org/10.1088/1751-8113/49/29/295301} {\bibfield  {journal}
  {\bibinfo  {journal} {J. Phys. A: Math. Theor.}\ }\textbf {\bibinfo {volume}
  {49}},\ \bibinfo {pages} {295301} (\bibinfo {year} {2016})}\BibitemShut
  {NoStop}%
\bibitem [{\citenamefont {Babbush}\ \emph {et~al.}(2018)\citenamefont
  {Babbush}, \citenamefont {Berry}, \citenamefont {Sanders}, \citenamefont
  {Kivlichan}, \citenamefont {Scherer}, \citenamefont {Wei}, \citenamefont
  {Love},\ and\ \citenamefont {Aspuru-Guzik}}]{babbush_exponentially_2018}%
  \BibitemOpen
  \bibfield  {author} {\bibinfo {author} {\bibfnamefont {R.}~\bibnamefont
  {Babbush}}, \bibinfo {author} {\bibfnamefont {D.~W.}\ \bibnamefont {Berry}},
  \bibinfo {author} {\bibfnamefont {Y.~R.}\ \bibnamefont {Sanders}}, \bibinfo
  {author} {\bibfnamefont {I.~D.}\ \bibnamefont {Kivlichan}}, \bibinfo {author}
  {\bibfnamefont {A.}~\bibnamefont {Scherer}}, \bibinfo {author} {\bibfnamefont
  {A.~Y.}\ \bibnamefont {Wei}}, \bibinfo {author} {\bibfnamefont {P.~J.}\
  \bibnamefont {Love}},\ and\ \bibinfo {author} {\bibfnamefont
  {A.}~\bibnamefont {Aspuru-Guzik}},\ }\href
  {https://doi.org/10.1088/2058-9565/aa9463} {\bibfield  {journal} {\bibinfo
  {journal} {Quantum Sci. Technol.}\ }\textbf {\bibinfo {volume} {3}},\
  \bibinfo {pages} {015006} (\bibinfo {year} {2018})},\ \Eprint
  {https://arxiv.org/abs/1506.01029} {1506.01029} \BibitemShut {NoStop}%
\bibitem [{\citenamefont {Steudtner}\ and\ \citenamefont
  {Wehner}(2018)}]{steudtner_fermion--qubit_2018}%
  \BibitemOpen
  \bibfield  {author} {\bibinfo {author} {\bibfnamefont {M.}~\bibnamefont
  {Steudtner}}\ and\ \bibinfo {author} {\bibfnamefont {S.}~\bibnamefont
  {Wehner}},\ }\href {https://doi.org/10.1088/1367-2630/aac54f} {\bibfield
  {journal} {\bibinfo  {journal} {New J. Phys.}\ }\textbf {\bibinfo {volume}
  {20}},\ \bibinfo {pages} {063010} (\bibinfo {year} {2018})}\BibitemShut
  {NoStop}%
\bibitem [{\citenamefont {Kirby}\ \emph {et~al.}(2021)\citenamefont {Kirby},
  \citenamefont {Fuller}, \citenamefont {Hadfield},\ and\ \citenamefont
  {Mezzacapo}}]{kirby_second-quantized_2021}%
  \BibitemOpen
  \bibfield  {author} {\bibinfo {author} {\bibfnamefont {W.}~\bibnamefont
  {Kirby}}, \bibinfo {author} {\bibfnamefont {B.}~\bibnamefont {Fuller}},
  \bibinfo {author} {\bibfnamefont {C.}~\bibnamefont {Hadfield}},\ and\
  \bibinfo {author} {\bibfnamefont {A.}~\bibnamefont {Mezzacapo}},\ }\href
  {http://arxiv.org/abs/2109.14465} {\bibfield  {journal} {\bibinfo  {journal}
  {{arXiv}:2109.14465 [quant-ph]}\ } (\bibinfo {year} {2021})},\ \Eprint
  {https://arxiv.org/abs/2109.14465} {2109.14465} \BibitemShut {NoStop}%
\bibitem [{\citenamefont {Shee}\ \emph {et~al.}(2021)\citenamefont {Shee},
  \citenamefont {Tsai}, \citenamefont {Hong}, \citenamefont {Cheng},\ and\
  \citenamefont {Goan}}]{shee_qubit-efficient_2021}%
  \BibitemOpen
  \bibfield  {author} {\bibinfo {author} {\bibfnamefont {Y.}~\bibnamefont
  {Shee}}, \bibinfo {author} {\bibfnamefont {P.-K.}\ \bibnamefont {Tsai}},
  \bibinfo {author} {\bibfnamefont {C.-L.}\ \bibnamefont {Hong}}, \bibinfo
  {author} {\bibfnamefont {H.-C.}\ \bibnamefont {Cheng}},\ and\ \bibinfo
  {author} {\bibfnamefont {H.-S.}\ \bibnamefont {Goan}},\ }\href
  {http://arxiv.org/abs/2110.04112} {\bibfield  {journal} {\bibinfo  {journal}
  {arXiv:2110.04112 [cond-mat, physics:physics, physics:quant-ph]}\ } (\bibinfo
  {year} {2021})},\ \bibinfo {note} {arXiv: 2110.04112}\BibitemShut {NoStop}%
\bibitem [{\citenamefont {Rathi}\ \emph {et~al.}(2020)\citenamefont {Rathi},
  \citenamefont {Ludlow},\ and\ \citenamefont {Verdonk}}]{Rathi2020-ki}%
  \BibitemOpen
  \bibfield  {author} {\bibinfo {author} {\bibfnamefont {P.~C.}\ \bibnamefont
  {Rathi}}, \bibinfo {author} {\bibfnamefont {R.~F.}\ \bibnamefont {Ludlow}},\
  and\ \bibinfo {author} {\bibfnamefont {M.~L.}\ \bibnamefont {Verdonk}},\
  }\href@noop {} {\bibfield  {journal} {\bibinfo  {journal} {J. Med. Chem.}\
  }\textbf {\bibinfo {volume} {63}},\ \bibinfo {pages} {8778} (\bibinfo {year}
  {2020})}\BibitemShut {NoStop}%
\bibitem [{\citenamefont {Watanabe}\ \emph {et~al.}(2018)\citenamefont
  {Watanabe}, \citenamefont {Tanaka}, \citenamefont {Yuki}, \citenamefont
  {Hirai},\ and\ \citenamefont {Yamamoto}}]{Watanabe2018-lp}%
  \BibitemOpen
  \bibfield  {author} {\bibinfo {author} {\bibfnamefont {K.}~\bibnamefont
  {Watanabe}}, \bibinfo {author} {\bibfnamefont {M.}~\bibnamefont {Tanaka}},
  \bibinfo {author} {\bibfnamefont {S.}~\bibnamefont {Yuki}}, \bibinfo {author}
  {\bibfnamefont {M.}~\bibnamefont {Hirai}},\ and\ \bibinfo {author}
  {\bibfnamefont {Y.}~\bibnamefont {Yamamoto}},\ }\href@noop {} {\bibfield
  {journal} {\bibinfo  {journal} {J. Clin. Biochem. Nutr.}\ }\textbf {\bibinfo
  {volume} {62}},\ \bibinfo {pages} {20} (\bibinfo {year} {2018})}\BibitemShut
  {NoStop}%
\bibitem [{\citenamefont {Hanwell}\ \emph {et~al.}(2012)\citenamefont
  {Hanwell}, \citenamefont {Curtis}, \citenamefont {Lonie}, \citenamefont
  {Vandermeersch}, \citenamefont {Zurek},\ and\ \citenamefont
  {Hutchison}}]{Hanwell2012-nf}%
  \BibitemOpen
  \bibfield  {author} {\bibinfo {author} {\bibfnamefont {M.~D.}\ \bibnamefont
  {Hanwell}}, \bibinfo {author} {\bibfnamefont {D.~E.}\ \bibnamefont {Curtis}},
  \bibinfo {author} {\bibfnamefont {D.~C.}\ \bibnamefont {Lonie}}, \bibinfo
  {author} {\bibfnamefont {T.}~\bibnamefont {Vandermeersch}}, \bibinfo {author}
  {\bibfnamefont {E.}~\bibnamefont {Zurek}},\ and\ \bibinfo {author}
  {\bibfnamefont {G.~R.}\ \bibnamefont {Hutchison}},\ }\href@noop {} {\bibfield
   {journal} {\bibinfo  {journal} {J. Cheminform.}\ }\textbf {\bibinfo {volume}
  {4}},\ \bibinfo {pages} {17} (\bibinfo {year} {2012})}\BibitemShut {NoStop}%
\bibitem [{\citenamefont {Halgren}(1996{\natexlab{a}})}]{Halgren1996-hg}%
  \BibitemOpen
  \bibfield  {author} {\bibinfo {author} {\bibfnamefont {T.~A.}\ \bibnamefont
  {Halgren}},\ }\href@noop {} {\bibfield  {journal} {\bibinfo  {journal} {J.
  Comput. Chem.}\ }\textbf {\bibinfo {volume} {17}},\ \bibinfo {pages} {616}
  (\bibinfo {year} {1996}{\natexlab{a}})}\BibitemShut {NoStop}%
\bibitem [{\citenamefont {Halgren}(1996{\natexlab{b}})}]{Halgren1996-jm}%
  \BibitemOpen
  \bibfield  {author} {\bibinfo {author} {\bibfnamefont {T.~A.}\ \bibnamefont
  {Halgren}},\ }\href@noop {} {\bibfield  {journal} {\bibinfo  {journal} {J.
  Comput. Chem.}\ }\textbf {\bibinfo {volume} {17}},\ \bibinfo {pages} {520}
  (\bibinfo {year} {1996}{\natexlab{b}})}\BibitemShut {NoStop}%
\bibitem [{\citenamefont {Halgren}(1996{\natexlab{c}})}]{Halgren1996-nn}%
  \BibitemOpen
  \bibfield  {author} {\bibinfo {author} {\bibfnamefont {T.~A.}\ \bibnamefont
  {Halgren}},\ }\href@noop {} {\bibfield  {journal} {\bibinfo  {journal} {J.
  Comput. Chem.}\ }\textbf {\bibinfo {volume} {17}},\ \bibinfo {pages} {553}
  (\bibinfo {year} {1996}{\natexlab{c}})}\BibitemShut {NoStop}%
\bibitem [{\citenamefont {Halgren}(1996{\natexlab{d}})}]{Halgren1996-ql}%
  \BibitemOpen
  \bibfield  {author} {\bibinfo {author} {\bibfnamefont {T.~A.}\ \bibnamefont
  {Halgren}},\ }\href@noop {} {\bibfield  {journal} {\bibinfo  {journal} {J.
  Comput. Chem.}\ }\textbf {\bibinfo {volume} {17}},\ \bibinfo {pages} {490}
  (\bibinfo {year} {1996}{\natexlab{d}})}\BibitemShut {NoStop}%
\bibitem [{\citenamefont {Halgren}\ and\ \citenamefont
  {Nachbar}(1996)}]{Halgren1996-zp}%
  \BibitemOpen
  \bibfield  {author} {\bibinfo {author} {\bibfnamefont {T.~A.}\ \bibnamefont
  {Halgren}}\ and\ \bibinfo {author} {\bibfnamefont {R.~B.}\ \bibnamefont
  {Nachbar}},\ }\href@noop {} {\bibfield  {journal} {\bibinfo  {journal} {J.
  Comput. Chem.}\ }\textbf {\bibinfo {volume} {17}},\ \bibinfo {pages} {587}
  (\bibinfo {year} {1996})}\BibitemShut {NoStop}%
\bibitem [{\citenamefont {Becke}(1993)}]{Becke1993-mi}%
  \BibitemOpen
  \bibfield  {author} {\bibinfo {author} {\bibfnamefont {A.~D.}\ \bibnamefont
  {Becke}},\ }\href@noop {} {\bibfield  {journal} {\bibinfo  {journal} {J.
  Chem. Phys}\ }\textbf {\bibinfo {volume} {98}},\ \bibinfo {pages} {5648}
  (\bibinfo {year} {1993})}\BibitemShut {NoStop}%
\bibitem [{\citenamefont {Stephens}\ \emph {et~al.}(1994)\citenamefont
  {Stephens}, \citenamefont {Devlin}, \citenamefont {Chabalowski},\ and\
  \citenamefont {Frisch}}]{Stephens1994-nk}%
  \BibitemOpen
  \bibfield  {author} {\bibinfo {author} {\bibfnamefont {P.~J.}\ \bibnamefont
  {Stephens}}, \bibinfo {author} {\bibfnamefont {F.~J.}\ \bibnamefont
  {Devlin}}, \bibinfo {author} {\bibfnamefont {C.~F.}\ \bibnamefont
  {Chabalowski}},\ and\ \bibinfo {author} {\bibfnamefont {M.~J.}\ \bibnamefont
  {Frisch}},\ }\href@noop {} {\bibfield  {journal} {\bibinfo  {journal} {The
  Journal of physical chemistry}\ }\textbf {\bibinfo {volume} {98}},\ \bibinfo
  {pages} {11623} (\bibinfo {year} {1994})}\BibitemShut {NoStop}%
\bibitem [{\citenamefont {Head-Gordon}\ \emph {et~al.}(1988)\citenamefont
  {Head-Gordon}, \citenamefont {Pople},\ and\ \citenamefont
  {Frisch}}]{Head-Gordon1988-ez}%
  \BibitemOpen
  \bibfield  {author} {\bibinfo {author} {\bibfnamefont {M.}~\bibnamefont
  {Head-Gordon}}, \bibinfo {author} {\bibfnamefont {J.~A.}\ \bibnamefont
  {Pople}},\ and\ \bibinfo {author} {\bibfnamefont {M.~J.}\ \bibnamefont
  {Frisch}},\ }\href@noop {} {\bibfield  {journal} {\bibinfo  {journal} {Chem.
  Phys. Lett.}\ }\textbf {\bibinfo {volume} {153}},\ \bibinfo {pages} {503}
  (\bibinfo {year} {1988})}\BibitemShut {NoStop}%
\bibitem [{\citenamefont {Du}\ \emph {et~al.}(2020)\citenamefont {Du},
  \citenamefont {Hsieh}, \citenamefont {Liu},\ and\ \citenamefont
  {Tao}}]{PhysRevResearch.2.033125}%
  \BibitemOpen
  \bibfield  {author} {\bibinfo {author} {\bibfnamefont {Y.}~\bibnamefont
  {Du}}, \bibinfo {author} {\bibfnamefont {M.-H.}\ \bibnamefont {Hsieh}},
  \bibinfo {author} {\bibfnamefont {T.}~\bibnamefont {Liu}},\ and\ \bibinfo
  {author} {\bibfnamefont {D.}~\bibnamefont {Tao}},\ }\href
  {https://doi.org/10.1103/PhysRevResearch.2.033125} {\bibfield  {journal}
  {\bibinfo  {journal} {Phys. Rev. Research}\ }\textbf {\bibinfo {volume}
  {2}},\ \bibinfo {pages} {033125} (\bibinfo {year} {2020})}\BibitemShut
  {NoStop}%
\bibitem [{\citenamefont {Sim}\ \emph {et~al.}(2019)\citenamefont {Sim},
  \citenamefont {Johnson},\ and\ \citenamefont
  {Aspuru‐Guzik}}]{sim_expressibility_2019}%
  \BibitemOpen
  \bibfield  {author} {\bibinfo {author} {\bibfnamefont {S.}~\bibnamefont
  {Sim}}, \bibinfo {author} {\bibfnamefont {P.~D.}\ \bibnamefont {Johnson}},\
  and\ \bibinfo {author} {\bibfnamefont {A.}~\bibnamefont {Aspuru‐Guzik}},\
  }\href {https://doi.org/10.1002/qute.201900070} {\bibfield  {journal}
  {\bibinfo  {journal} {Adv Quantum Tech}\ }\textbf {\bibinfo {volume} {2}},\
  \bibinfo {pages} {1900070} (\bibinfo {year} {2019})}\BibitemShut {NoStop}%
\bibitem [{\citenamefont {McClean}\ \emph {et~al.}(2018)\citenamefont
  {McClean}, \citenamefont {Boixo}, \citenamefont {Smelyanskiy}, \citenamefont
  {Babbush},\ and\ \citenamefont {Neven}}]{mcclean_barren_2018}%
  \BibitemOpen
  \bibfield  {author} {\bibinfo {author} {\bibfnamefont {J.~R.}\ \bibnamefont
  {McClean}}, \bibinfo {author} {\bibfnamefont {S.}~\bibnamefont {Boixo}},
  \bibinfo {author} {\bibfnamefont {V.~N.}\ \bibnamefont {Smelyanskiy}},
  \bibinfo {author} {\bibfnamefont {R.}~\bibnamefont {Babbush}},\ and\ \bibinfo
  {author} {\bibfnamefont {H.}~\bibnamefont {Neven}},\ }\href
  {https://doi.org/10.1038/s41467-018-07090-4} {\bibfield  {journal} {\bibinfo
  {journal} {Nat Commun}\ }\textbf {\bibinfo {volume} {9}},\ \bibinfo {pages}
  {4812} (\bibinfo {year} {2018})}\BibitemShut {NoStop}%
\bibitem [{\citenamefont {Zhang}\ \emph {et~al.}(2020)\citenamefont {Zhang},
  \citenamefont {Hsieh}, \citenamefont {Liu},\ and\ \citenamefont
  {Tao}}]{https://doi.org/10.48550/arxiv.2011.06258}%
  \BibitemOpen
  \bibfield  {author} {\bibinfo {author} {\bibfnamefont {K.}~\bibnamefont
  {Zhang}}, \bibinfo {author} {\bibfnamefont {M.-H.}\ \bibnamefont {Hsieh}},
  \bibinfo {author} {\bibfnamefont {L.}~\bibnamefont {Liu}},\ and\ \bibinfo
  {author} {\bibfnamefont {D.}~\bibnamefont {Tao}},\ }\href
  {https://doi.org/10.48550/ARXIV.2011.06258} {\bibinfo {title} {Toward
  trainability of quantum neural networks}} (\bibinfo {year}
  {2020})\BibitemShut {NoStop}%
\bibitem [{\citenamefont {Zhang}\ \emph {et~al.}(2021)\citenamefont {Zhang},
  \citenamefont {Hsieh}, \citenamefont {Liu},\ and\ \citenamefont
  {Tao}}]{https://doi.org/10.48550/arxiv.2112.15002}%
  \BibitemOpen
  \bibfield  {author} {\bibinfo {author} {\bibfnamefont {K.}~\bibnamefont
  {Zhang}}, \bibinfo {author} {\bibfnamefont {M.-H.}\ \bibnamefont {Hsieh}},
  \bibinfo {author} {\bibfnamefont {L.}~\bibnamefont {Liu}},\ and\ \bibinfo
  {author} {\bibfnamefont {D.}~\bibnamefont {Tao}},\ }\href
  {https://doi.org/10.48550/ARXIV.2112.15002} {\bibinfo {title} {Toward
  trainability of deep quantum neural networks}} (\bibinfo {year}
  {2021})\BibitemShut {NoStop}%
\bibitem [{\citenamefont {Zhang}\ \emph {et~al.}(2022)\citenamefont {Zhang},
  \citenamefont {Hsieh}, \citenamefont {Liu},\ and\ \citenamefont
  {Tao}}]{kay_2022}%
  \BibitemOpen
  \bibfield  {author} {\bibinfo {author} {\bibfnamefont {K.}~\bibnamefont
  {Zhang}}, \bibinfo {author} {\bibfnamefont {M.-H.}\ \bibnamefont {Hsieh}},
  \bibinfo {author} {\bibfnamefont {L.}~\bibnamefont {Liu}},\ and\ \bibinfo
  {author} {\bibfnamefont {D.}~\bibnamefont {Tao}},\ }\href
  {https://doi.org/10.48550/ARXIV.2203.09376} {\bibinfo {title} {Gaussian
  initializations help deep variational quantum circuits escape from the barren
  plateau}} (\bibinfo {year} {2022})\BibitemShut {NoStop}%
\bibitem [{\citenamefont {Amovilli}\ \emph {et~al.}(1998)\citenamefont
  {Amovilli}, \citenamefont {Barone}, \citenamefont {Cammi}, \citenamefont
  {Canc{\`e}s}, \citenamefont {Cossi}, \citenamefont {Mennucci}, \citenamefont
  {Pomelli},\ and\ \citenamefont {Tomasi}}]{Amovilli1998-ii}%
  \BibitemOpen
  \bibfield  {author} {\bibinfo {author} {\bibfnamefont {C.}~\bibnamefont
  {Amovilli}}, \bibinfo {author} {\bibfnamefont {V.}~\bibnamefont {Barone}},
  \bibinfo {author} {\bibfnamefont {R.}~\bibnamefont {Cammi}}, \bibinfo
  {author} {\bibfnamefont {E.}~\bibnamefont {Canc{\`e}s}}, \bibinfo {author}
  {\bibfnamefont {M.}~\bibnamefont {Cossi}}, \bibinfo {author} {\bibfnamefont
  {B.}~\bibnamefont {Mennucci}}, \bibinfo {author} {\bibfnamefont {C.~S.}\
  \bibnamefont {Pomelli}},\ and\ \bibinfo {author} {\bibfnamefont
  {J.}~\bibnamefont {Tomasi}},\ }in\ \href@noop {} {\emph {\bibinfo {booktitle}
  {Advances in Quantum Chemistry}}},\ Vol.~\bibinfo {volume} {32},\ \bibinfo
  {editor} {edited by\ \bibinfo {editor} {\bibfnamefont {P.-O.}\ \bibnamefont
  {L{\"o}wdin}}}\ (\bibinfo  {publisher} {Academic Press},\ \bibinfo {year}
  {1998})\ pp.\ \bibinfo {pages} {227--261}\BibitemShut {NoStop}%
\bibitem [{\citenamefont {Marenich}\ \emph {et~al.}(2009)\citenamefont
  {Marenich}, \citenamefont {Cramer},\ and\ \citenamefont
  {Truhlar}}]{Marenich2009-vg}%
  \BibitemOpen
  \bibfield  {author} {\bibinfo {author} {\bibfnamefont {A.~V.}\ \bibnamefont
  {Marenich}}, \bibinfo {author} {\bibfnamefont {C.~J.}\ \bibnamefont
  {Cramer}},\ and\ \bibinfo {author} {\bibfnamefont {D.~G.}\ \bibnamefont
  {Truhlar}},\ }\href@noop {} {\bibfield  {journal} {\bibinfo  {journal} {The
  Journal of Physical Chemistry B}\ }\textbf {\bibinfo {volume} {113}},\
  \bibinfo {pages} {6378} (\bibinfo {year} {2009})}\BibitemShut {NoStop}%
\bibitem [{\citenamefont {Klamt}\ and\ \citenamefont
  {Sch{\"u}{\"u}rmann}(1993)}]{Klamt1993-ra}%
  \BibitemOpen
  \bibfield  {author} {\bibinfo {author} {\bibfnamefont {A.}~\bibnamefont
  {Klamt}}\ and\ \bibinfo {author} {\bibfnamefont {G.}~\bibnamefont
  {Sch{\"u}{\"u}rmann}},\ }\href@noop {} {\bibfield  {journal} {\bibinfo
  {journal} {J. Chem. Soc. Perkin Trans. 2}\ ,\ \bibinfo {pages} {799}}
  (\bibinfo {year} {1993})}\BibitemShut {NoStop}%
\bibitem [{\citenamefont {Grant}\ \emph {et~al.}(2019)\citenamefont {Grant},
  \citenamefont {Wossnig}, \citenamefont {Ostaszewski},\ and\ \citenamefont
  {Benedetti}}]{grant_initialization_2019}%
  \BibitemOpen
  \bibfield  {author} {\bibinfo {author} {\bibfnamefont {E.}~\bibnamefont
  {Grant}}, \bibinfo {author} {\bibfnamefont {L.}~\bibnamefont {Wossnig}},
  \bibinfo {author} {\bibfnamefont {M.}~\bibnamefont {Ostaszewski}},\ and\
  \bibinfo {author} {\bibfnamefont {M.}~\bibnamefont {Benedetti}},\ }\bibfield
  {journal} {\bibinfo  {journal} {Quantum}\ }\textbf {\bibinfo {volume} {3}},\
  \href {https://doi.org/10.22331/q-2019-12-09-214} {10.22331/q-2019-12-09-214}
  (\bibinfo {year} {2019})\BibitemShut {NoStop}%
\bibitem [{\citenamefont {McArdle}\ \emph {et~al.}(2019)\citenamefont
  {McArdle}, \citenamefont {Jones}, \citenamefont {Endo}, \citenamefont {Li},
  \citenamefont {Benjamin},\ and\ \citenamefont
  {Yuan}}]{mcardle_variational_2019}%
  \BibitemOpen
  \bibfield  {author} {\bibinfo {author} {\bibfnamefont {S.}~\bibnamefont
  {McArdle}}, \bibinfo {author} {\bibfnamefont {T.}~\bibnamefont {Jones}},
  \bibinfo {author} {\bibfnamefont {S.}~\bibnamefont {Endo}}, \bibinfo {author}
  {\bibfnamefont {Y.}~\bibnamefont {Li}}, \bibinfo {author} {\bibfnamefont
  {S.~C.}\ \bibnamefont {Benjamin}},\ and\ \bibinfo {author} {\bibfnamefont
  {X.}~\bibnamefont {Yuan}},\ }\href
  {https://doi.org/10.1038/s41534-019-0187-2} {\bibfield  {journal} {\bibinfo
  {journal} {npj Quantum Inf}\ }\textbf {\bibinfo {volume} {5}},\ \bibinfo
  {pages} {75} (\bibinfo {year} {2019})}\BibitemShut {NoStop}%
\end{thebibliography}%

\appendix

\section{Noise Model}
\label{appendix:Noise Model}

For the noisy simulation of VQE in this work, we implemented the 8-qubit noise model using the Qiskit package. We have employed the thermal relaxation errors for the simulation where the relaxation time constants are shown in Table~\ref{tab:RelaxationConstantsAcetoneKeto} and Table~\ref{tab:RelaxationConstantsAcetoneEnol} and the gate time and measurement time are shown in Table~\ref{tab:gateTime}.

\begin{table}[hbt!]
\centering
\caption{Relaxation time constants of the 8-qubit noise model used for noisy VQE simulation of acetone.}
\begin{ruledtabular}
\begin{tabular}{ccccccccc}
 & Qubit 0 & Qubit 1 & Qubit 2 & Qubit 3 & Qubit 4 & Qubit 5 & Qubit 6 & Qubit 7 \\
\hline
$T_1$ (ms) & 194.24	& 230.20	& 218.81	& 212.36	& 189.67	& 198.52	& 230.69	& 153.56 \\
$T_2$ (ms) & 188.28	& 219.84	& 226.43	& 165.82	& 236.03	& 204.83	& 196.17	& 166.16 \\
\end{tabular}
\end{ruledtabular}
\label{tab:RelaxationConstantsAcetoneKeto}
\end{table}

\begin{table}[hbt!]
\centering
\caption{Relaxation time constants of the 8-qubit noise model used for noisy VQE simulation of propen-2-ol.}
\begin{ruledtabular}
\begin{tabular}{ccccccccc}
 & Qubit 0 & Qubit 1 & Qubit 2 & Qubit 3 & Qubit 4 & Qubit 5 & Qubit 6 & Qubit 7 \\
\hline

$T_1$ (ms) & 134.45	& 147.61	& 208.15	& 211.67	& 170.22	& 208.24	& 200.66	& 242.95 \\
$T_2$ (ms) & 60.85	& 234.66	& 178.45	& 210.15	& 224.01	& 178.00	& 105.58	& 136.21 \\

\end{tabular}
\end{ruledtabular}
\label{tab:RelaxationConstantsAcetoneEnol}
\end{table}

\begin{table}[hbt!]
\centering
\caption{Gate time and measurement time in nanoseconds of the 8-qubit noise models used for noisy VQE simulation of acetone and propen-2-ol.}
\begin{ruledtabular}
\begin{tabular}{cccccc}
$U_1$ gate & $U_2$ gate & $U_3$ gate & CNOT gate & Reset time & Measurement time \\
\hline

0 & 50 & 100 & 200 & 1000 & 1000 \\ 

\end{tabular}
\end{ruledtabular}
\label{tab:gateTime}
\end{table}
\pagebreak

%\newpage
\section{Natural Orbital Occupancy}
\label{appendix:Molecular Orbital Occupancy}
Natural orbital occupancies were used to select active molecule orbital (MOs) sets in this work. The natural occupancy reflects the occupation state of a given orbital. With the natural orbital occupation numbers, we reduce the MO sets by removing those orbitals that are nearly virtual (natural orbital occupancy close to 0) or fully occupied (natural orbital occupancy close to 2), which are not likely involved in the tautomerization process.

Table~\ref{Ace_eigenenergy_occupancy_table} and Table~\ref{Eda_eigenenergy_occupancy_table} show the MO energies and the natural orbital occupancies of acetone and Edaravone systems using MP2 calculations. 

\begin{table}[hbt!]
    \centering\ 
    \caption{The eigenvalues of the MOs and the occupancies of the natural orbitals of the acetone tautomers calculated at the MP2/STO-3G level.}
    
\begin{ruledtabular}
\begin{tabular}{lrrrr}
{} & \multicolumn{2}{c}{Keto} & \multicolumn{2}{c}{Enol} \\
\cline{2-3} \cline{4-5}
MO &  Eigenvalue (Hatree) &  Occupancy &  Eigenvalue (Hatree) &  Occupancy \\
\hline
0  &            -20.26834 &   2.000004 &              -20.28742 &     2.000003 \\
1  &            -11.12074 &   2.000002 &              -11.08992 &     2.000002 \\
2  &            -11.04974 &   2.000000 &              -11.05175 &     2.000000 \\
3  &            -11.04853 &   2.000000 &              -10.98477 &     2.000000 \\
4  &             -1.32547 &   1.997842 &               -1.32047 &     1.998189 \\
5  &             -0.98289 &   1.993911 &               -0.99634 &     1.995010 \\
6  &             -0.91791 &   1.992408 &               -0.89618 &     1.992262 \\
7  &             -0.68101 &   1.991211 &               -0.71645 &     1.990954 \\
8  &             -0.59048 &   1.987791 &               -0.61208 &     1.988330 \\
9  &             -0.57718 &   1.987463 &               -0.58038 &     1.987414 \\
10 &             -0.57548 &   1.987179 &               -0.55659 &     1.987216 \\
11 &             -0.52423 &   1.987075 &               -0.50819 &     1.987117 \\
12 &             -0.49429 &   1.986979 &               -0.46796 &     1.985933 \\
13 &             -0.48573 &   1.983741 &               -0.45614 &     1.982110 \\
14 &             -0.39341 &   1.978252 &               -0.43099 &     1.979630 \\
15 &             -0.32185 &   1.942575 &               -0.26688 &     1.950616 \\
16 &              0.29921 &   0.061559 &                0.33083 &     0.054016 \\
17 &              0.61656 &   0.021915 &                0.56090 &     0.020035 \\
18 &              0.66220 &   0.016135 &                0.61885 &     0.016292 \\
19 &              0.67844 &   0.015970 &                0.69870 &     0.015365 \\
20 &              0.68052 &   0.012180 &                0.71459 &     0.013645 \\
21 &              0.71925 &   0.011860 &                0.71780 &     0.012069 \\
22 &              0.73087 &   0.011671 &                0.72110 &     0.011897 \\
23 &              0.73720 &   0.011361 &                0.75222 &     0.011141 \\
24 &              0.81362 &   0.010535 &                0.93500 &     0.010535 \\
25 &              1.02163 &   0.010382 &                1.04183 &     0.010220 \\
\end{tabular}
\end{ruledtabular}
    \label{Ace_eigenenergy_occupancy_table}
\end{table}

\begin{table}[hbt!]
    \centering\ 
    \caption{The eigenvalues of the MOs and the occupancies of the natural orbitals of the Edaravone tautomers calculated at the MP2/STO-3G level.}
\begin{ruledtabular}
\scriptsize
\begin{tabular}{lrrrrrr}
{} & \multicolumn{2}{c}{Keto} & \multicolumn{2}{c}{Enol} & \multicolumn{2}{c}{Amine}\\
\cline{2-3} \cline{4-5} \cline{6-7}
MO &  Eigenvalue &  Occupancy &  Eigenvalue &  Occupancy &  Eigenvalue &  Occupancy \\
\hline
0  &            -20.26132 &   2.000007 &              -20.3286 &     2.000007 &              -20.21207 &     2.000004 \\
1  &            -15.41341 &   2.000004 &              -15.4416 &     2.000002 &              -15.41433 &     2.000002 \\
2  &            -15.39275 &   2.000002 &              -15.3295 &     2.000002 &              -15.39921 &     2.000002 \\
3  &            -11.17214 &   2.000002 &               -11.13800 &     2.000002 &              -11.14758 &     2.000002 \\
4  &            -11.09193 &   2.000002 &              -11.0931 &     2.000002 &              -11.10969 &     2.000001 \\
5  &            -11.08892 &   2.000001 &              -11.0503 &     2.000001 &              -11.08820 &     2.000001 \\
6  &            -11.08030 &   2.000001 &              -11.0441 &     2.000001 &              -11.08076 &     2.000001 \\
7  &            -11.07000 &   2.000001 &              -11.0357 &     2.000001 &              -11.03183 &     2.000001 \\
8  &            -11.02927 &   2.000001 &              -11.0351 &     2.000001 &              -11.03030 &     2.000001 \\
9  &            -11.02916 &   2.000000 &               -11.03500 &     2.000001 &              -11.02411 &     2.000000 \\
10 &            -11.02174 &   2.000000 &              -11.0335 &     2.000000 &              -11.02210 &     2.000000 \\
11 &            -11.02090 &   2.000000 &              -11.0315 &     2.000000 &              -11.02121 &     2.000000 \\
12 &            -11.02028 &   2.000000 &              -10.9921 &     2.000000 &              -11.01432 &     2.000000 \\
13 &             -1.35162 &   1.998050 &              -1.36683 &     1.998025 &               -1.32861 &     1.998176 \\
14 &             -1.28962 &   1.995627 &              -1.29727 &     1.995976 &               -1.27162 &     1.994969 \\
15 &             -1.11989 &   1.993057 &              -1.11338 &     1.994566 &               -1.11756 &     1.993101 \\
16 &             -1.06485 &   1.992505 &              -1.05489 &     1.992798 &               -1.06206 &     1.992443 \\
17 &             -0.99215 &   1.992335 &              -0.98405 &     1.992418 &               -1.00022 &     1.992341 \\
18 &             -0.96551 &   1.991838 &              -0.96177 &     1.991945 &               -0.96884 &     1.991857 \\
19 &             -0.95409 &   1.990887 &              -0.95266 &     1.990953 &               -0.95565 &     1.990873 \\
20 &             -0.91721 &   1.990643 &               -0.90150 &     1.990852 &               -0.91324 &     1.990719 \\
21 &             -0.79819 &   1.989859 &              -0.79539 &     1.989632 &               -0.78523 &     1.990324 \\
22 &             -0.77423 &   1.988589 &             -0.77676 &     1.988703 &               -0.77646 &     1.989263 \\
23 &             -0.72782 &   1.987920 &              -0.73127 &     1.988055 &               -0.73674 &     1.988787 \\
24 &             -0.69806 &   1.987380 &              -0.69621 &     1.987876 &               -0.71001 &     1.988546 \\
25 &             -0.64911 &   1.987330 &              -0.68063 &     1.987247 &               -0.66793 &     1.987863 \\
26 &             -0.62823 &   1.987190 &              -0.63355 &     1.987126 &               -0.63605 &     1.987103 \\
27 &             -0.62162 &   1.986792 &              -0.60903 &     1.986818 &               -0.61366 &     1.986961 \\
28 &             -0.58933 &   1.986318 &              -0.59742 &     1.986296 &               -0.58833 &     1.986617 \\
29 &             -0.56595 &   1.985577 &              -0.57066 &     1.985343 &               -0.56420 &     1.986337 \\
30 &             -0.55252 &   1.985491 &              -0.55774 &     1.983740 &               -0.56325 &     1.985574 \\
31 &             -0.54924 &   1.983691 &              -0.54279 &     1.983326 &               -0.55100 &     1.983833 \\
32 &             -0.53899 &   1.983332 &              -0.53714 &     1.983290 &               -0.53788 &     1.983331 \\
33 &             -0.52916 &   1.983207 &              -0.52954 &     1.981034 &               -0.53218 &     1.983047 \\
34 &             -0.52595 &   1.980687 &              -0.51471 &     1.980603 &               -0.52911 &     1.980885 \\
35 &             -0.51468 &   1.979081 &              -0.49153 &     1.979242 &               -0.50496 &     1.979233 \\
36 &             -0.49568 &   1.978694 &              -0.47726 &     1.978879 &               -0.48472 &     1.978964 \\
37 &             -0.45479 &   1.978138 &              -0.46014 &     1.977825 &               -0.45968 &     1.978061 \\
38 &             -0.44884 &   1.977486 &              -0.44551 &     1.977524 &               -0.45128 &     1.976938 \\
39 &             -0.43422 &   1.975835 &               -0.44170 &     1.974787 &               -0.43474 &     1.975575 \\
40 &             -0.41856 &   1.972720 &              -0.43297 &     1.971615 &               -0.42470 &     1.974069 \\
41 &             -0.39887 &   1.968219 &              -0.35506 &     1.968096 &               -0.35961 &     1.968380 \\
42 &             -0.32380 &   1.947555 &              -0.33103 &     1.958506 &               -0.31246 &     1.957051 \\
43 &             -0.32217 &   1.941835 &              -0.28577 &     1.941071 &               -0.28196 &     1.939416 \\
44 &             -0.28022 &   1.938252 &              -0.24767 &     1.937216 &               -0.26851 &     1.938023 \\
45 &             -0.22348 &   1.933366 &              -0.24342 &     1.934966 &               -0.21386 &     1.934162 \\
46 &              0.25533 &   0.068063 &               0.23986 &     0.073750 &                0.23389 &     0.071535 \\
47 &              0.25768 &   0.065302 &               0.26492 &     0.064523 &                0.26903 &     0.065091 \\
48 &              0.27132 &   0.064662 &               0.32588 &     0.064330 &                0.28245 &     0.062220 \\
49 &              0.35220 &   0.059913 &               0.38423 &     0.051273 &                0.40600 &     0.048977 \\
50 &              0.51155 &   0.032174 &               0.49939 &     0.031852 &                0.50366 &     0.032236 \\
51 &              0.51424 &   0.026326 &               0.53225 &     0.026128 &                0.52337 &     0.025914 \\
52 &              0.54395 &   0.023414 &               0.53709 &     0.022501 &                0.57065 &     0.022931 \\
53 &              0.59018 &   0.021915 &               0.60062 &     0.021388 &                0.57568 &     0.022116 \\
54 &              0.61087 &   0.021247 &               0.60867 &     0.020487 &                0.61116 &     0.020147 \\
55 &              0.65256 &   0.019565 &               0.64232 &     0.019322 &                0.62894 &     0.018357 \\
56 &              0.67183 &   0.018364 &               0.66297 &     0.018339 &                0.65775 &     0.018193 \\
57 &              0.67780 &   0.018049 &               0.68513 &     0.018043 &                0.66583 &     0.018053 \\
58 &              0.67839 &   0.017572 &               0.69013 &     0.017615 &                0.68247 &     0.017503 \\
59 &              0.68719 &   0.017058 &               0.70808 &     0.016812 &                0.69122 &     0.016896 \\
60 &              0.71836 &   0.016353 &               0.72472 &     0.015515 &                0.70617 &     0.016056 \\
61 &              0.72364 &   0.015473 &               0.72612 &     0.015122 &                0.71606 &     0.015525 \\
62 &              0.72700 &   0.014330 &               0.73009 &     0.014612 &                0.73061 &     0.014222 \\
63 &              0.73249 &   0.012992 &               0.74037 &     0.014018 &                0.73815 &     0.013586 \\
64 &              0.73670 &   0.012567 &               0.76168 &     0.013661 &                0.76211 &     0.013127 \\
65 &              0.77141 &   0.012296 &               0.81055 &     0.012333 &                0.77341 &     0.012328 \\
66 &              0.85449 &   0.012056 &               0.85656 &     0.011997 &                0.87823 &     0.012037 \\
67 &              0.87814 &   0.012016 &               0.88338 &     0.011977 &                0.88962 &     0.011974 \\
68 &              0.89312 &   0.011976 &               0.88839 &     0.011952 &                0.89827 &     0.011929 \\
69 &              0.90300 &   0.011815 &               0.90412 &     0.011807 &                0.90475 &     0.011797 \\
70 &              0.93342 &   0.011758 &               0.94604 &     0.011744 &                0.92936 &     0.011747 \\
71 &              0.98338 &   0.011587 &                1.00630 &     0.011687 &                0.97650 &     0.011703 \\
72 &              1.04928 &   0.011067 &               1.08665 &     0.010419 &                1.09233 &     0.010538 \\
73 &              1.09713 &   0.010395 &               1.11121 &     0.010313 &                1.12981 &     0.010254 \\
74 &              1.15351 &   0.010190 &               1.14982 &     0.010110 &                1.15709 &     0.010169 \\
\end{tabular}
\end{ruledtabular}
    \label{Eda_eigenenergy_occupancy_table}
\end{table}

\section{Selective combination of active molecular Orbitals}
\label{appendix:Active_MO}

A selection of active molecular orbital sets were chosen and their corresponding ground state energies were calculated at the level of CCSD theory for all tautomers of the two example.

Table~\ref{tab:Acetone_activeMO} and Table~\ref{tab:Edavarone_activeMO} show the relative CCSD energies for several active-space MO of the acetone and Edaravone systems.

\begin{table}[hbt!]
    \centering\ 
    \caption{Active space selection of molecular orbitals for acetone. Relative ground state energies of the keto, and enol forms of acetone were calculated for selected combinations of molecular orbitals. Listed ground state energies are relative to the keto form. The values are shown in kcal/mol.}
\begin{ruledtabular}
\begin{tabular}{lccccc}
Active MO &  \# of MO &  \# of SO &  \# of e &  Keto &    Enol \\
\hline
0-25      &       26 &       52 &      32 &     0 &  24.070 \\
10-19     &       10 &       20 &      12 &     0 &  18.716 \\
11-20     &       10 &       20 &      10 &     0 &  22.478 \\
11-19     &        9 &       18 &      10 &     0 &  21.260 \\
12-20     &        9 &       18 &       8 &     0 &  21.374 \\
12-19     &        8 &       16 &       8 &     0 &  19.888 \\
13-19     &        7 &       14 &       6 &     0 &  20.325 \\
13-18     &        6 &       12 &       6 &     0 &  20.565 \\
14-19     &        6 &       12 &       4 &     0 &  23.766 \\
\end{tabular}
\end{ruledtabular}
    \label{tab:Acetone_activeMO}
\end{table}
\pagebreak

\begin{table}[hbt!]
    \centering\ 
    \caption{Active space selection of molecular orbitals for Edaravone. Relative ground state energies of the keto, enol and amine forms of Edaravone were calculated for selected combinations of molecular orbitals. Listed ground state energies are relative to the keto form. The values are shown in kcal/mol.}
\begin{ruledtabular}
\begin{tabular}{lcccccc}
Active MO &  \# of MO &  \# of SO &  \# of e &  Keto &    Enol &   Amine \\
\hline
0-74      &       75 &      150 &      92 &     0 &  13.726 &  25.947 \\
41-50     &       10 &       20 &      10 &     0 & -12.793 &   4.904 \\
41-49     &        9 &       18 &      10 &     0 & -12.959 &   6.556 \\
42-50     &        9 &       18 &       8 &     0 & -12.650 &  23.293 \\
42-49     &        8 &       16 &       8 &     0 & -12.874 &  22.420 \\
43-49     &        7 &       14 &       6 &     0 &   8.583 &  23.892 \\
43-48     &        6 &       12 &       6 &     0 &   5.667 &  18.618 \\
44-49     &        6 &       12 &       4 &     0 &   9.197 &  25.398 \\
\end{tabular}
\end{ruledtabular}
    \label{tab:Edavarone_activeMO}
\end{table}

\end{document}